\documentclass[a4paper,10pt,reprint,
bibnotes,amsmath,amssymb,aps,
pre]{revtex4-2}
\usepackage[utf8]{inputenc}
\usepackage{subcaption}
\usepackage{graphicx}

\begin{document}

\title{The structure of clusters formed by Stockmayer supracolloidal magnetic polymers}
\author{Ekaterina V. Novak*, Elena S. Pyanzina}
\affiliation{Ural Federal University, Lenin Av. 51, 620000, Ekaterinburg, Russia }%

\author{Pedro A. S{\'a}nchez}
\affiliation{%
Helmholtz-Zentrum Dresden-Rossendorf, Bautzner Landstrasse 400, 01328 Dresden, Germany\\
 Ural Federal University, Lenin Av. 51, 620000, Ekaterinburg, Russia}

\author{Sofia S. Kantorovich}
\affiliation{%
 University of Vienna, Sensengasse 8, 1090, Vienna, Austria\\
 Ural Federal University, Lenin Av. 51, 620000, Ekaterinburg, Russia}

\begin{abstract}
Unlike Stockmayer fluids, that prove to undergo gas-liquid transition on cooling, the system of dipolar hard or soft spheres without any additional central attraction so far has not been shown to  have a critical point. Instead, in the latter, one observes diverse self-assembly scenarios. Crosslinking dipolar soft spheres into supracolloidal magnetic polymer-like structures (SMPs) changes the self-assembly behaviour. Moreover, aggregation in  systems of SMPs strongly depends on the constituent topology. For Y- and X-shaped SMPs, under the same conditions in which dipolar hard spheres would form chains, the formation of very large loose gel-like clusters was observed [Journal  of  Molecular Liquids, 271, 631  (2018)].  In this work, using molecular dynamics simulations, we investigate self-assembly in suspensions of four topologically different SMPs -- chains, rings, X and Y -- monomers in which interact via Stockmayer potential.  As expected, compact drop-like clusters are formed by SMPs in all cases if the central isotropic attraction is introduced, however, their shape and internal structure turn out to depend on the SMPs topology.
\end{abstract}

\maketitle

\section{\label{sec:level1}Introduction}
Dipolar interactions alone seem to not lead to the classical vapour-liquid phase transition in systems of spherical magnetic particles \cite{1996-roij,1996-sear,1999-levin,teixeira00a,weis93a,weis93b,safrannew,rovigatti11a,2013-kantorovich-prl}. Instead, magnetic dipole-dipole interaction is responsible for extensive self-assembly phenomenon. Thorough investigations of dipolar systems carried out for more than 30 years showed that the most probable cluster topologies observed in dipolar hard sphere systems are chains, rings, Y- and X-junctions \cite{2000-camp-prl,2006-klokkenburg,2017-ronti}. On decreasing temperature and growing concentration, those structures of magnetic particles form loose networks \cite{rovigatti11a,2015-kantorovich-pccp1}.

Modern experimental techniques allow the stabilisation of dipolar clusters by polymer crosslinking, forming supracolloidal magnetic polymer-like structures \cite{2006-korth,2007-keng,2015-bharti,2017-hongy}. In general, the behaviour of chain-like magnetic SMPs has been actively investigated in experiment \cite{2005-dreyfus,2008-choi,2008-erglis-jpcm,2008-zhou}, theory \cite{2005-cebers,2009-belovs-pre,2011-javaitis} and coarse-grained computer simulations \cite{2011-sanchez-sm,2016-cerda-pccp,2016-wei,2019-kuznetsov}. All these works agree that SMPs represent a promise for potential medical and microfluidics application \cite{2011-devicente,2010-park}. In a recent study, we generalised the analysis of SMPs to the rest of basic structures formed by self-assembly of dipolar spheres \cite{rozhkov17a,2018-novak}. The interest to study rings-, X- and Y-like SMPs is twofold. First, once synthesised, such structures can be useful building blocks for novel magnetic soft materials. Second, they offer a new approach to the theoretical study of self-assembled structures.  In-silico investigations \cite{2018-novak} showed basically no self-assembly in suspensions of ring-shaped SMPs; chain-like SMPs formed linear clusters, whose size distribution exhibited an exponential decay. In contrast, Y and X SMPs form very large but rather loose clusters with up to 90 per cent of system SMPs connected in them. Such clusters can be considered as precursors for a phase transition. These results agree well with the phase behaviour observed for patchy particles of different valency \cite{bianchi2011patchy}, where the authors show that the gas-liquid critical point shifts towards lower temperature if the valency decreases. In fact, linear SMPs can be associated with the patchy particles of valency two, whereas Y- and X-like SMPs exhibit behaviour analogous to patchy particles with valency three and four, respectively.

Another actively studied system whose behaviour is largely defined by dipole-dipole interactions is Stockmayer fluid.  Additionally to dipolar forces, Stockmayer particles experience short-range isotropic attraction. The properties of these systems were investigated in detail by many authors and it was found that Stockmayer fluid undergo vapour-liquid phase transition on cooling and concentration increase \cite{1989-smit,1993-vanleeuwen,1992-panagiotopoulos,1995-stevens,1981-adams}. Thus, colloidal particles form compact isotropic clusters.

The idea of the present study is to combine our knowledge about self-assembly of SMPs and Stockmayer fluids and elucidate the self-assembly of SMPs of different topologies, when magnetic particles (SMP monomers) composing them experience an additional central attraction. 

The structure of the manuscript is the following. First, we introduce the model and describe methods. When discussing results, we start with discussing macroscopic characteristics, such as cluster size distributions and cluster shape description.  In order to explain the differences brought up by topology into the cluster shape, as the next step, we study the structure of clusters and report the orientation of individual SMPs, as well as their dipole moments in the cluster. Next, we go down to the particle (monomer) resolution and analyse the neighbourhoods of monomers in the cluster and monomer dipolar orientations. Finally, we compare the properties of the clusters formed by SMPs to those found in a non-crosslinked Stockmayer fluid.

\section{Model and simulation details}\label{sec-model}
For modelling a dispersion of SMPs, we used a bead-spring model with a system of reduced units. We assume that a SMP consists of monodisperse ferromagnetic spherical particles of diameter $\sigma=1$ and mass $m=1$, with a permanent magnetic moment represented by a point dipole, $\vec \mu$, placed in the particle centre. The long-range magnetic interaction between any pair $i$, $j$ of magnetic particles (SMP monomers) is described by the dipole-dipole potential:
\begin{equation}
U_{dd}(\vec r_{ij})=\frac{\vec{\mu}_{i}\cdot\vec{\mu}_{j}}{r^{3}}-\frac{3\left[\vec{\mu}_{i}\cdot\vec{r}_{ij}\right]\left[\vec{\mu}_{j}\cdot\vec{r}_{ij}\right]}{r^{5}},
\label{eq:dipdip}
\end{equation}
where $\vec \mu_i$ and $\vec \mu_j$ are their respective dipole moments, $\vec r_{ij} = \vec r_i - \vec r_j$ is the displacement vector connecting their centres and $r=\left | \vec r _{ij}\right |$. 

To model the central attraction between the monomers of SMP, Lennard-Jones potential is used that, considering $\sigma=1$ and setting energy scale to unity, can be written as:
\begin{equation}
U_{\mathrm{{LJ}}}(r)=4 \left ( r^{-12}-r^{-6} \right ). 
\label{eq:LJ}
\end{equation}

\begin{figure}[h!]
\centering{\includegraphics[width=4.5cm]{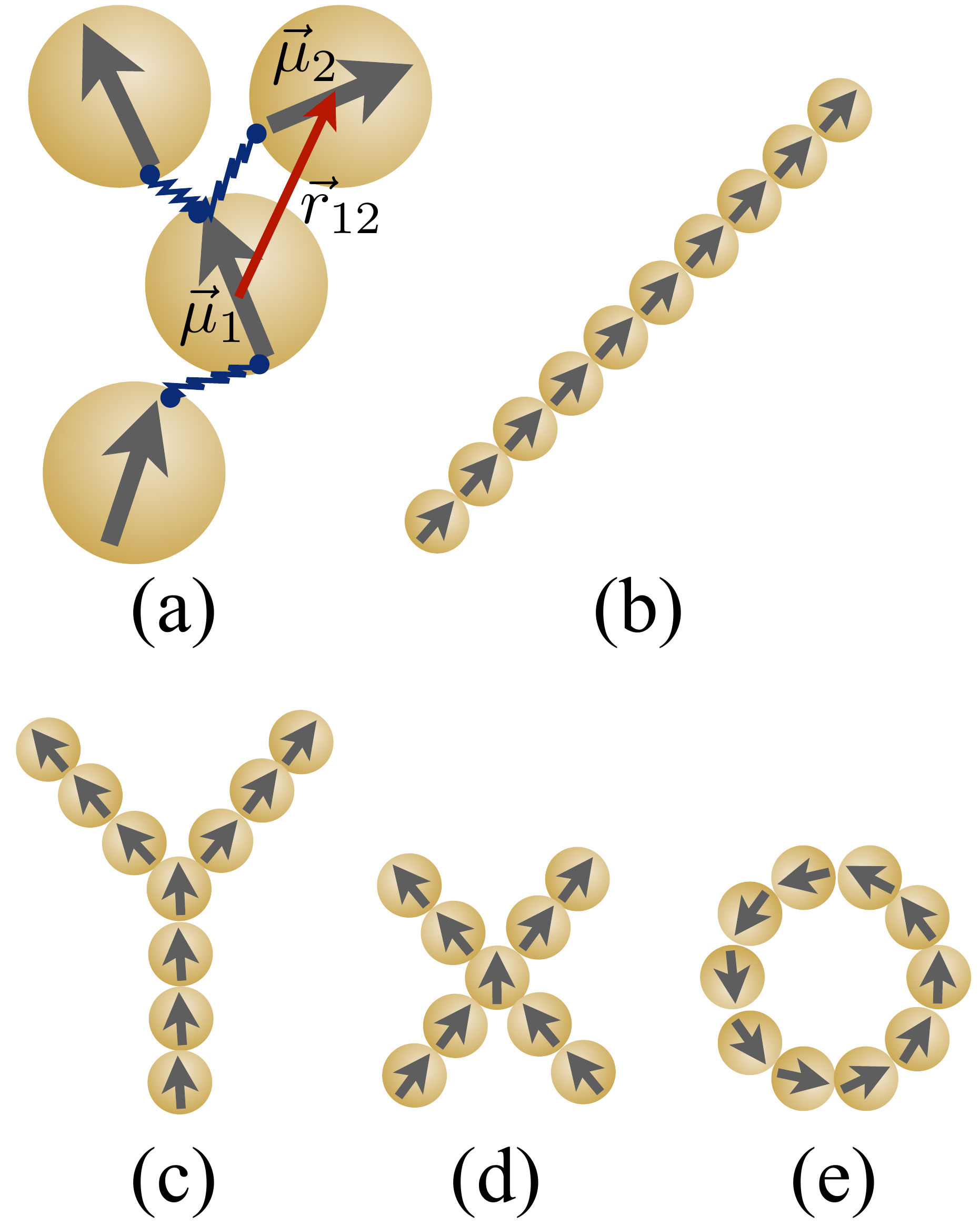}}
\caption{Explanation of the model. (a) Crosslinking and magnetic moments. (b) chain-like SMPs; (c) Y-like SMPs; (d) X-like SMPs; (e) ring-like SMPs}
\label{fig:shapes}
\end{figure}

The bonding between crosslinked monomers within every polymer is modelled as a pair potential with two terms. The first term is a simple harmonic spring whose ends are attached to the monomer surfaces, as shown in Fig. \ref{fig:shapes} (a). The spring attachment points are placed at the projection points of the head and the tail of the central dipole moment. The second term corresponds to a FENE potential that limits the maximum extension of the bond. Therefore, the bonding potential is defined as:
\begin{equation}
   U_{S}(\vec r_{ij}) = \frac{K}{2} \left [ \left( \vec r_{ij} - \frac{1}{2}(\hat{\mu}_i+\hat{\mu}_j) \right )^2 -\right.
   \label{eq:bondmodel}
\end{equation}
  
 $$\left . - \frac{r^2_0}{2}\ln\left[1-\left(\frac {\vec r_{ij}}{r_0} \right)^2 \right] \right],$$
where $K$ is the energy scale of the interaction, $\hat{\mu}_i=\vec \mu_i / \left | \vec \mu_i\right |$ and $\hat{\mu}_j=\vec \mu_j / \left | \vec \mu_j\right |$ are the unitary vectors parallel to each associated dipole moment and $r_0$ is the maximum allowed extension for the bond. We take $K=30$ and $r_0=1.5$ in reduced units in accordance with our previous studies\cite{2017-novak-jmmmb, rozhkov17a}.

We performed molecular dynamics simulations in the canonical ensemble, using a Langevin thermostat  in order to approximate implicitly the effects of the thermal fluctuations of the background fluid. Periodic boundary conditions were used.  The simulations were performed in the ESPResSo 3.2.0 package \cite{2006-limbach}. The initial simulation box contained 512 identical magnetic polymers with size either $L = 10$ for chain-like, Y-like and ring-like SMPs or $L=9$ for X-structures in their basic configurations as shown in Figure\ \ref{fig:shapes} (b)-(e). The dimensionless reduced concentration of monomers was always fixed to $\rho^{\ast} = N\sigma/V = 0.05$, where the number of monomers, $N$, and the volume of the simulation box, $V$, are identical for all but X-like SMPs. SMPs were initially placed parallel to each other at random positions. Magnetic moment of each monomer $\mu$ in dimensionless units was fixed so that $\mu^2 = 5$. The system was first equilibrated at high $T=4$ to assure random initial configuration before switching on magnetic interactions and central attraction. Afterwards, before the production runs were performed, the system was re-equilibrated at $T=1$ for $9 \cdot 10^5$ integration steps, using a time step $\delta t =5 \cdot 10 ^{-3}$. Finally, a production cycle of $3 \cdot 10^6$ steps was performed, in which the system configurations were measured at intervals of $10 ^5$ steps. Both the starting point and the length of the production cycle were chosen so that the total energy of the system has reached the constant value and started fluctuating around its mean value by less then five per cent already before the starting point and kept this property throughout the production. The long range magnetic interactions were calculated using the dipolar-P$^3$M algorithm \cite{2008-cerda-jcp}.

\begin{figure}[h]
   \includegraphics[width=0.75\columnwidth]{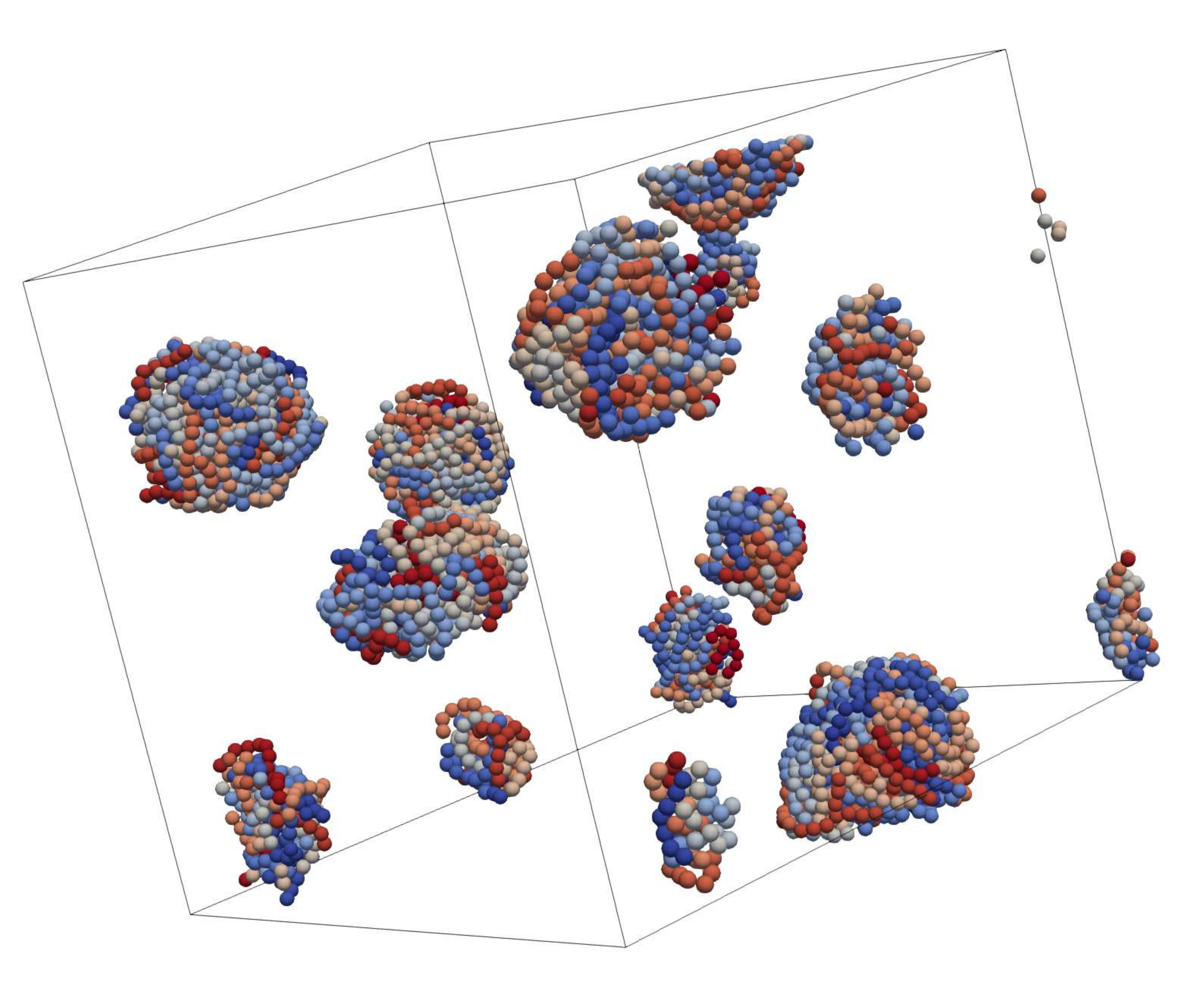}
   \caption{Typical simulation snapshot. Clusters formed by Y-like SMPs. Different colours help to distinguish  different SMPs.}
   \label{fig:snap_solution}
\end{figure}

A typical simulation snapshot during the production is presented in Fig. \ref{fig:snap_solution}. This very snapshot is taken from the system of Y-like SMPs, but it is impossible to visually spot the difference between the snapshots taken from other suspensions.  It can be seen that the system is composed by quasi-spherical aggregates containing multiple SMPs. Besides this apparent overall resemblance, we want to determine whether the clusters have the same internal structure independently from the SMPs topology. For this we performed systematic cluster analysis of the obtained aggregates. In all cases we identify the non permanent connections between the monomers by means of a combination of distance or distance and energy criteria. In both cases the two monomers are considered to be connected if their centre-to-centre distance is smaller than $r_{ij} \le 2^{1/6}$; in the second case, additionally their dipole-dipole pair energy, given by expression \ref{eq:dipdip}, $U_{dd}(\vec r_{cut}; \vec \mu_1, \vec \mu_2)<0$, should be negative.

\section{Results and Discussions}\label{sec-res}
In order to thoroughly investigate the internal structure of the clusters we will increase the resolution step by step from individual clusters, through SMPs within them to monomers. In other words, first, the cluster shapes for each SMP topology will be analysed. Next, we will study the orientation and positioning of individual SMPs within clusters and the orientations of their magnetic moments. Finally, going to the individual particle level, we will elucidate the neighbourhoods of monomers and the orientations of their dipole moments within the clusters. 

\subsection{Cluster distributions and shapes}

\begin{figure*}[th!]
    \begin{subfigure}[b]{0.22\textwidth}
        \includegraphics[width=\textwidth]{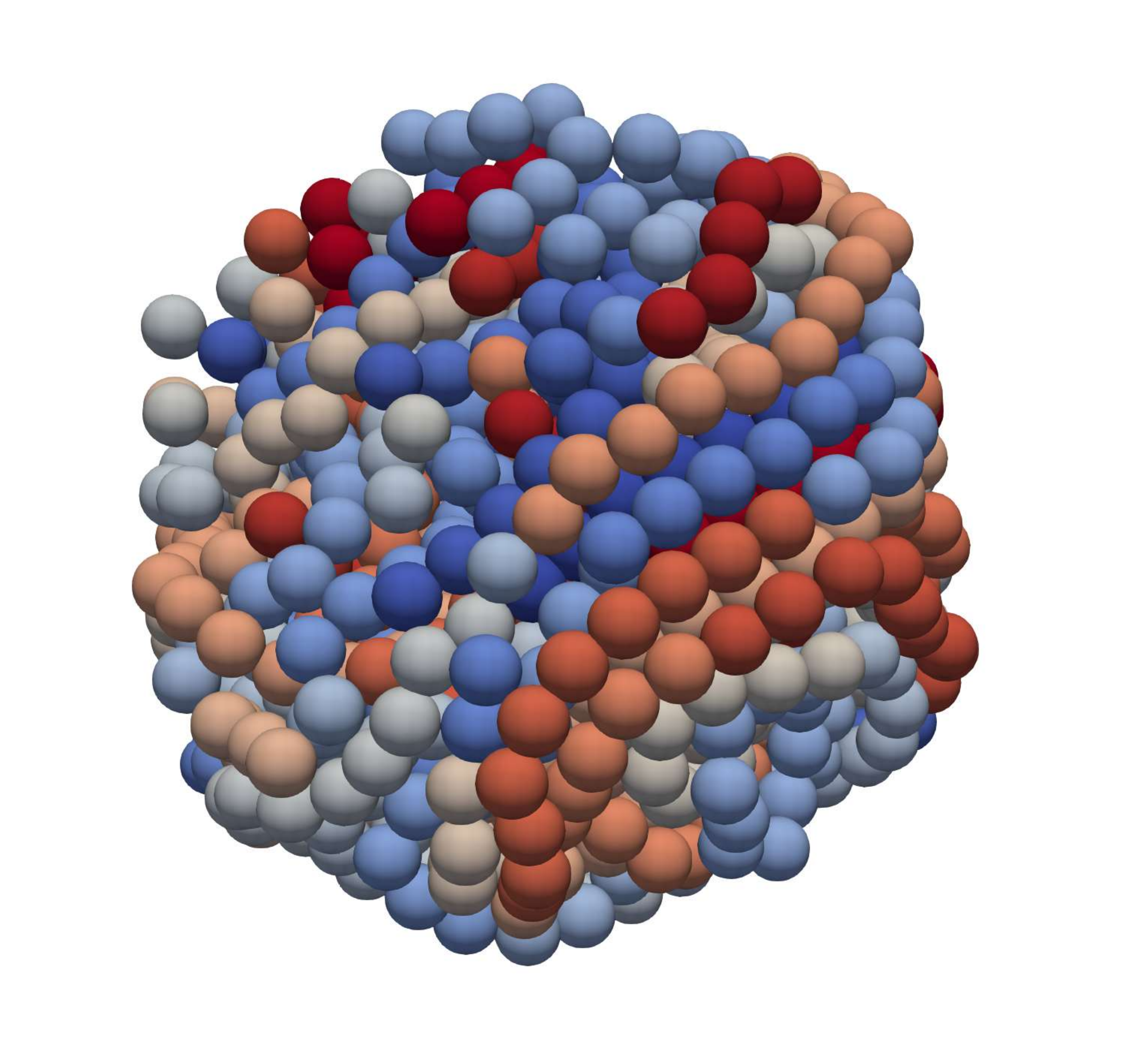} 
        \label{fig:snap-ch}
    \end{subfigure}
    \begin{subfigure}[b]{0.22\textwidth}
        \includegraphics[width=\textwidth]{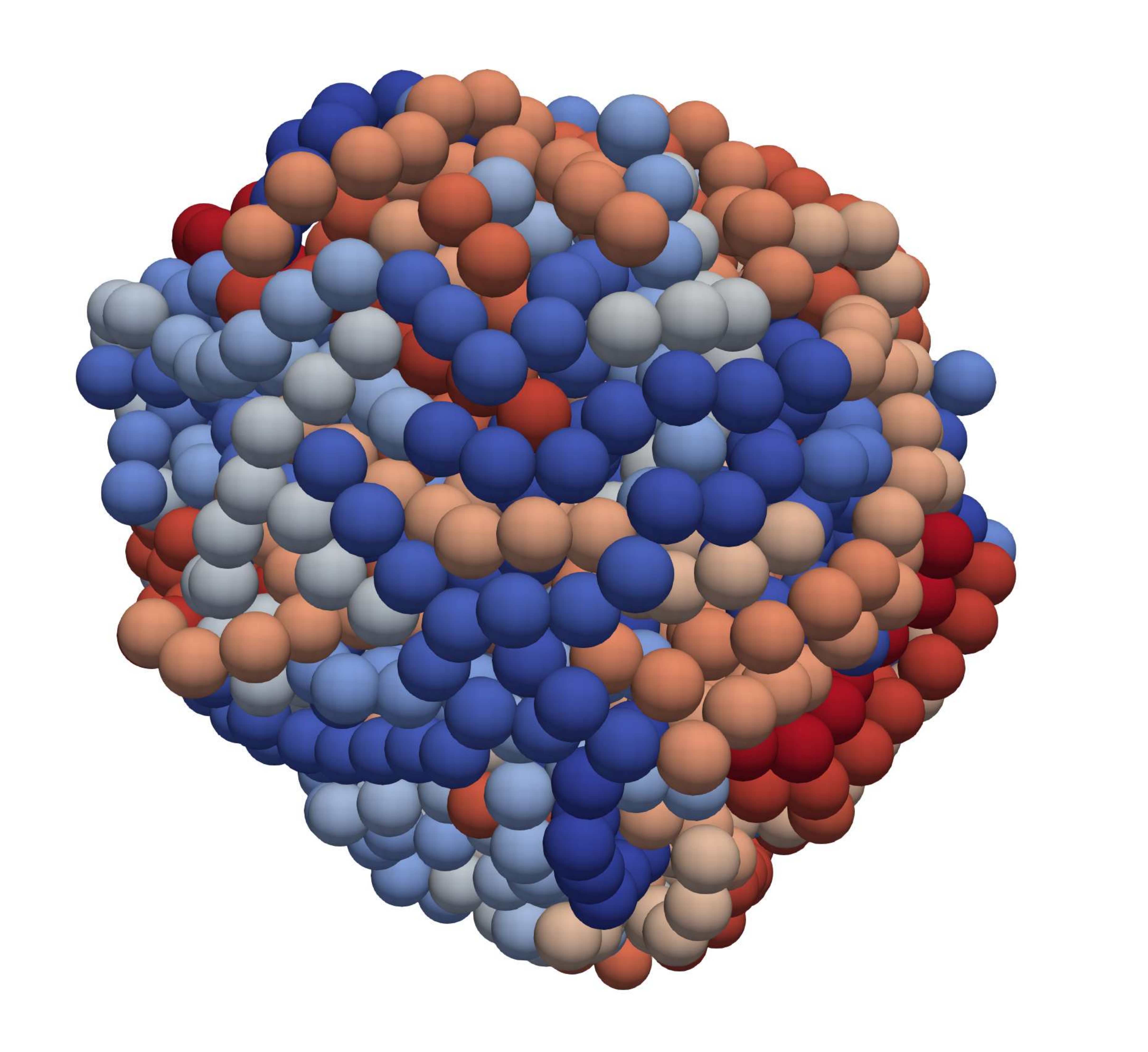}
        \label{fig:snap-y}
    \end{subfigure}
      \begin{subfigure}[b]{0.22\textwidth}
        \includegraphics[width=\textwidth]{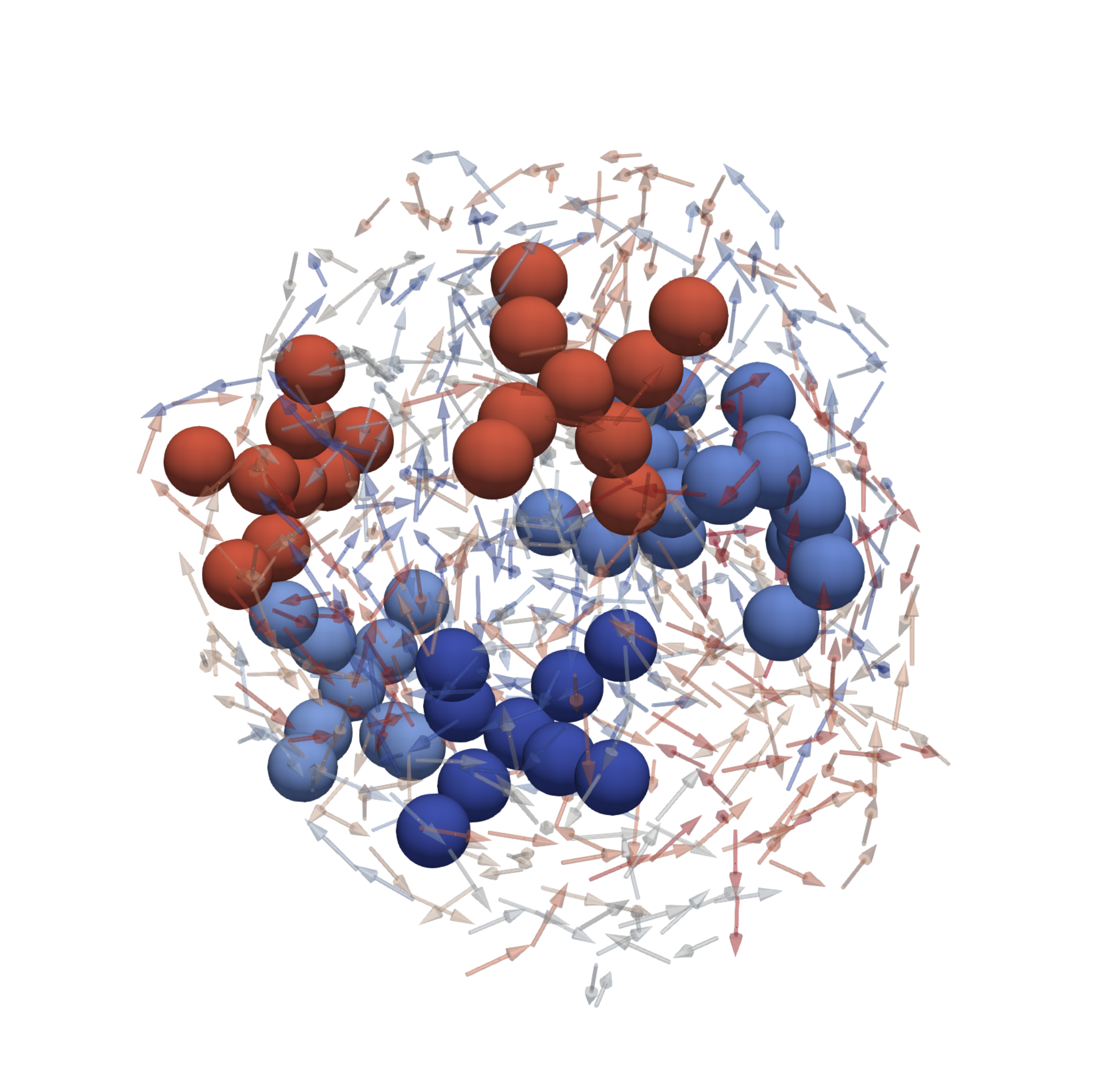}
        \label{fig:snap-x}
    \end{subfigure}
     \begin{subfigure}[b]{0.22\textwidth}
        \includegraphics[width=\textwidth]{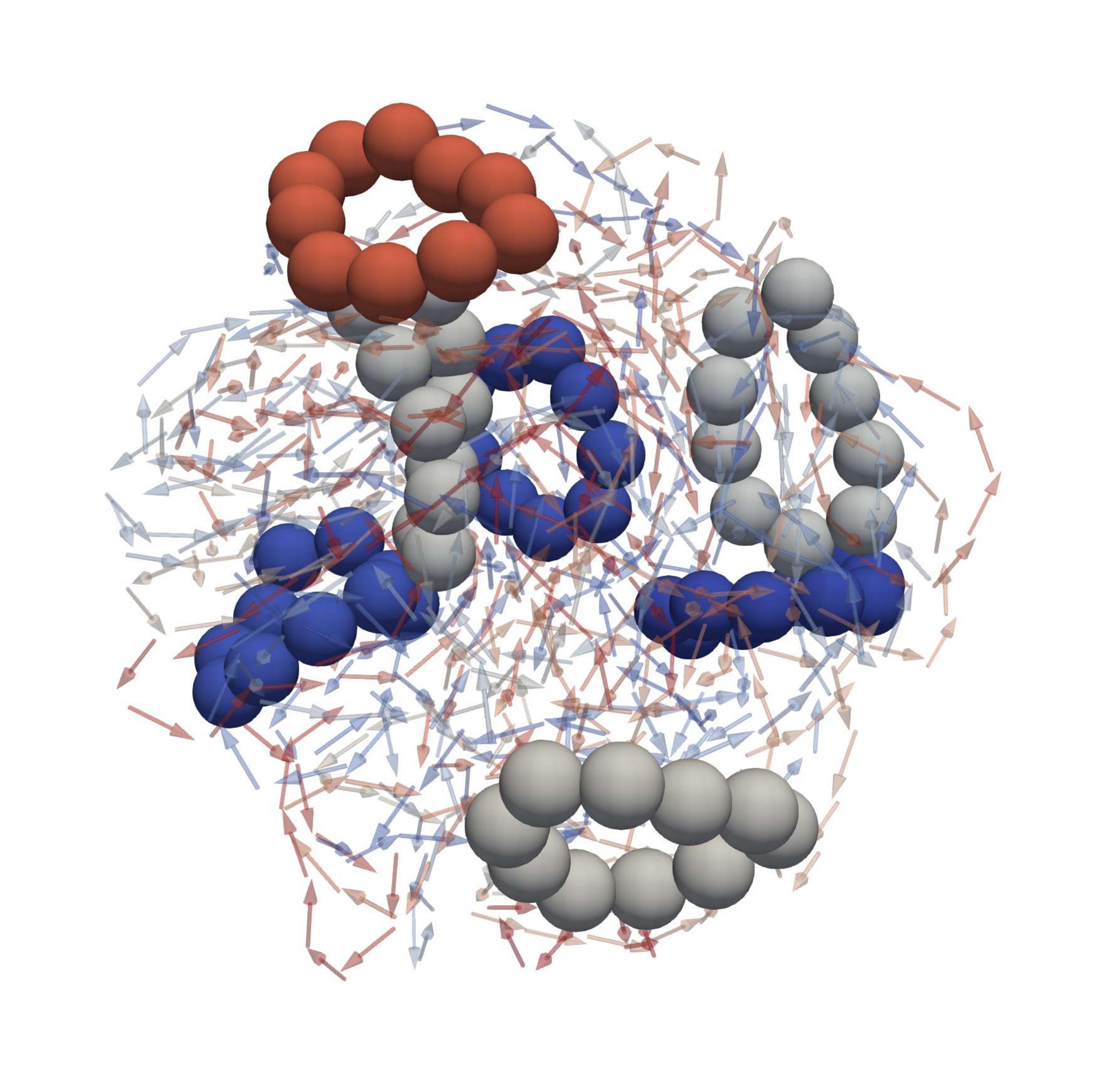}
        \label{fig:snap-r}
    \end{subfigure}
    \begin{subfigure}[b]{0.22\textwidth}
        \includegraphics[width=\textwidth]{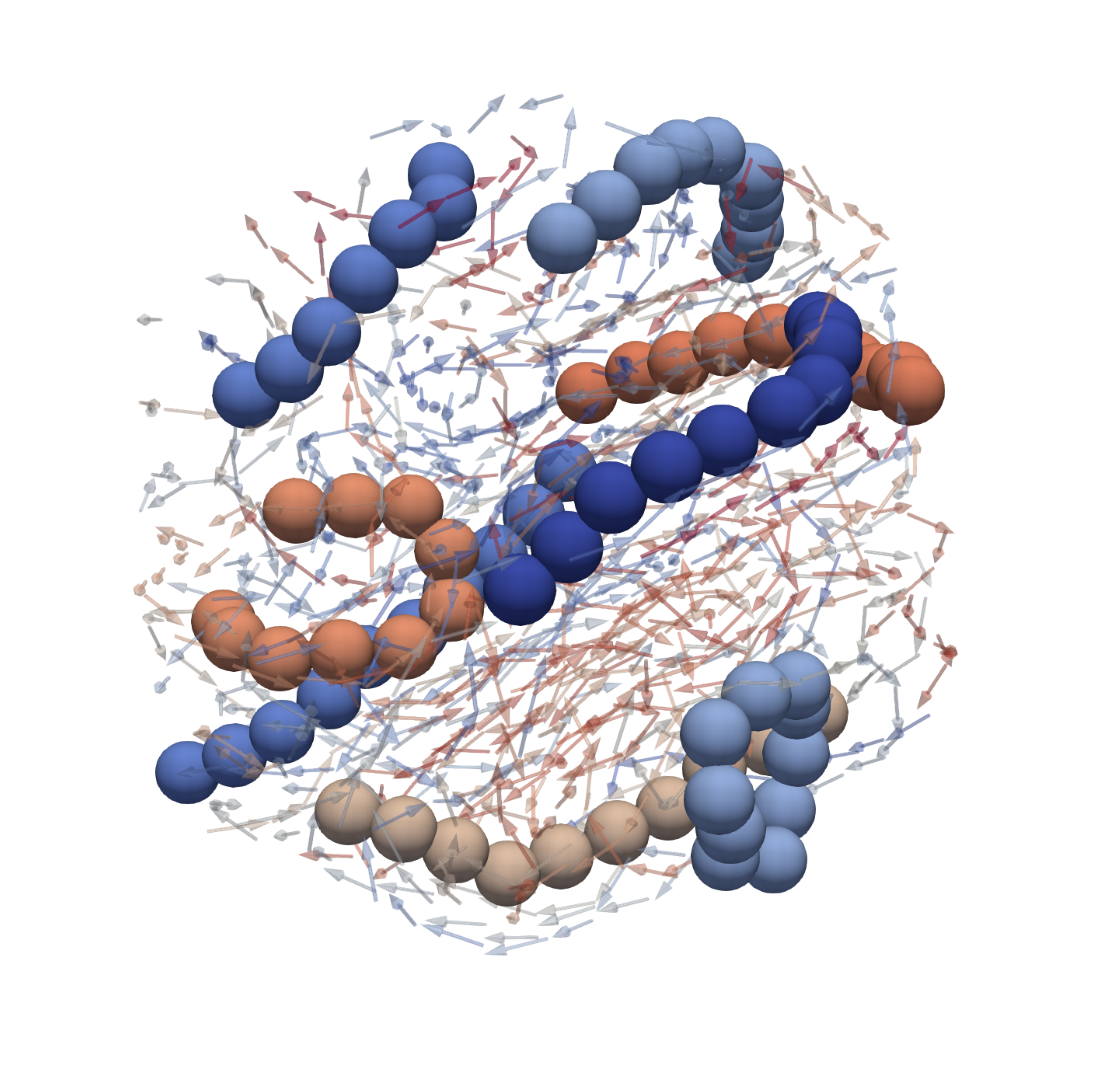}
        \caption{Chain-like SMPs}
        \label{fig:snap-sel-ch}
    \end{subfigure}
       \begin{subfigure}[b]{0.22\textwidth}
        \includegraphics[width=\textwidth]{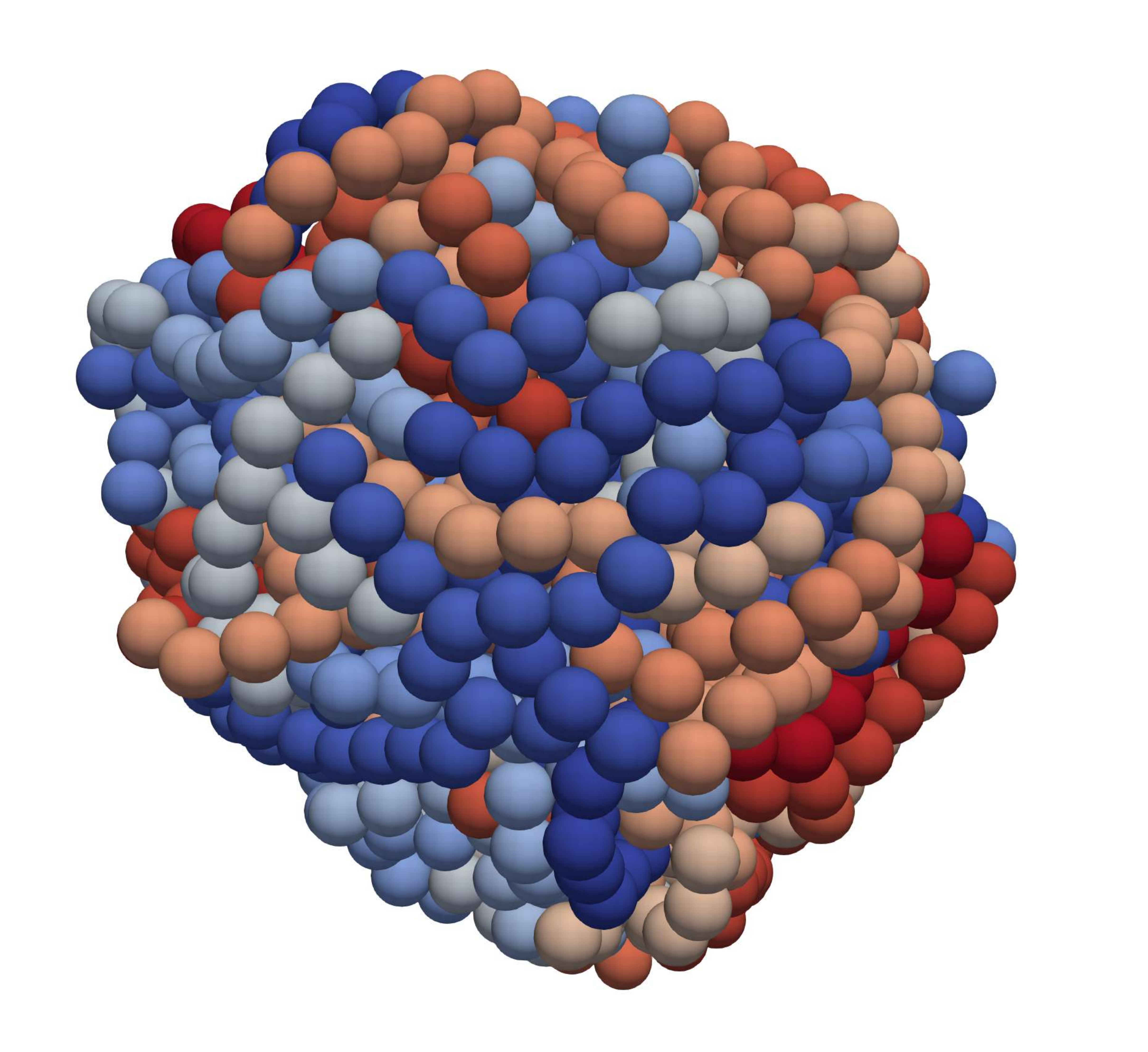}
        \caption{Y-like SMPs}
        \label{fig:snap-sel-y}
    \end{subfigure}
    \begin{subfigure}[b]{0.22\textwidth}
        \includegraphics[width=\textwidth]{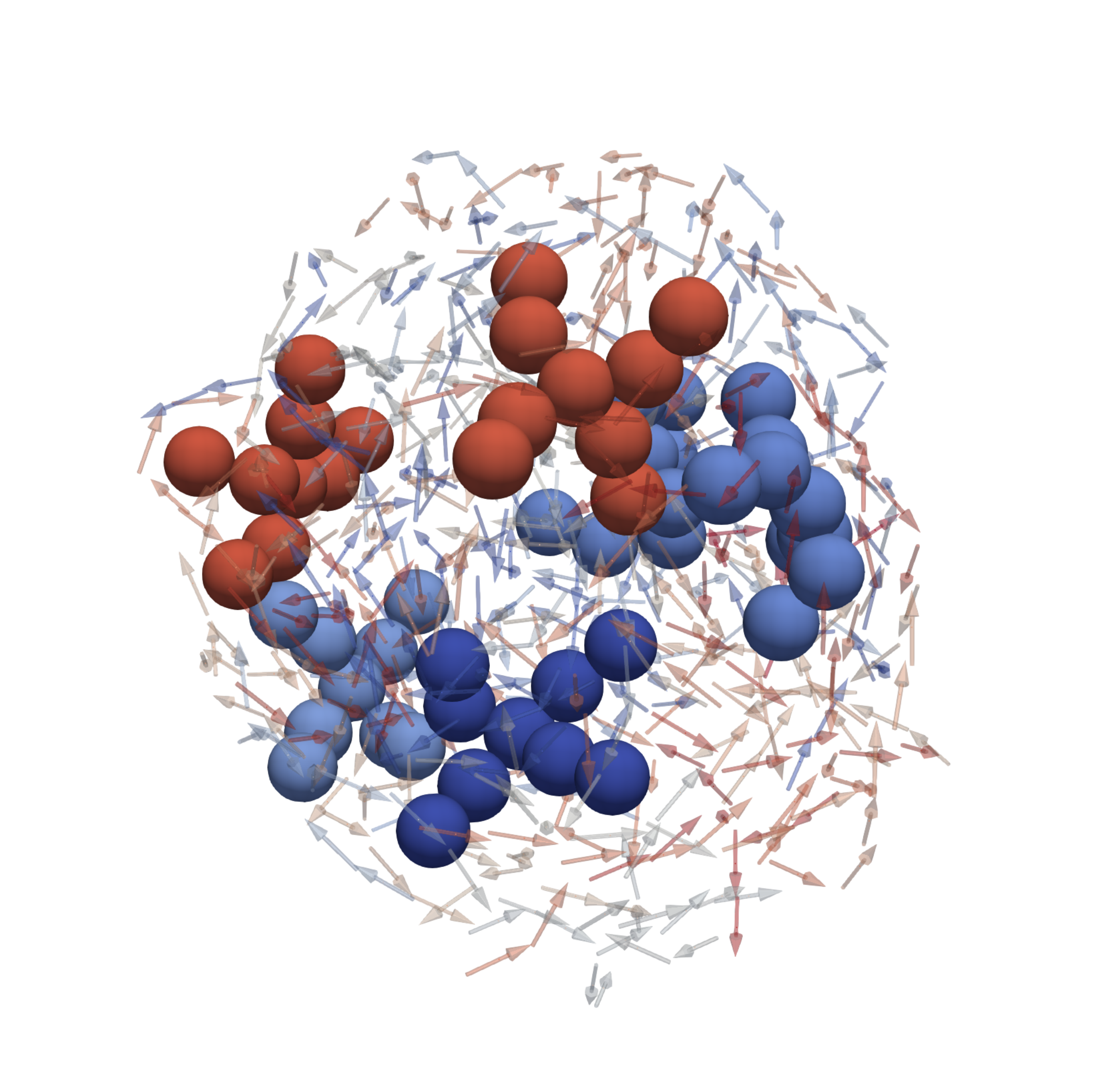}
        \caption{X-like SMPs}
        \label{fig:snap-sel-x}
    \end{subfigure}
      \begin{subfigure}[b]{0.22\textwidth}
        \includegraphics[width=\textwidth]{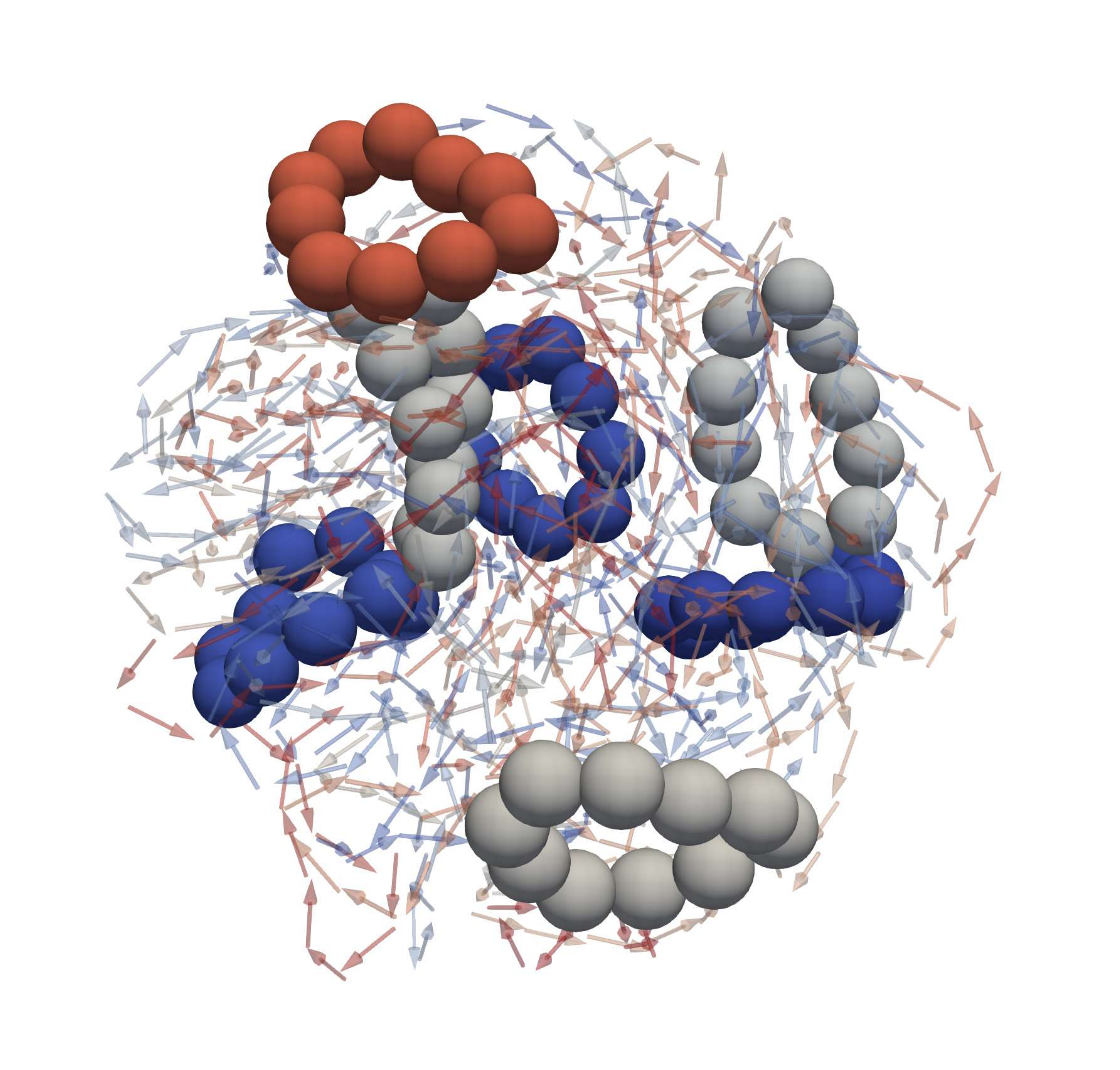}
        \caption{Ring-like SMPs}
        \label{fig:snap-sel-r}
    \end{subfigure}
       \caption{Snapshots of typical clusters. In the upper row all monomers are shown, whereas in the lower row only selected SMPs are visualised, the rest of the monomers are represented by arrows of their dipoles. Columns: (a) chain-like SMPs; (b) Y-like SMPs; (c) X-like SMPs; (d) ring-like SMPs}
       \label{fig:snaps-n-skel}
\end{figure*}

The size-distribution of clusters formed by chain-like SMPs ranges between 20 and 300 SMPs per cluster, with the maximum around 30. 
The clusters made of other SMP topologies have size distributions that are more narrow than for the case of chains and all of them range between 10 and 140. The maxima of the cluster size distributions, in case of Y-, X- and ring-like SMPs, occur in the same region as for clusters formed by chain-like SMPs, around 30. It is worth noting that below all results are obtained through averaging over all clusters independently from their size. Such an approach is validated not only by the increased statistics, but also by the following fact: the comparison between the aforementioned averaging procedure and the  averaging in which only overlapping size regions near distribution maxima (between 20 and 50 for all SMPs topologies), does not reveal any qualitative changes in the results.    

\begin{figure}[hb]
    \begin{subfigure}[b]{0.49\columnwidth}
        \includegraphics[width=\textwidth]{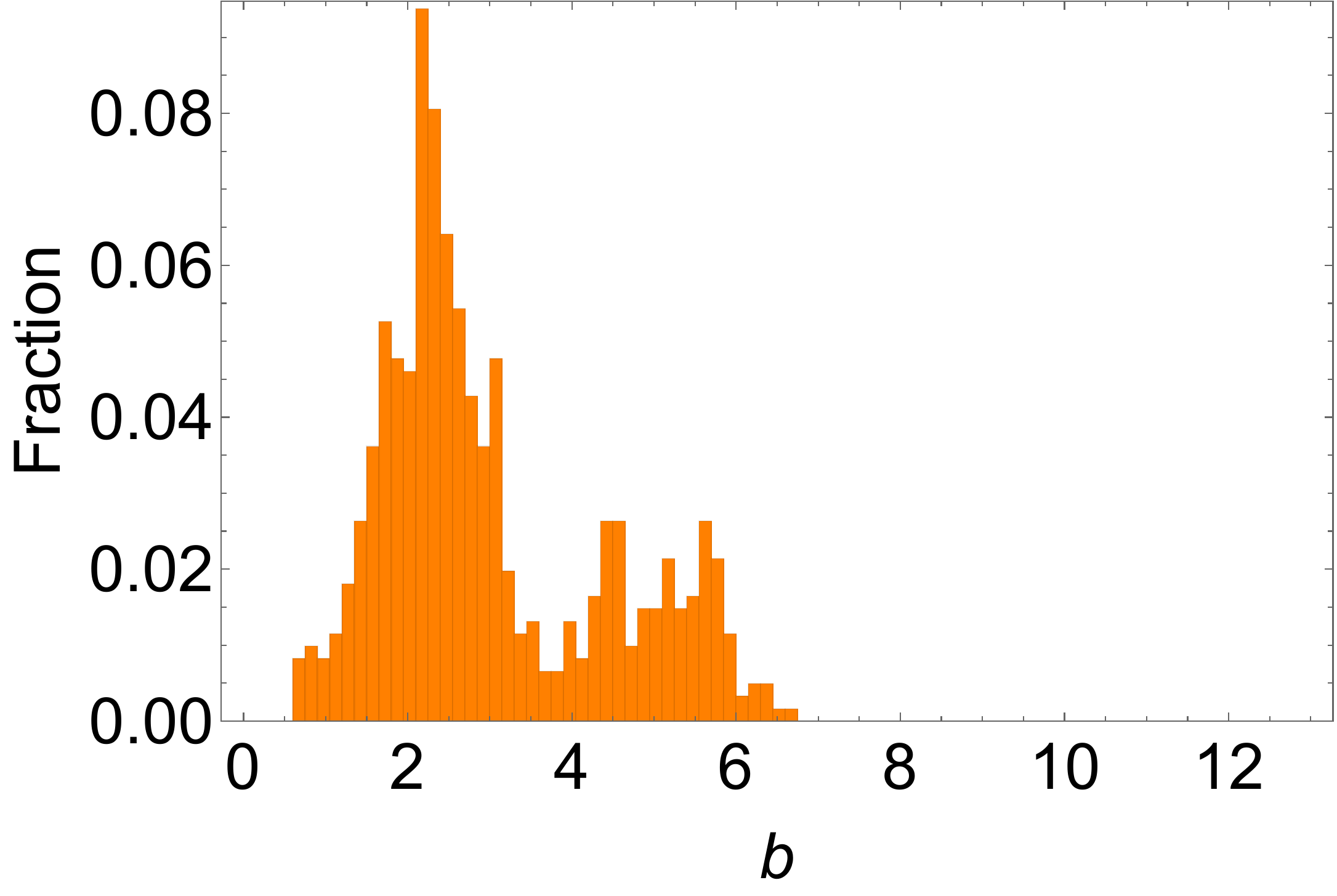}
        \caption{Chain-like SMPs}
        \label{fig:asph-chain}
    \end{subfigure}
    \begin{subfigure}[b]{0.49\columnwidth}
        \includegraphics[width=\textwidth]{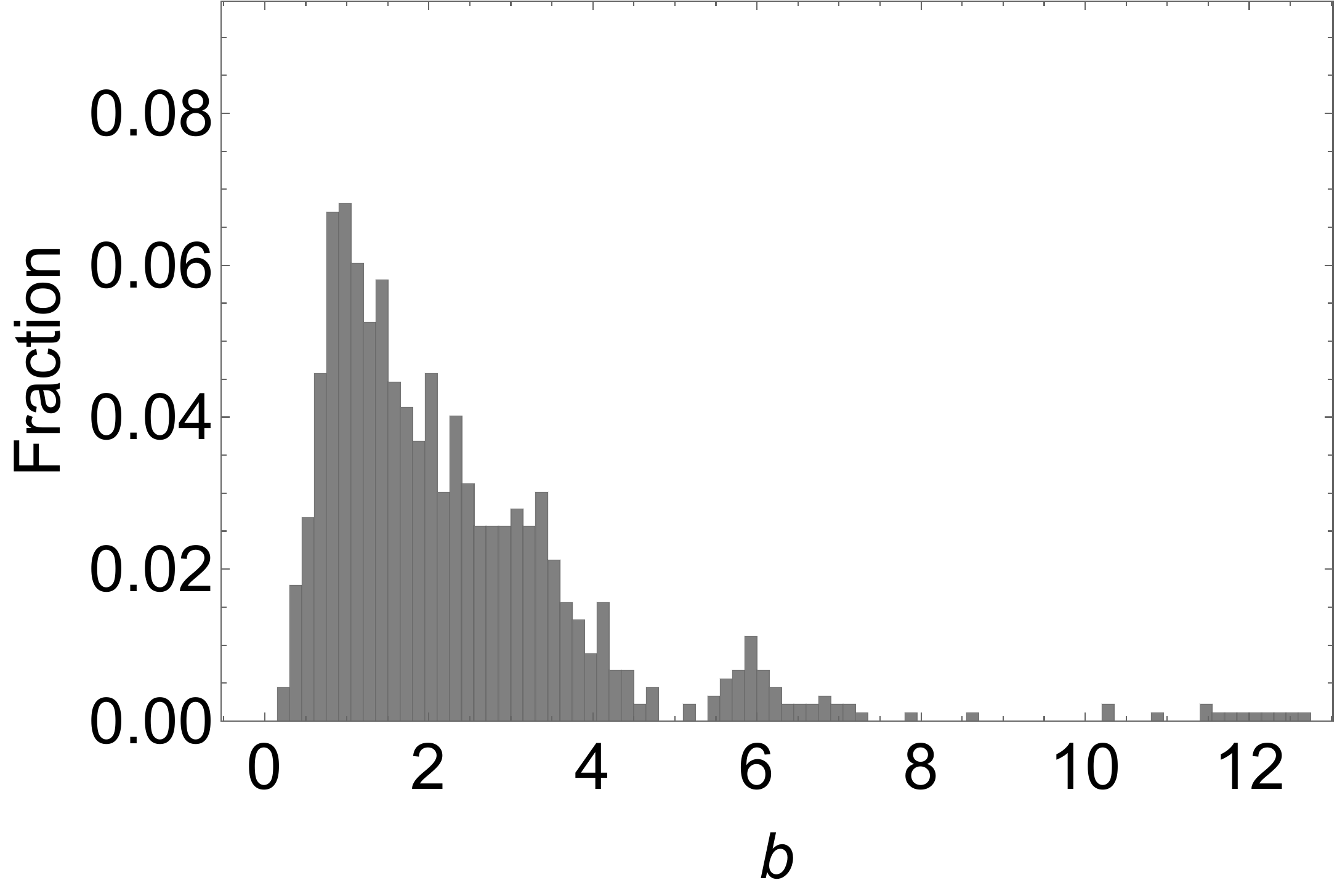}
        \caption{Y-like SMPs}
        \label{fig:asph-y}
    \end{subfigure}
     \begin{subfigure}[b]{0.49\columnwidth}
        \includegraphics[width=\textwidth]{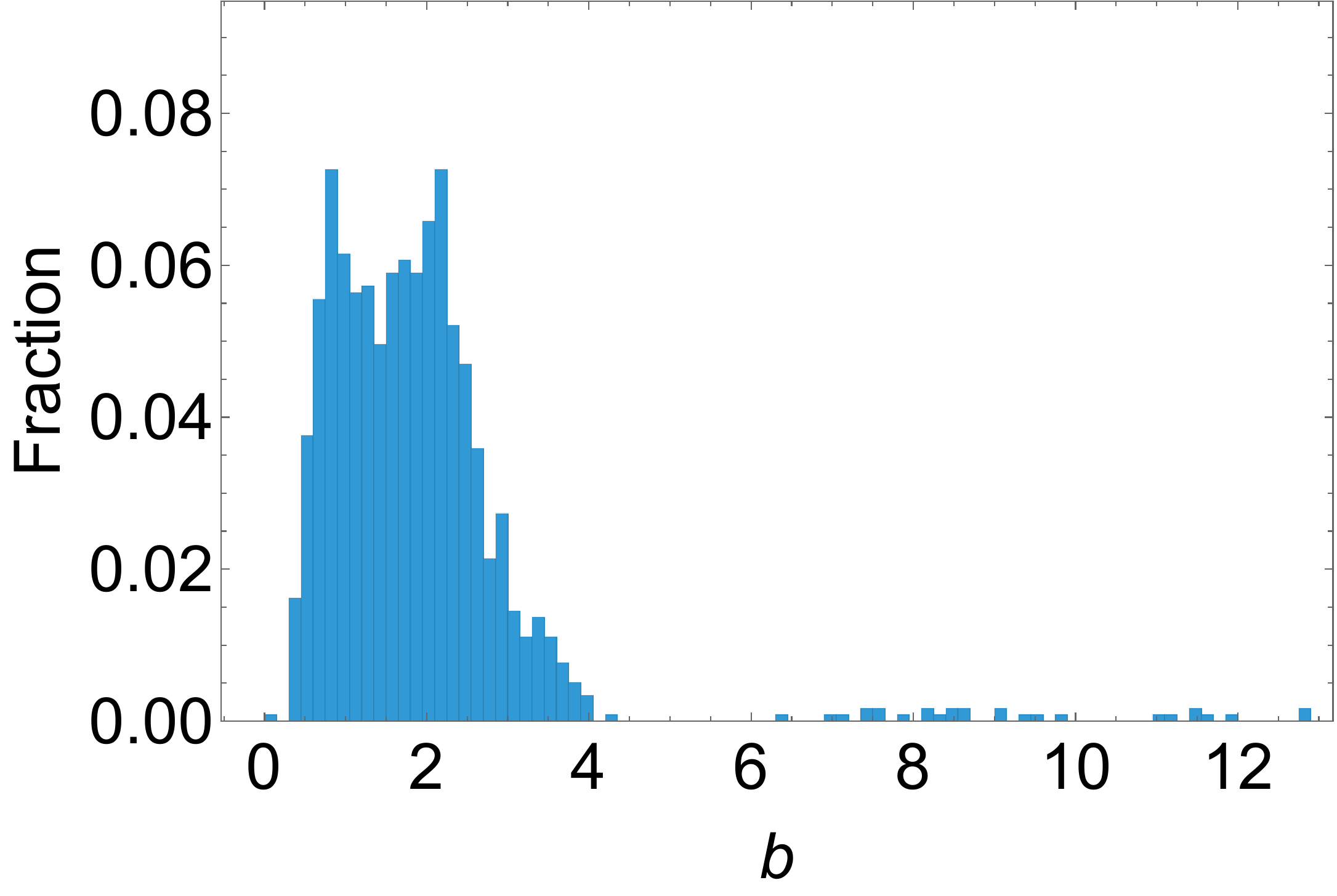}
        \caption{X-like SMPs}
        \label{fig:asph-x}
    \end{subfigure}
      \begin{subfigure}[b]{0.49\columnwidth}
        \includegraphics[width=\textwidth]{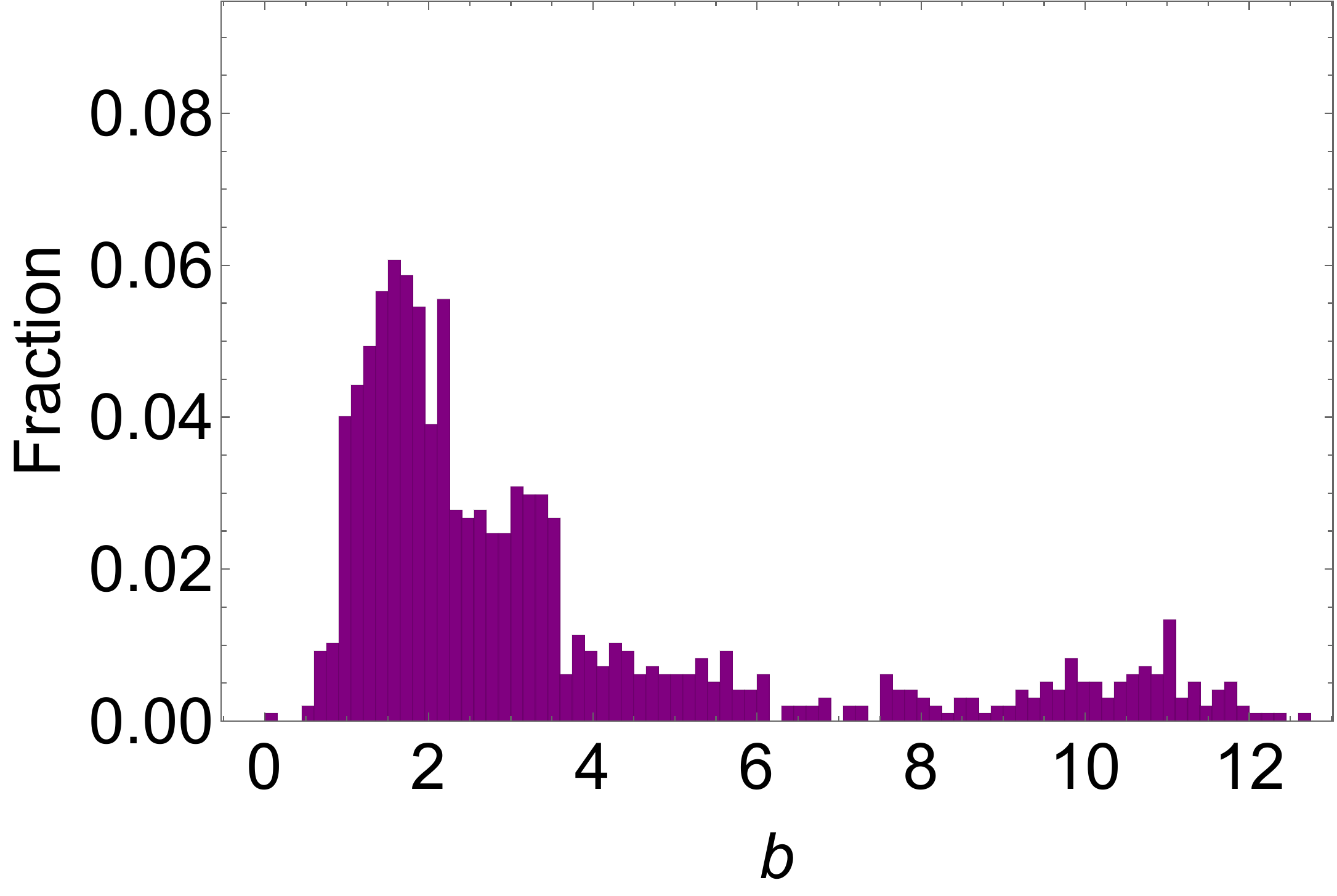}
        \caption{Ring-like SMPs}
        \label{fig:asph-rings}
    \end{subfigure}
       \caption{Histograms showing the fraction of clusters with a given asphericity $b$ calculated  from \eqref{eq:asph}. The values are averaged over all production runs. Subfigures are for clusters formed by (a) chain-like SMPs; (b) Y-like SMPs; (c) X-like SMPs and (d) by ring-like SMPs.}
       \label{fig:asphericity}
\end{figure}

As discussed above, visual inspection of Fig. \ref{fig:snap_solution} does not reveal any differences between clusters formed by different SMPs. Zoomed-in clusters with the size near maximum of the distribution are shown in Fig. \ref{fig:snaps-n-skel} that  consists of four columns. Each column corresponds to the topology of cluster-forming SMPs. In the upper row all monomers in the clusters visualised, whereas in the lower row we see only several selected SMPs while all other monomers are transparent and are only represented by their dipole moments. Looking at these clusters in the upper row and taking into account that all four of them are composed by approximately the same number of SMPs it seems that they have a slightly different shape. In order to quantify these differences we calculate two shape descriptors: the asphericity and the relative shape anisotropy. The asphericity $b$ is defined by
\begin{equation}
   b = {\lambda}_3^2 -\frac {1}{2} \left({\lambda}_1^2 + {\lambda}_2^2 \right ),
   \label{eq:asph}
\end{equation}
where $\lambda_{1(2,3)}$ are eigenvalues of the gyration tensor of a cluster. Parameter $b=0$ if the distribution of particles is spherically symmetric. In Fig. \ref{fig:asphericity} we plot the histograms that show the fraction of clusters formed by each SMP topology with a given value of $b$.  Clearly, none of the formed clusters is perfectly  spherical. The histograms which exhibit maximum close to zero are those for X- and Y-shaped SMP clusters as seen in Figs. \ref{fig:asph-x} and \ref{fig:asph-y} correspondingly. The histogram for clusters formed by chain-like SMPs (Fig. \ref{fig:asph-chain}) has a clear maximum at $b=2$ and a secondary maximum around 5. There are basically no clusters formed by chain-like SMPs with $b>7$. It is different for clusters formed by SMPs of any other topology: one can find several clusters with asphericity up to 12. The broadest distribution and the lowest sphericity can be found for clusters formed by ring-like SMPs, as shown in Fig. \ref{fig:asph-rings}.

\begin{figure}[ht!]
   \begin{subfigure}[b]{0.49\columnwidth}
        \includegraphics[width=\textwidth]{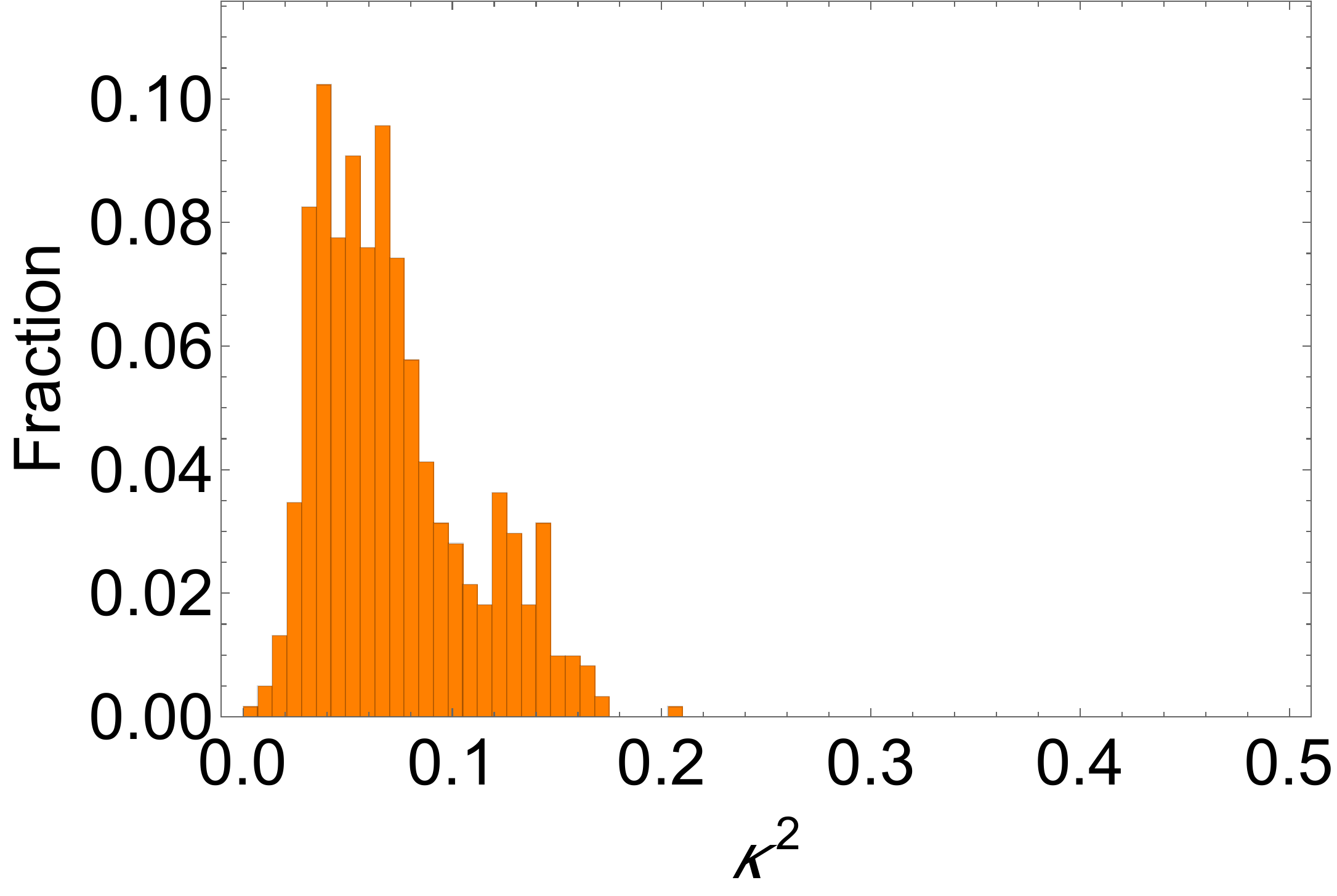}       
        \caption{Chain-like SMPs}
        \label{fig:anis-ch}
    \end{subfigure}
    \begin{subfigure}[b]{0.49\columnwidth}
        \includegraphics[width=\textwidth]{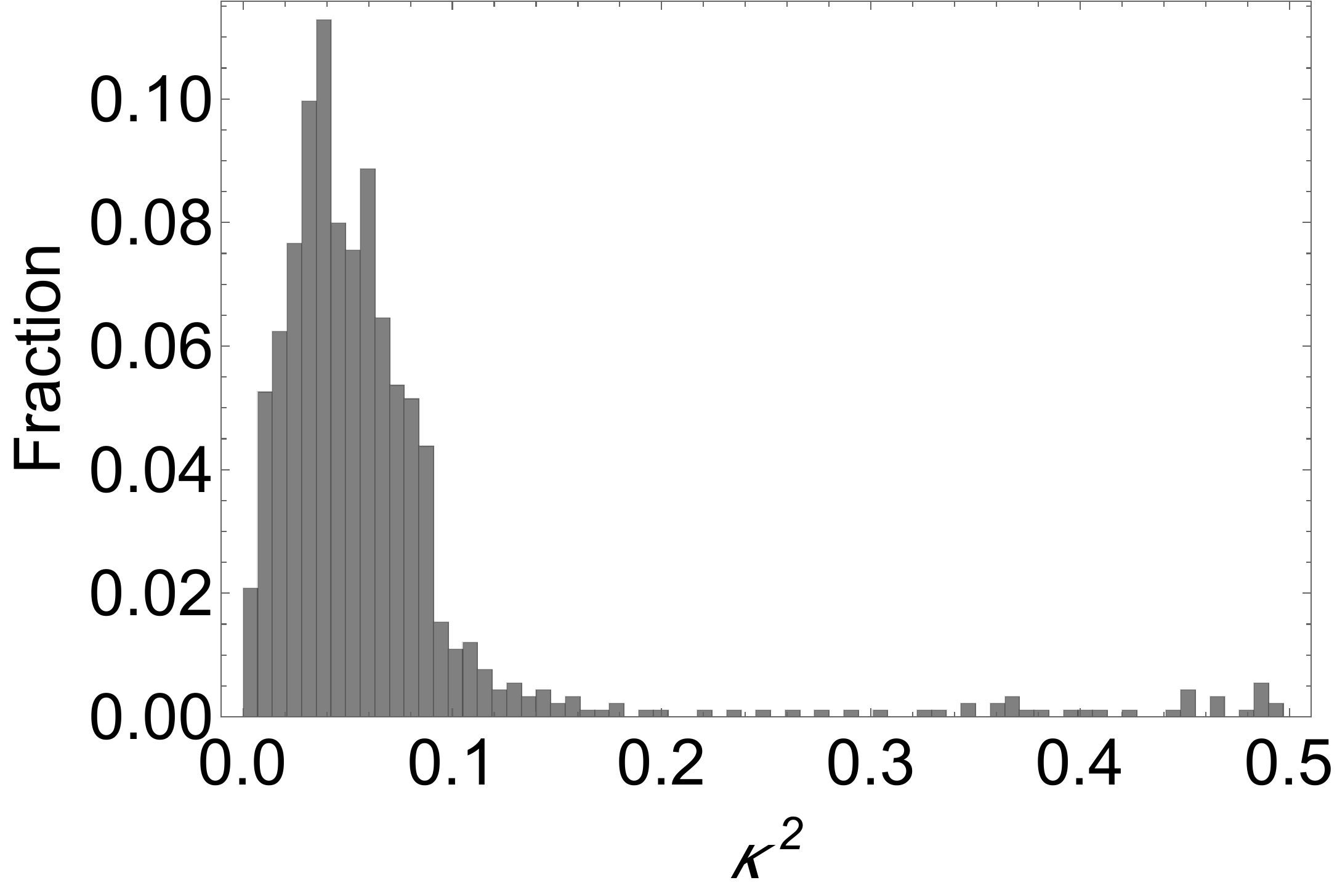}
        \caption{Y-like SMPs}
        \label{fig:anis-y}
    \end{subfigure}
     \begin{subfigure}[b]{0.49\columnwidth}
        \includegraphics[width=\textwidth]{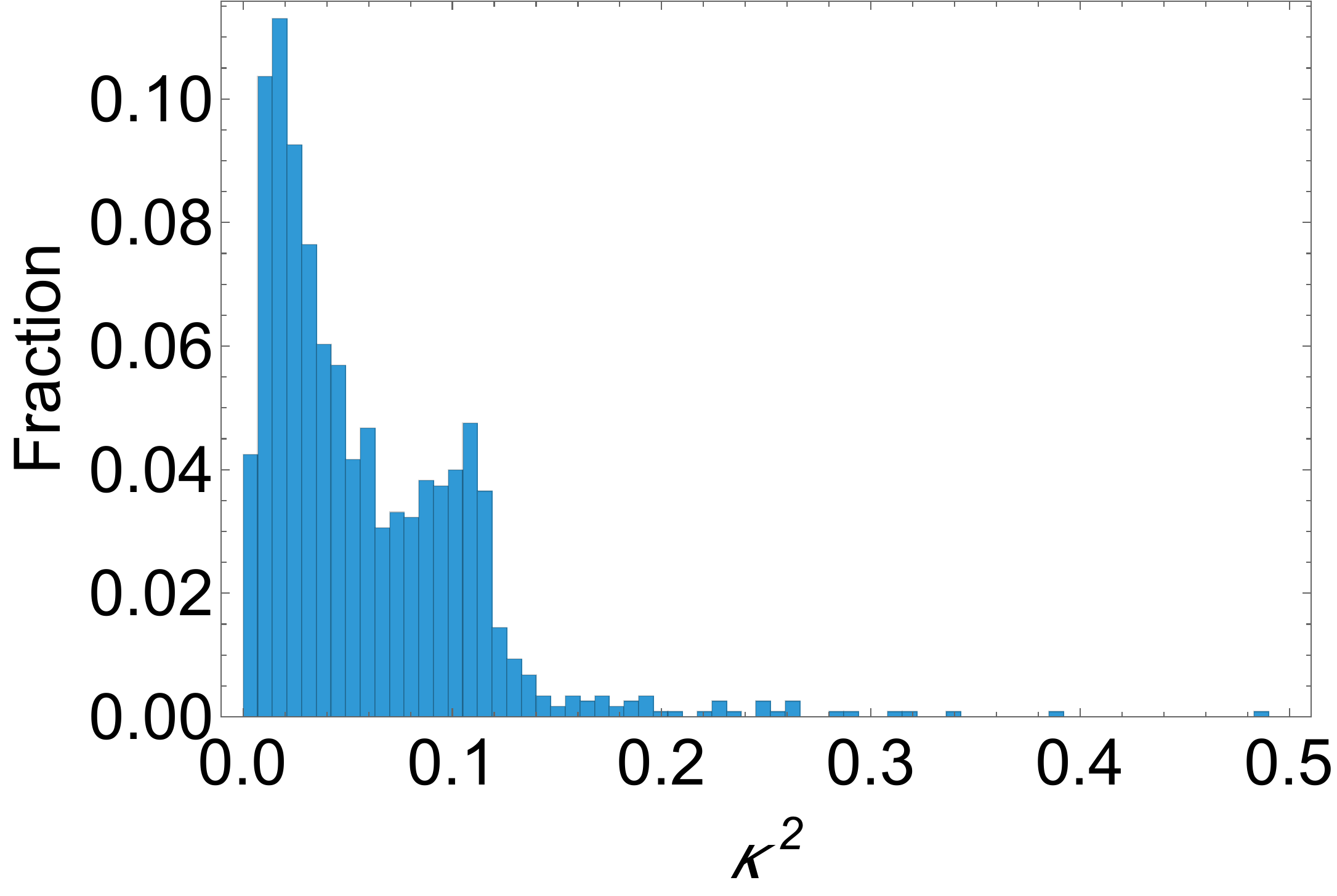}
        \caption{X-like SMPs}
        \label{fig:anis-x}
    \end{subfigure}
      \begin{subfigure}[b]{0.49\columnwidth}
        \includegraphics[width=\textwidth]{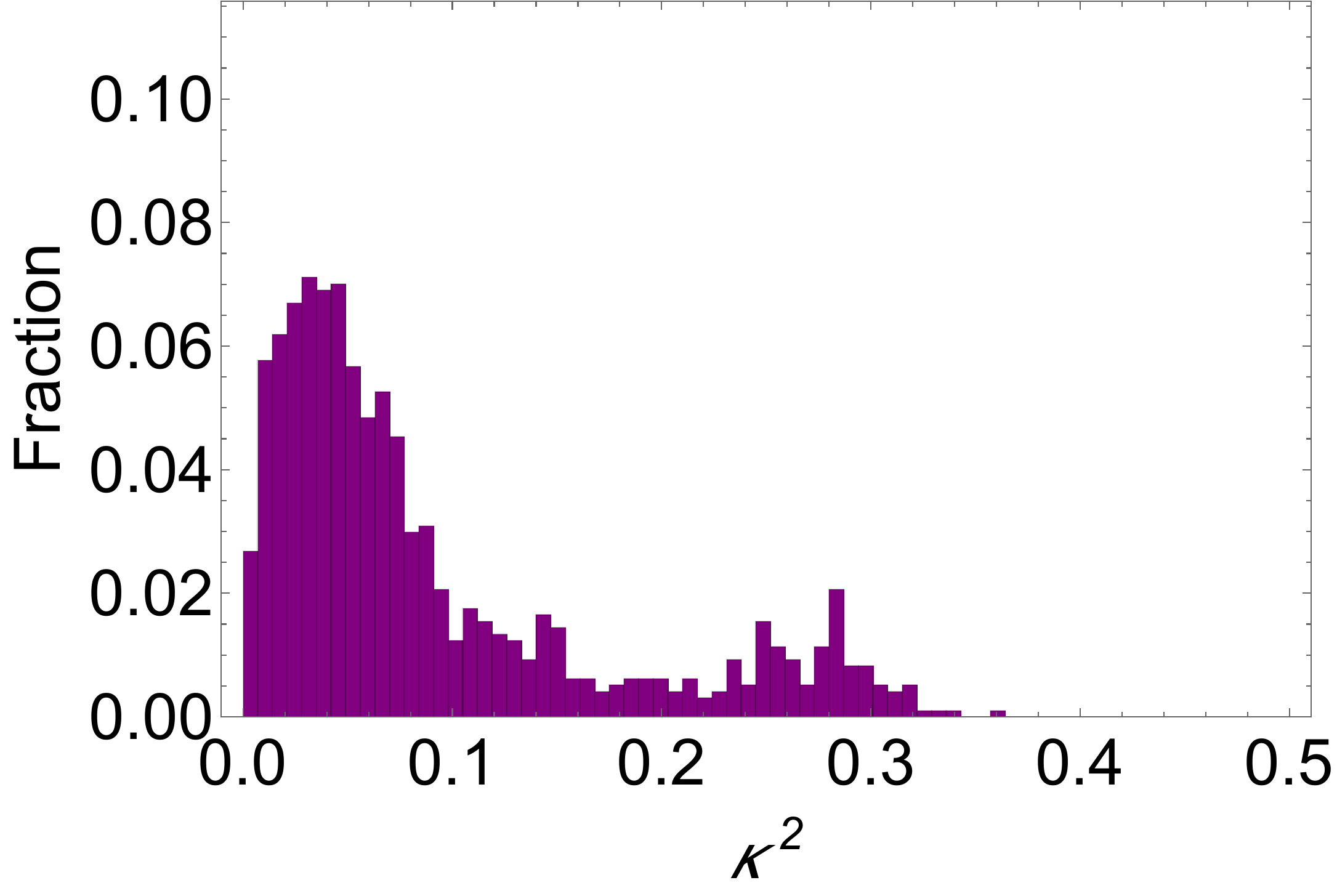}
        \caption{Ring-like SMP}
        \label{fig:anis-rings}
    \end{subfigure}
       \caption{Histograms showing the fraction of clusters with a given anisotropy $\kappa ^{2}$ calculated  from \eqref{eq:kappa}. The values are averaged over all production runs. Subfigures are for clusters formed by (a) chain-like SMPs; (b) Y-like SMPs; (c) X-like SMPs and (d) by ring-like SMPs.}
       \label{fig:anisotropy}
\end{figure}

In order to  shed more light on the shape of SMPs clusters we look at their relative shape anisotropy $\kappa ^{2}$. It is defined as
\begin{equation}
   \kappa ^{2} =  {\frac  {3}{2}}{\frac  {\lambda _{{1}}^{{4}}+\lambda _{{2}}^{{4}}+\lambda _{{3}}^{{4}}}{(\lambda _{{1}}^{{2}}+\lambda _{{2}}^{{2}}+\lambda _{{3}}^{{2}})^{{2}}}}-{\frac  {1}{2}},
   \label{eq:kappa}
\end{equation}
and it is bounded between zero and one. $\kappa ^{2} = 0$ only occurs if all particles are distributed spherically symmetric with respect to the centre of mass of the cluster, and $\kappa ^{2} = 1$ only occurs if all particles lie on a line.  We plot the resulting histograms in Fig. \ref{fig:anisotropy}. From all four subfigures one can immediately conclude that none of the clusters shows a tendency to elongate.  As it can be seen, for clusters formed by chain-like SMPs (Fig. \ref{fig:anis-ch}), the anisotropy shows the narrowest distribution with a clearly pronounced maximum  for  values $0.04 <\kappa ^{2} < 0.06$. For clusters formed by Y-like SMPs, shown in Fig. \ref{fig:anis-y} the maximum of the distribution is situated in nearly the same place, however the distribution seems to be broader. We find a rather broad distribution also for clusters formed by X-like and ring-like SMPs (Figs. \ref{fig:anis-x} and \ref{fig:anis-rings}), though with one difference: the distribution for clusters of ring-like SMPs is almost bimodal. The latter agrees well with the distribution of asphericity provided in Fig. \ref{fig:asph-rings}, where one can find a fraction of clusters whose asphericity is $10<b<13$.

So far, the outcome of our analysis can be summarised as follows. There are differences in the overall shape of clusters formed by SMPs with different topology. The differences are not large, but still significant. Moreover, the largest similarities are observed between clusters formed by Y-like and X-like SMPs, whereas those formed by chains and rings have clearly different shape characteristics. The next step is to explain observed features and look inside different clusters. 

\subsection{Clusters inside: SMP level}

First we address the orientation of SMPs within the clusters. Let us calculate the main axis of a SMP inside a cluster. It can be obtained as the eigenvector corresponding to the largest eigenvalue of the SMP gyration tensor. The next step is to calculate the vector, connecting the centre of mass of the cluster to that of a SMP. After that, one can define the angle $\alpha$ between the latter vector and SMP main axis. Average value of this angle will characterise different orientations of SMPs within the clusters. In case the structure of the cluster is onion-like, aforementioned angle $\alpha$ should be close to $\pi/2$, {\it i.e.} the major part of SMPs is oriented tangentially. On the other hand, if $\alpha \sim 0$, it means that SMPs are oriented radially. 
\begin{figure}[h]
      \begin{subfigure}[b]{0.49\columnwidth}
        \includegraphics[width=\textwidth]{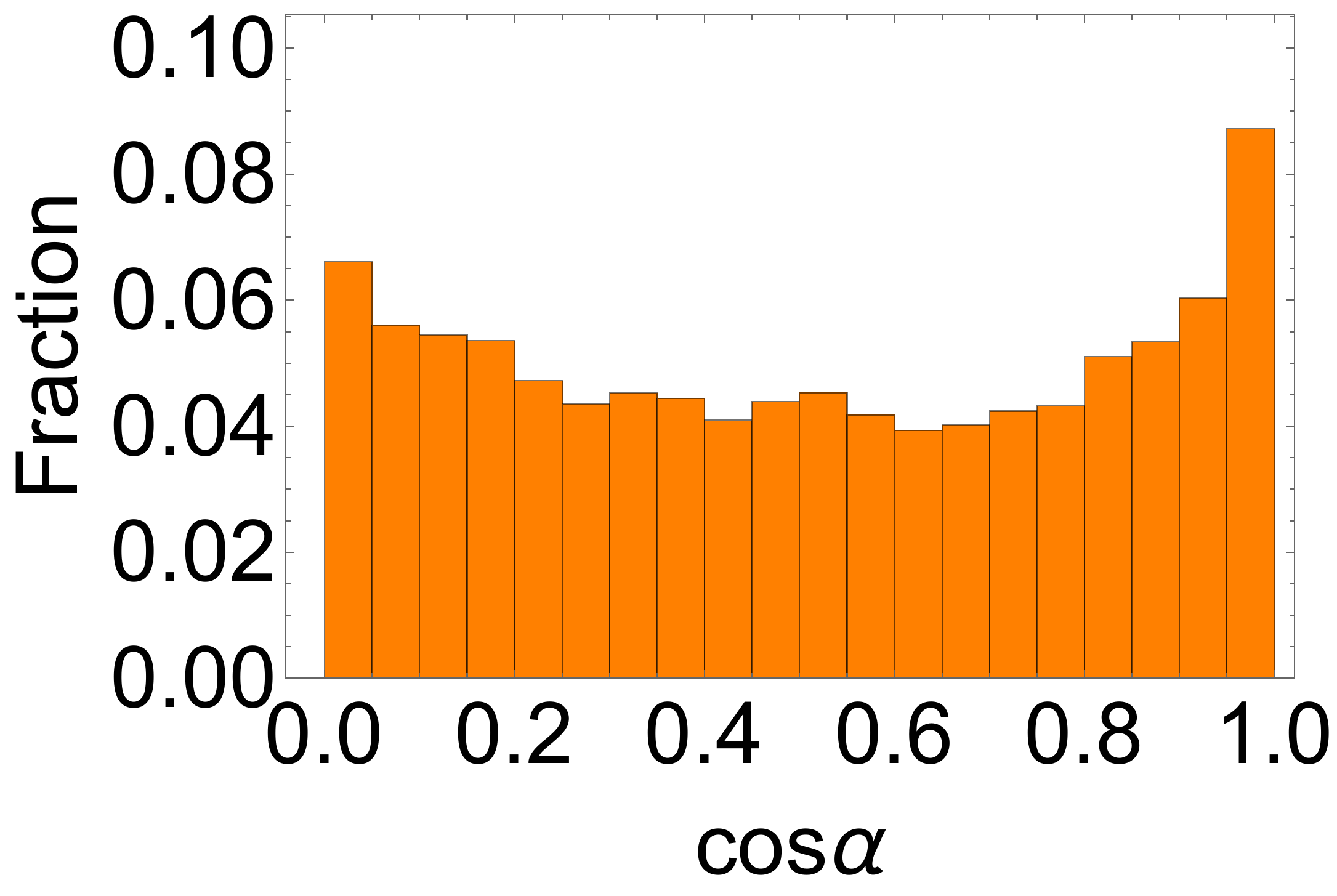}
        \caption{Chain-like SMPs}
        \label{fig:cos-alph-chain}
    \end{subfigure}
    \begin{subfigure}[b]{0.49\columnwidth}
        \includegraphics[width=\textwidth]{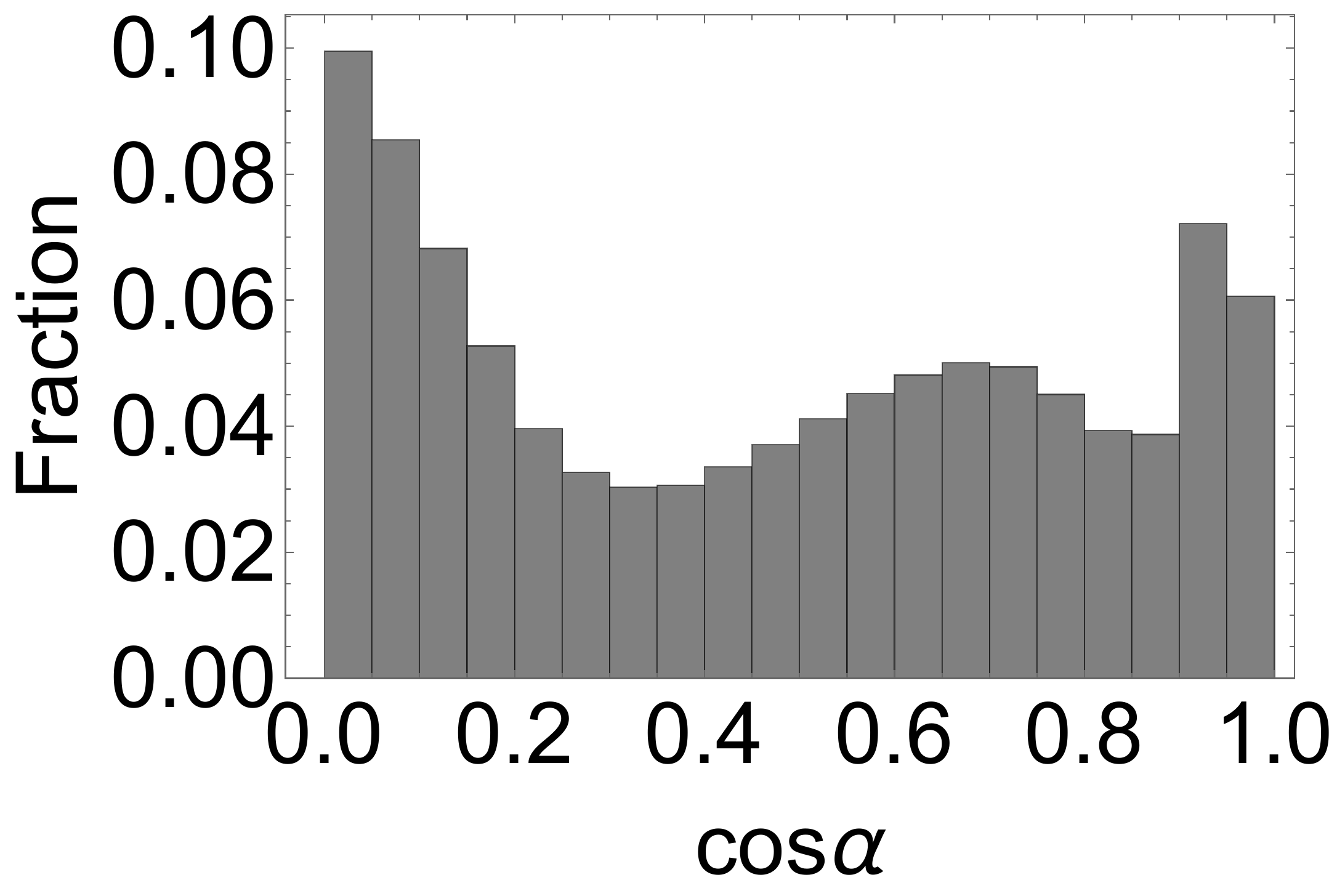}
        \caption{Y-like SMPs}
        \label{fig:cos-alph-y}
    \end{subfigure}
     \begin{subfigure}[b]{0.49\columnwidth}
        \includegraphics[width=\textwidth]{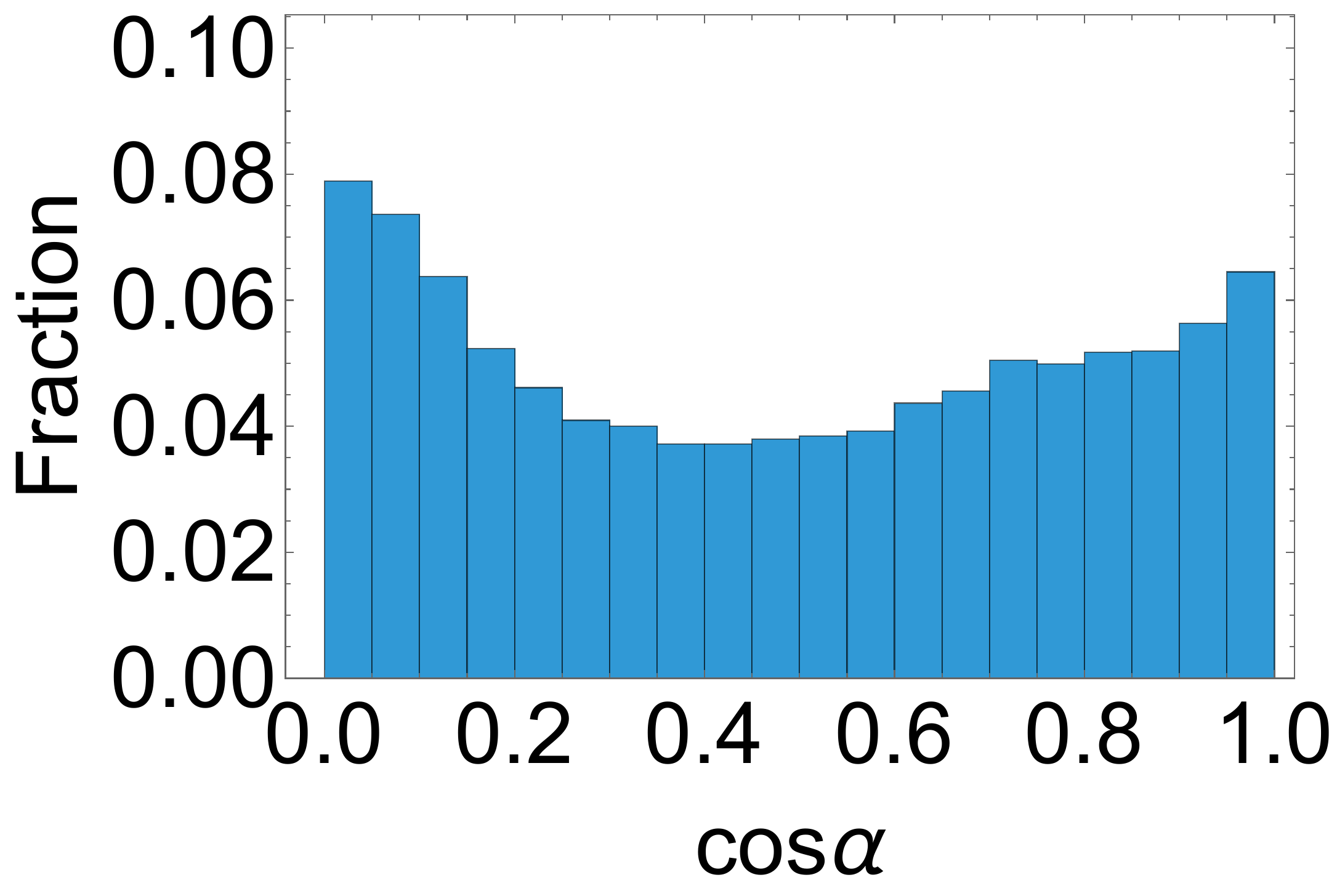}
        \caption{X-like SMPs}
        \label{fig:cos-alph-x}
    \end{subfigure}
      \begin{subfigure}[b]{0.49\columnwidth}
        \includegraphics[width=\textwidth]{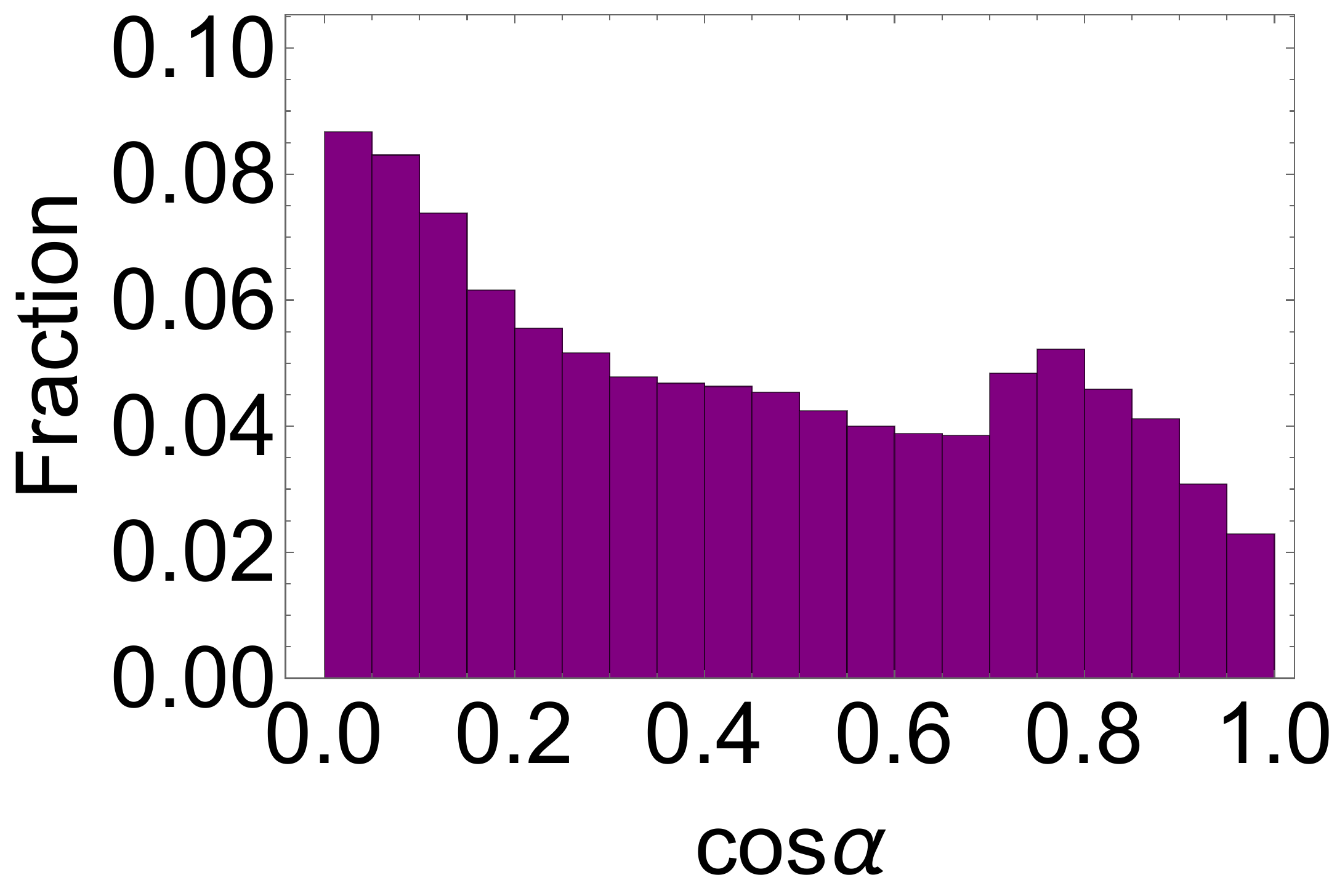}
        \caption{Ring-like SMPs}
        \label{fig:cos-alph-rings}
    \end{subfigure}
       \caption{Fraction of SMPs, whose main axis and the vector, connecting the centre of mass of the cluster and that of an SMP, form $\cos \alpha$.  The values are averaged over all production runs. Subfigures are for clusters formed by (a) chain-like SMPs; (b) Y-like SMPs; (c) X-like SMPs and (d) by ring-like SMPs.}
       \label{fig:cos-main-axis}
\end{figure}
In Fig. \ref{fig:cos-main-axis}, we plot the fraction of SMPs with a given topology as a function of $\cos \alpha$.  The plots are symmetric with respect to zero, that is why we show the data from 0 to 1. It is rather surprising that obtained distributions exhibit much clearer differences depending on the topology in comparison to shape characteristics studied in the previous section. It turns out that the chain-like SMPs (Fig. \ref{fig:cos-alph-chain}) can have practically any orientation inside the clusters with only mild preference to radial orientation. This finding clearly indicates that chain-like SMPs do not form onion-like clusters.  Actually, the only clusters that have an onion-like structures are those formed by rings. The distribution in Fig. \ref{fig:cos-alph-rings} is the only one that shows no maximum at $\cos \alpha = 1$. From Figs. \ref{fig:cos-alph-y} and \ref{fig:cos-alph-x} we can conclude that both Y-like and X-like SMPs with higher probability form onion-like clusters, however, there can be also seen the secondary maximum corresponding to radial orientation. With this information at hand, visual examination of the lower row of Fig. \ref{fig:snaps-n-skel} can reveal the same tendency: the clusters formed by ring-like SMPs have an onion internal structure, whereas X- and Y-like SMPs have mixed orientations with light preference towards tangential orientation. Most uniform orientation distribution can be found inside clusters formed by  chain-like SMPs.   

\begin{figure}[ht!]
    \begin{subfigure}[b]{0.49\columnwidth}
        \includegraphics[width=\textwidth]{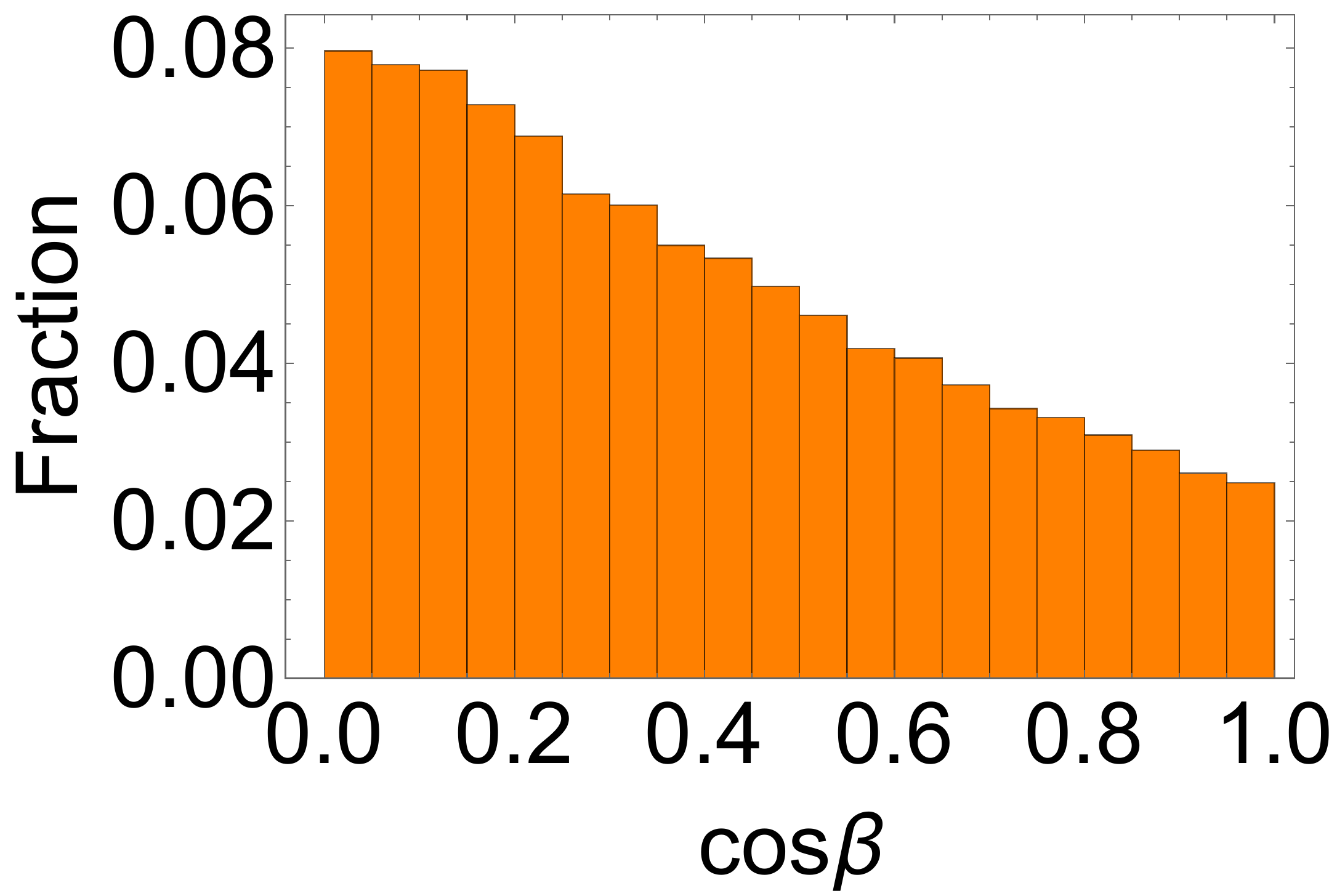}
        \caption{Chain-like SMPs}
        \label{fig:cos-beta-chain}
    \end{subfigure}
    \begin{subfigure}[b]{0.49\columnwidth}
        \includegraphics[width=\textwidth]{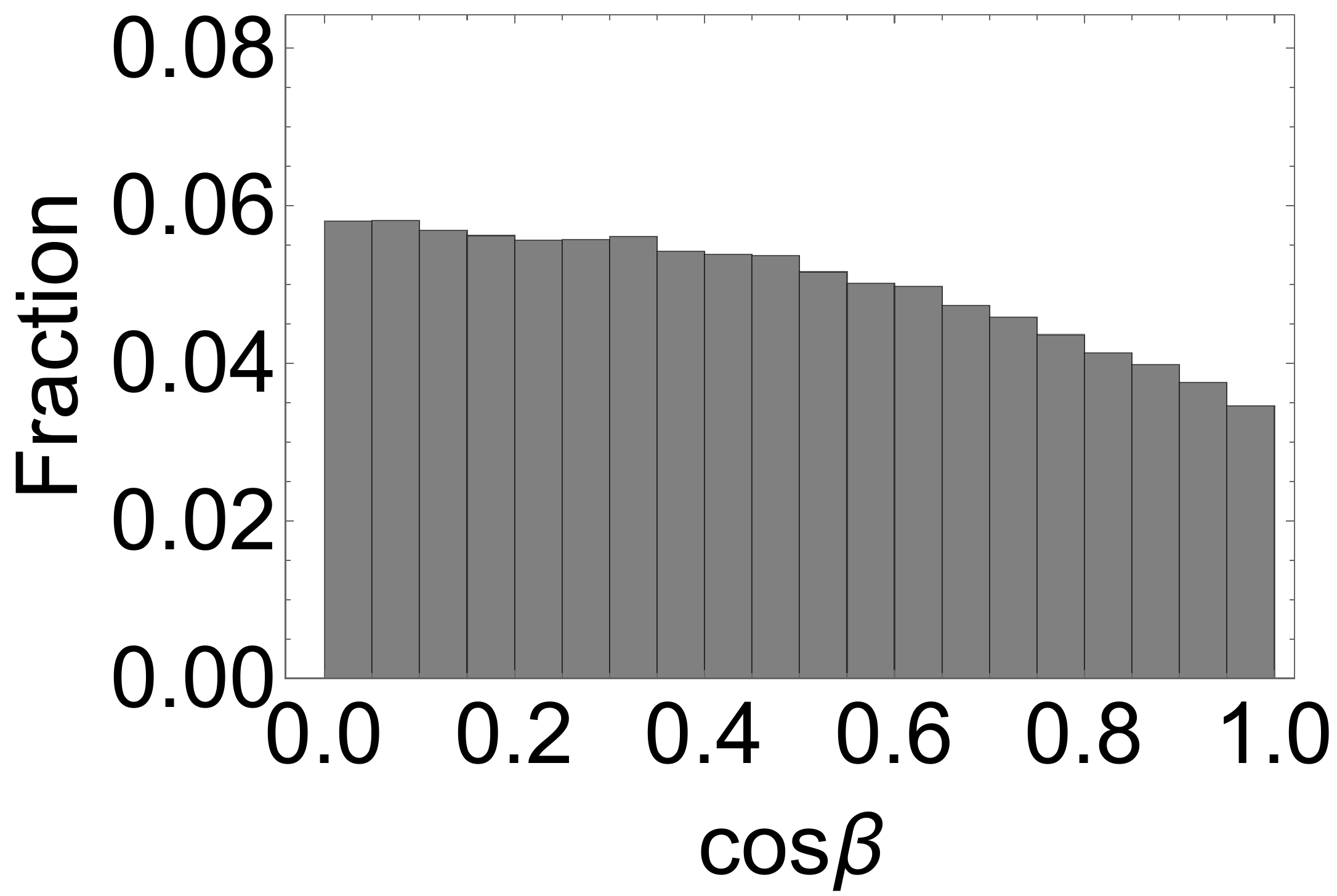}
        \caption{Y-like SMPs}
        \label{fig:cos-beta-y}
    \end{subfigure}
     \begin{subfigure}[b]{0.49\columnwidth}
        \includegraphics[width=\textwidth]{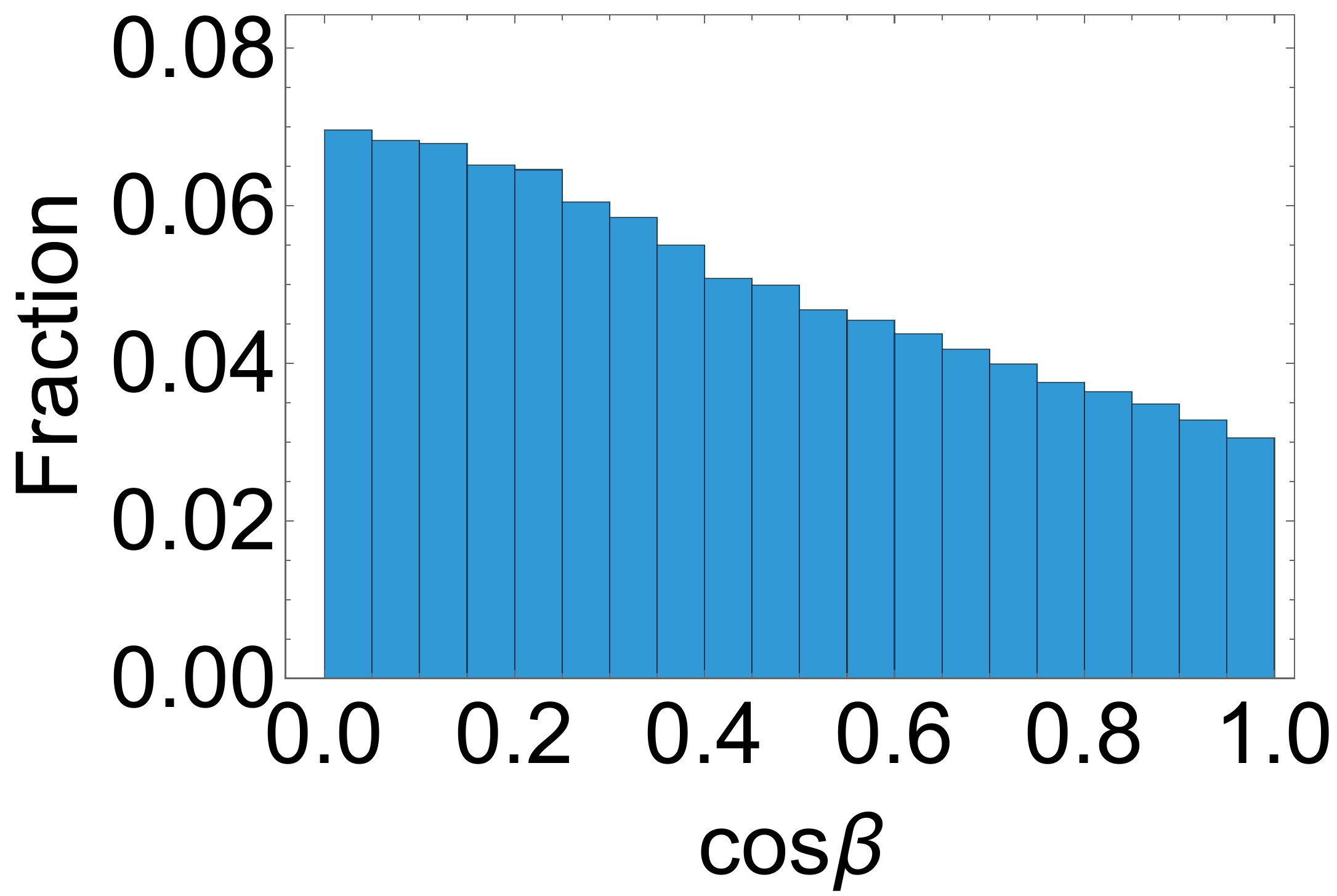}
        \caption{X-like SMPs}
        \label{fig:cos-beta-X}
    \end{subfigure}
      \begin{subfigure}[b]{0.49\columnwidth}
        \includegraphics[width=\textwidth]{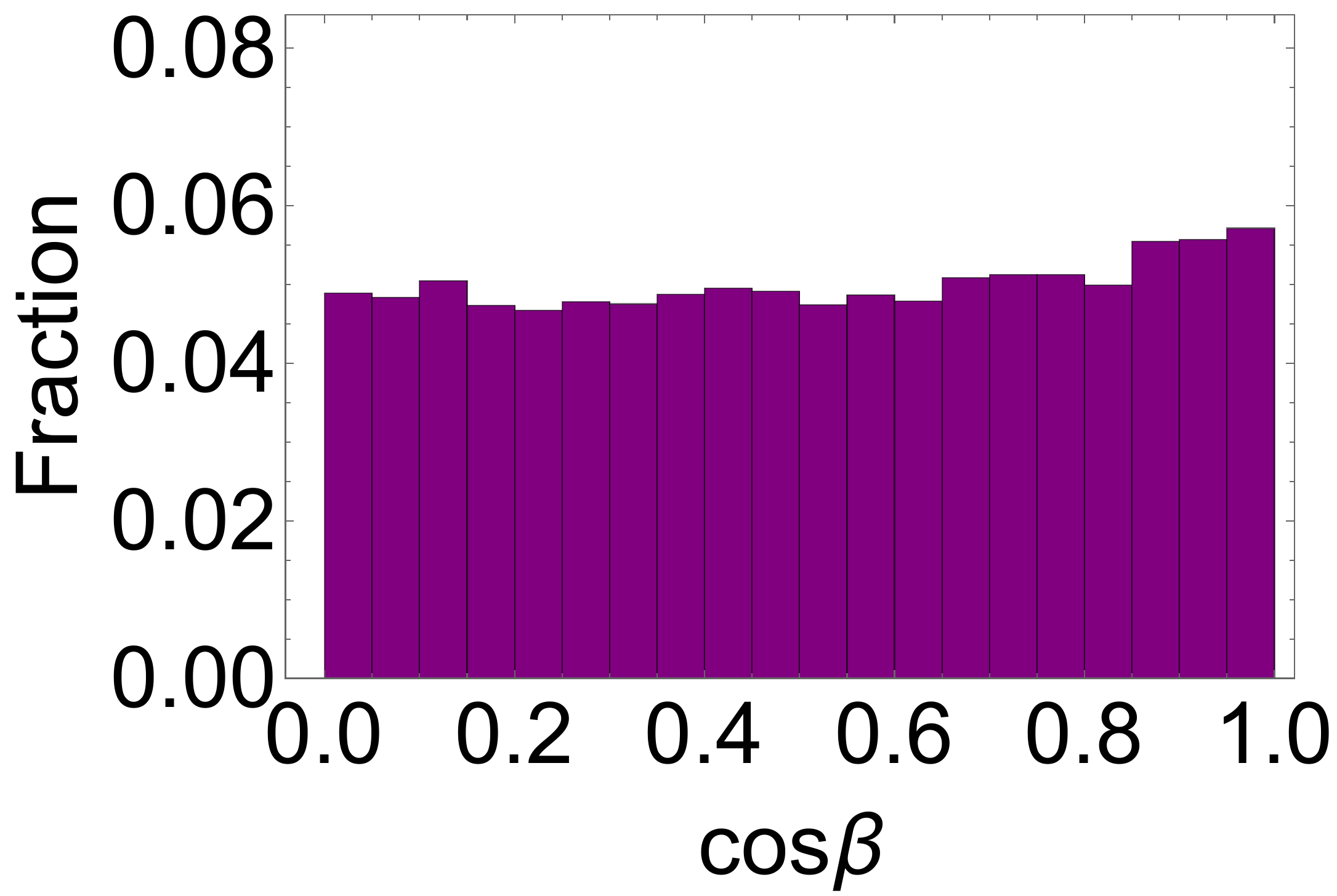}
        \caption{Ring-like SMPs}
        \label{fig:cos-beta-rings}
    \end{subfigure}
       \caption{Fraction of SMPs, whose magnetic moment $\vec{M}$, defined in Eq. \eqref{eq:magmom}, and the vector, connecting the centre of mass of the cluster and that of an SMP, form $\cos \beta$.  The values are averaged over all production runs. Subfigures are for clusters formed by (a) chain-like SMPs; (b) Y-like SMPs; (c) X-like SMPs and (d) by ring-like SMPs.}
       \label{fig:cos-mag-mom}
\end{figure}

Another interesting question is whether the orientation of an SMP inside the cluster is correlated with the orientation of its total dipole moment. So, as a second step, we computed the relative net magnetic moment of each SMP, $\vec{M}$, according to:
\begin{equation}
\vec{M}=\frac{1}{L\mu}\sum_{i=1}^L\vec{\mu}_i.
\label{eq:magmom}
\end{equation}

Analogously to $\alpha$, one can define angle $\beta$ between $\vec{M}$ of an SMP and the vector, connecting centres of mass of cluster and current SMP. In Fig. \ref{fig:cos-mag-mom} we plot the fraction of SMPs, whose magnetic moment forms angle $\beta$ with the vector connecting its centre of mass and the centre of mass of a cluster. 
Similar to Fig. \ref{fig:cos-main-axis}, here maximum at unity means the predominance of the radial orientation of $\vec{M}$ inside the cluster, whereas maximum at zero means the predominance of tangential orientation. Interestingly enough, Figs. \ref{fig:cos-beta-chain} -- \ref{fig:cos-beta-X} show that the orientation only weakly depends on the topology of SMPs: all the distributions monotonically decrease with growing $\cos \beta$, albeit for clusters made of Y-like SMPs the decay is weaker than for the others. The only uniform distribution of the dipole moment orientation can be found inside cluster made by ring-like SMPs, as seen in Fig. \ref{fig:cos-beta-rings}. However, one should keep in mind that the net dipole moment of a ring-like SMP is close to zero. As a result, one would not expect any kind of correlations between net dipoles of ring-like SMPs. 

Summarising this part of the analysis, we can say that the internal orientation of the SMPs in the formed clusters depends on their topology. Chain-like SMPs interlace and can have basically a random orientation inside the cluster, with high probability they can penetrate from surface to the centre of the cluster radially or as well oriented tangentially to the cluster surface. The rings, instead, do not mix and form clusters with onion structure and it is rather rare to find a ring oriented with its main axis radially inside the cluster. Y-like and X-like SMPs also have a preference to be tangential to the cluster surface, however both topologies might  also acquire radial orientation. The orientation of the net SMP dipole moment in clusters formed by all but ring-like SMPs is predominantly tangential and the probability of an orientation decreases monotonically, approaching the radial one. We find that for chain-, Y- and X-like SMPs, the probability of their dipoles to be oriented radially is at least twice lower than that of a tangential one. As for ring-like SMPs, their net dipole moment is infinitesimally small, as a result we find no preferred orientation of $\vec{M}$ in this case.

\subsection{Clusters inside: monomer level}
\begin{figure}[t!]
    \begin{subfigure}[b]{0.49\columnwidth}
        \includegraphics[width=\textwidth]{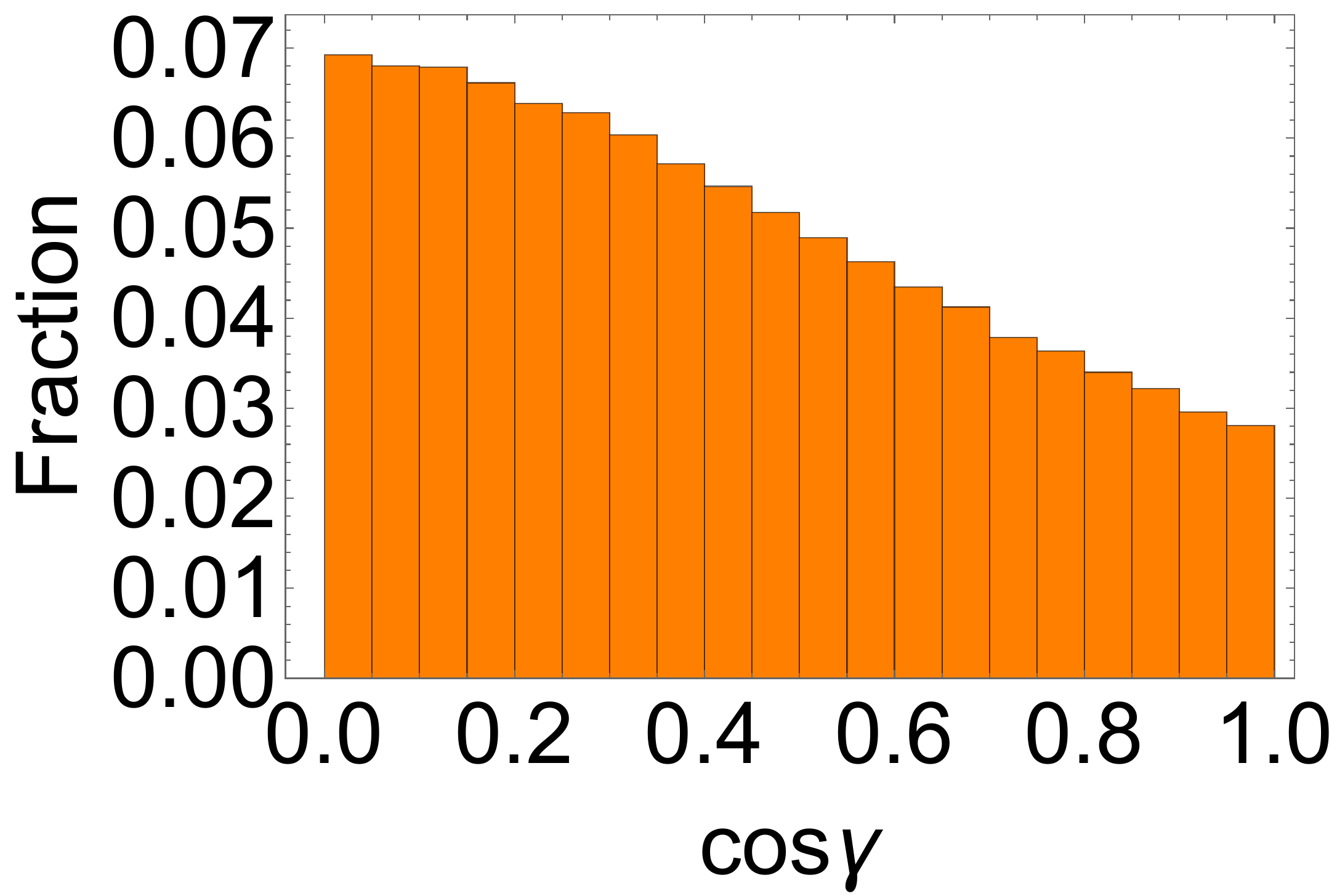}
        \caption{Chain-like SMPs}
        \label{fig:cos-gam-chains}
    \end{subfigure}
    \begin{subfigure}[b]{0.49\columnwidth}
        \includegraphics[width=\textwidth]{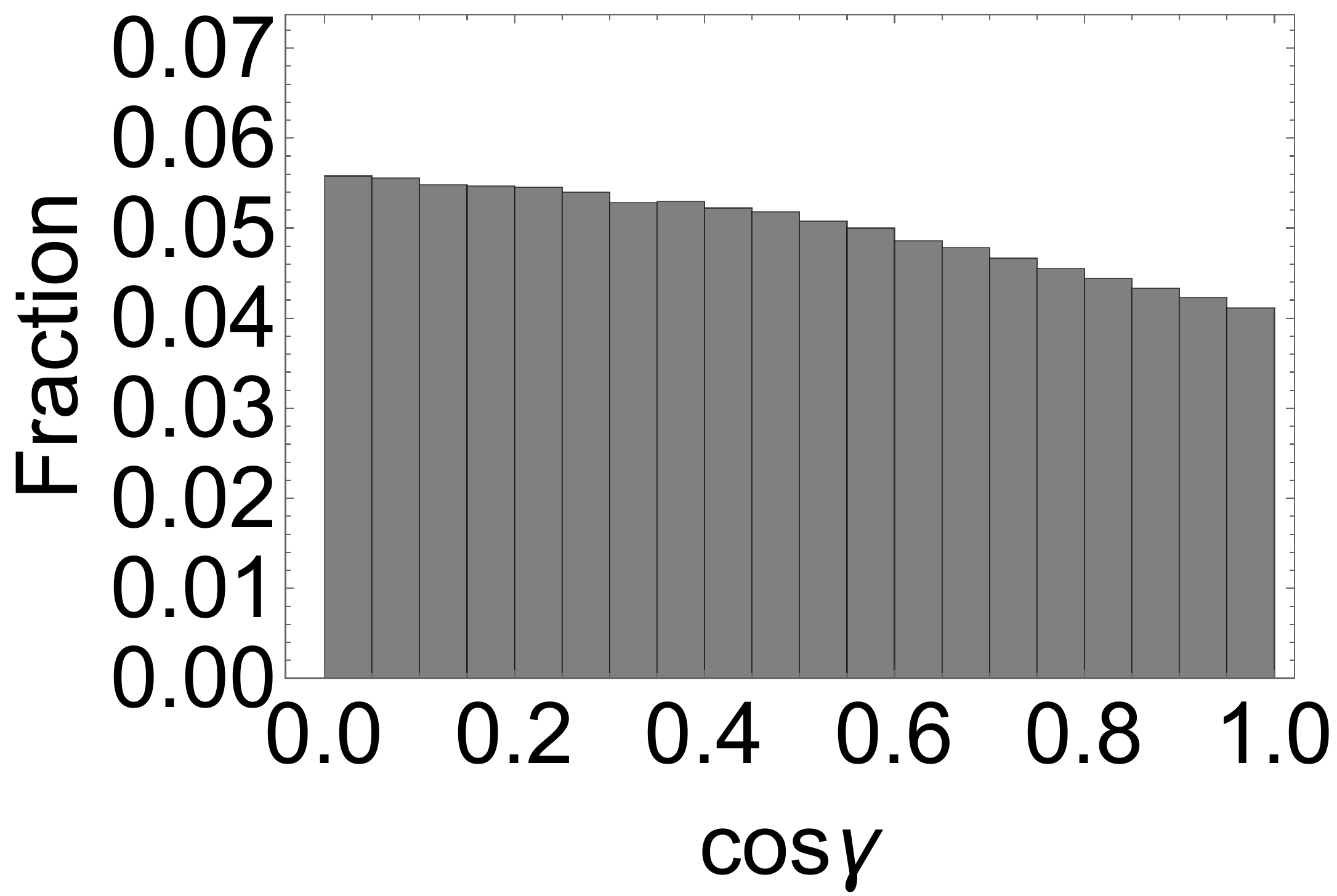}
        \caption{Y-like SMPs}
        \label{fig:cos-gam-Y}
    \end{subfigure}
     \begin{subfigure}[b]{0.49\columnwidth}
        \includegraphics[width=\textwidth]{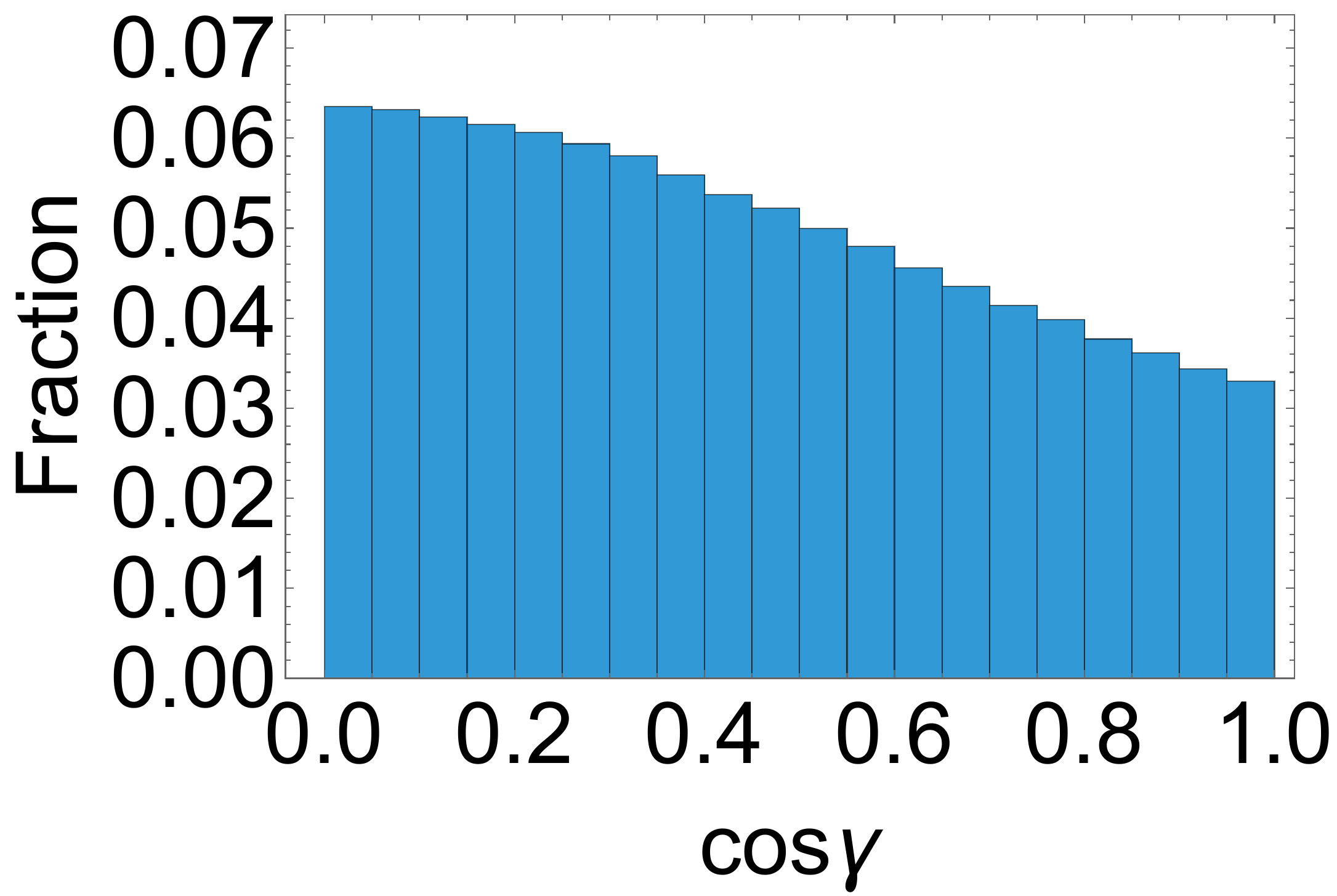}
        \caption{X-like SMPs}
        \label{fig:cos-gam-X}
    \end{subfigure}
      \begin{subfigure}[b]{0.49\columnwidth}
        \includegraphics[width=\textwidth]{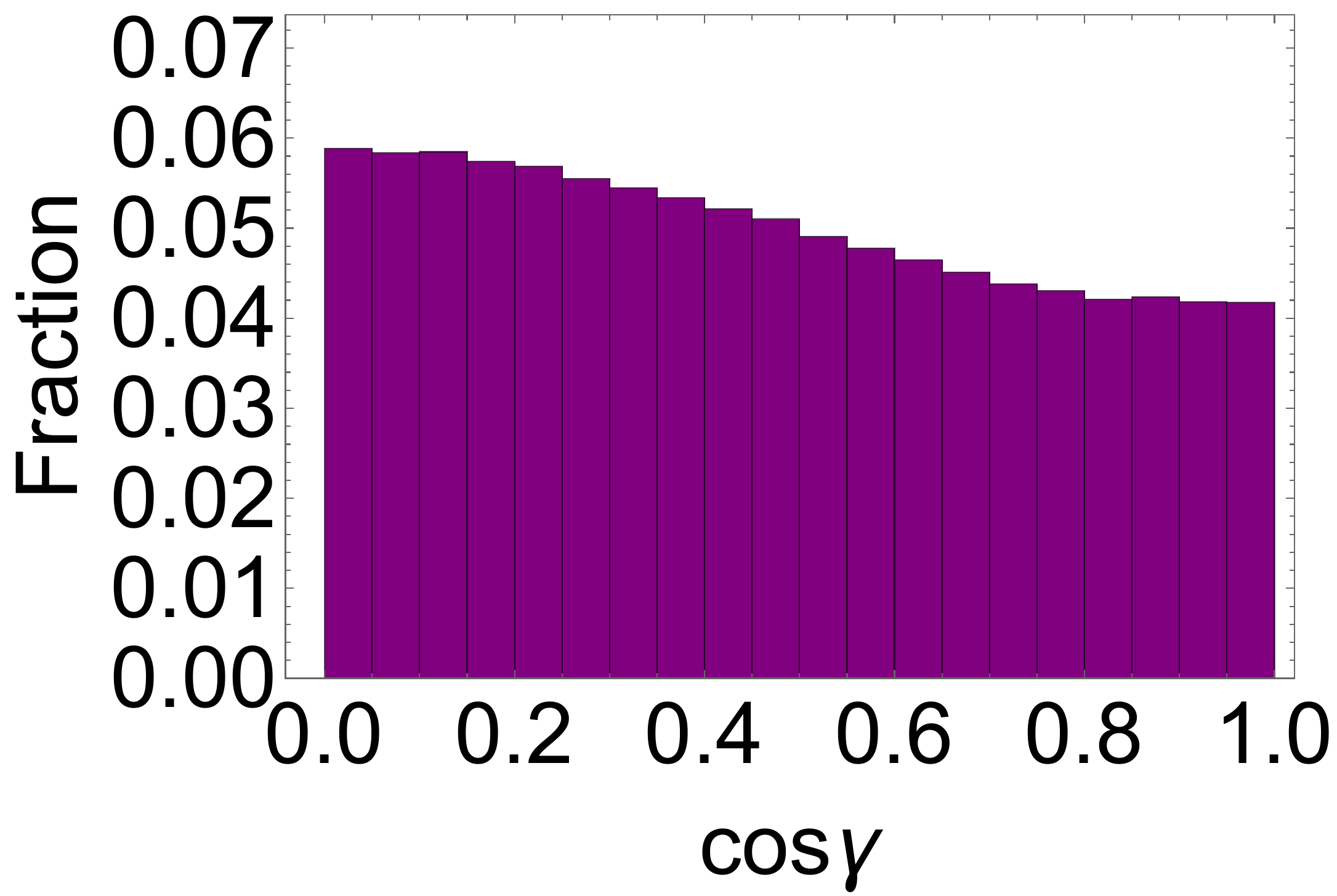}
        \caption{Ring-like SMPs}
        \label{fig:cos-gam-rings}
    \end{subfigure}
       \caption{Fraction of monomers, whose magnetic moment $\vec{\mu}$ and the vector, connecting the centre of mass of the cluster and that of an SMP, form $\cos \gamma$.  The values are averaged over all production runs. Subfigures are for clusters formed by (a) chain-like SMPs; (b) Y-like SMPs; (c) X-like SMPs and (d) by ring-like SMPs.}
       \label{fig:cos-mag-mom-mon}
\end{figure}

After having investigated the orientations of SMPs net magnetic moments, we  look at  individually magnetic moments $\vec{\mu}$ of monomers inside the clusters. In order to do that, for each monomer, we calculated the angle, $\gamma$, between $\vec{\mu}$ and the vector connecting the centre of the monomer to the centre of mass of the cluster. The results are presented in Fig. \ref{fig:cos-mag-mom-mon}. Here, it becomes clear that independently from the SMP topology, monomers in them have a preference to orient tangentially inside the clusters. This is an outcome of the optimisation of dipolar interactions. Topology of an SMP can slightly enhance or inhibit the tendency to be oriented tangentially. Thus, for chain-like SMPs shown in Fig. \ref{fig:cos-gam-chains}, the ratio between the fraction of monomers oriented tangentially and radially is larger than two, for monomers in X-like SMPs this ratio is slightly smaller than two (Fig. \ref{fig:cos-gam-X}). If one calculates the same ratio for monomers in Y-like and ring-like SMPs (Figs. \ref{fig:cos-gam-Y} and \ref{fig:cos-gam-rings} respectively) it will be around 40 per cent, {\it{i.e.}} much smaller than for other two topologies. 

Turning again to the lower row in Fig. \ref{fig:snaps-n-skel}, one can now indeed notice that the dipoles form vortexes. 

\begin{figure}[h!]
    \begin{subfigure}[b]{0.49\columnwidth}
        \includegraphics[width=\textwidth]{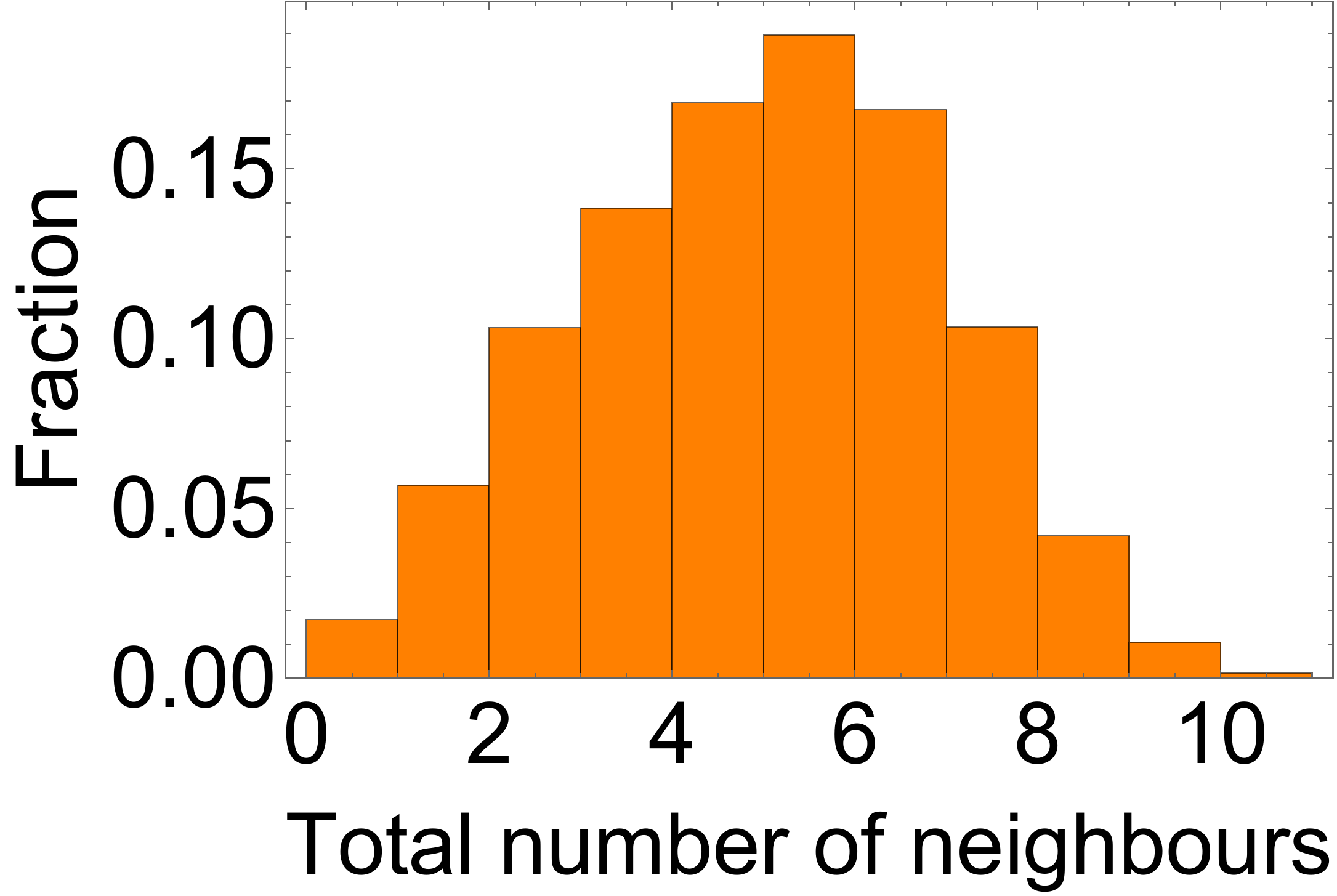}
        \caption{Chain-like SMPs}
        \label{fig:neigh-all-chain}
    \end{subfigure}
    \begin{subfigure}[b]{0.49\columnwidth}
        \includegraphics[width=\textwidth]{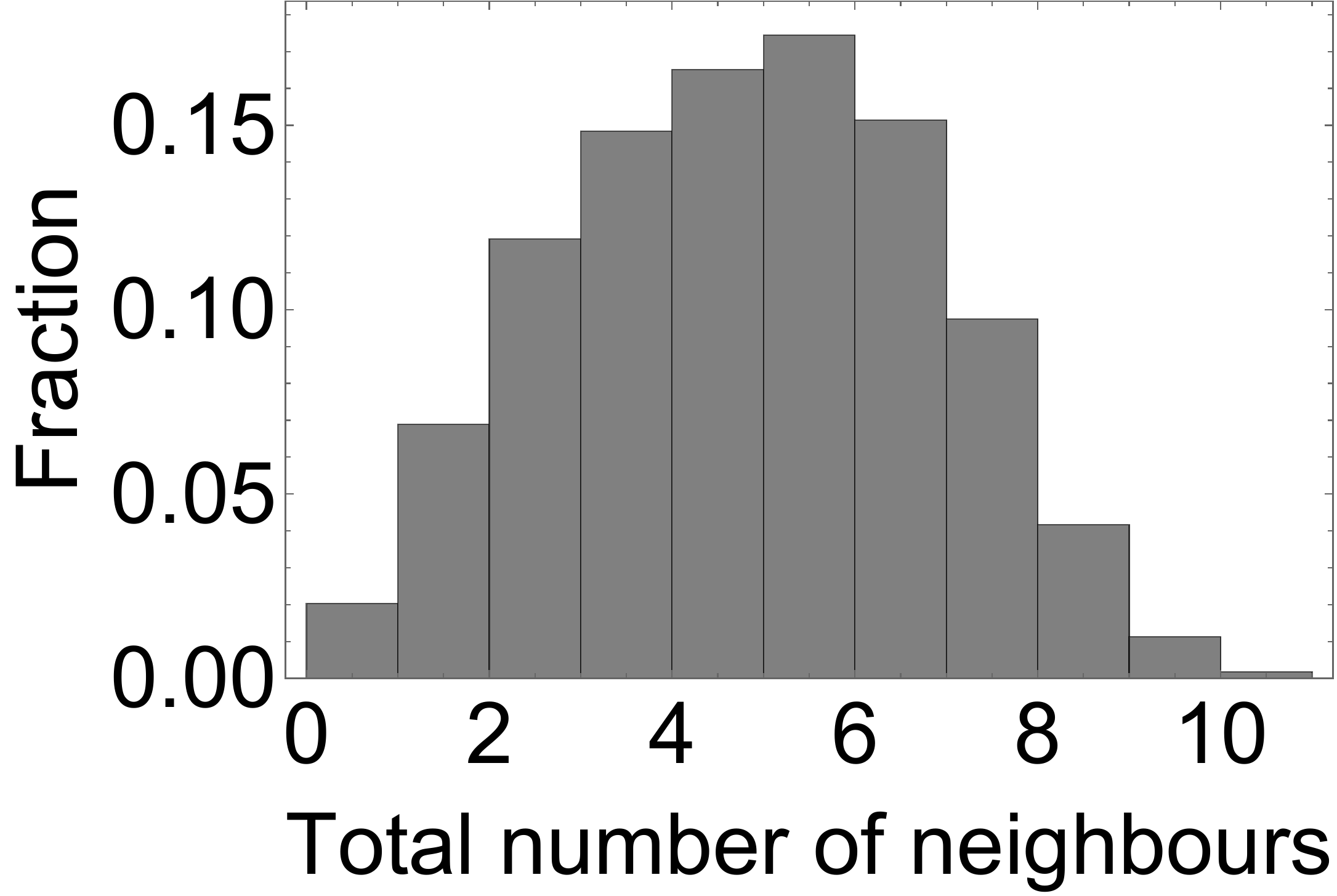}
        \caption{Y-like SMPs}
        \label{fig:neigh-all-y}
    \end{subfigure}
     \begin{subfigure}[b]{0.49\columnwidth}
        \includegraphics[width=\textwidth]{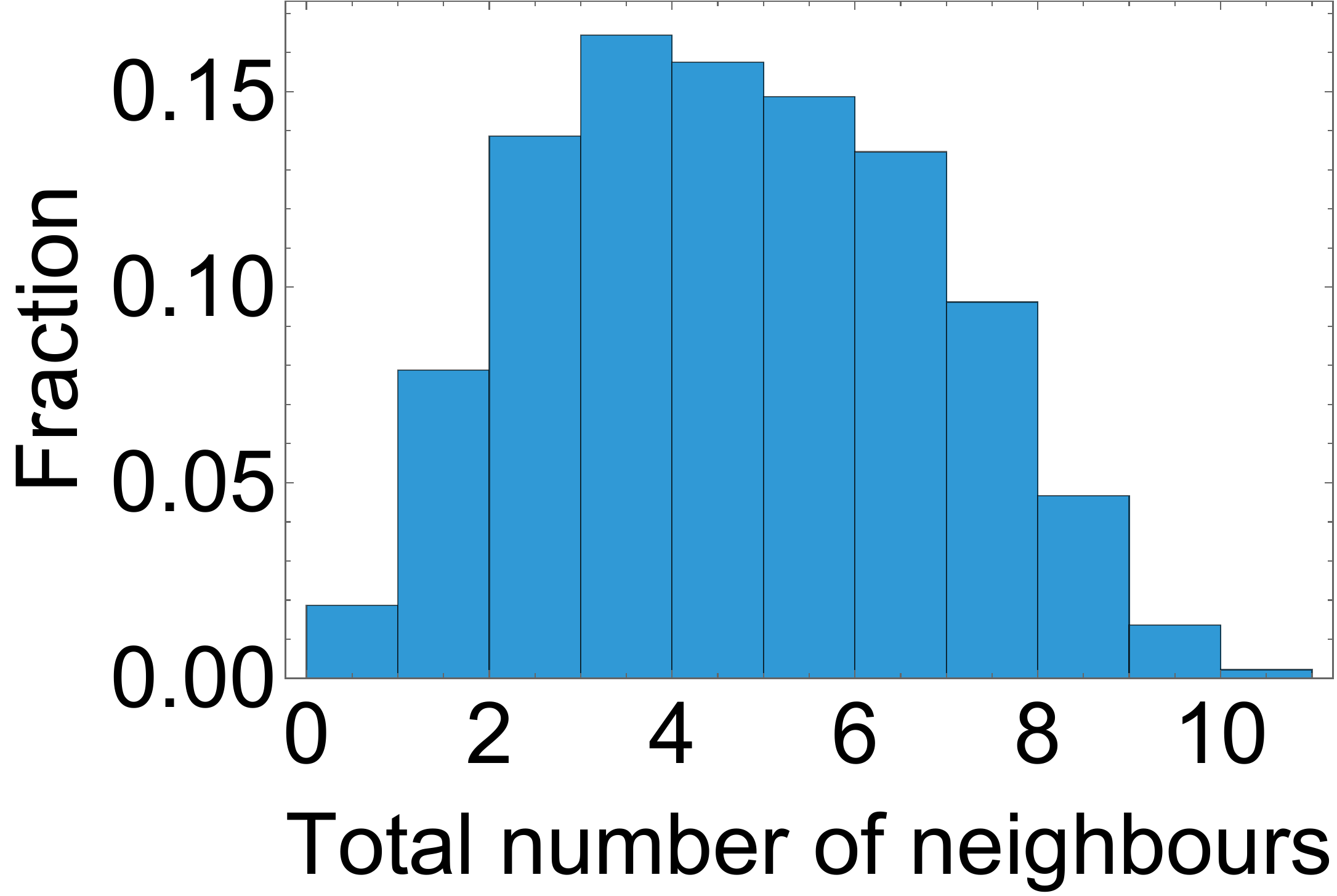}
        \caption{X-like SMPs}
        \label{fig:neigh-all-x}
    \end{subfigure}
      \begin{subfigure}[b]{0.49\columnwidth}
        \includegraphics[width=\textwidth]{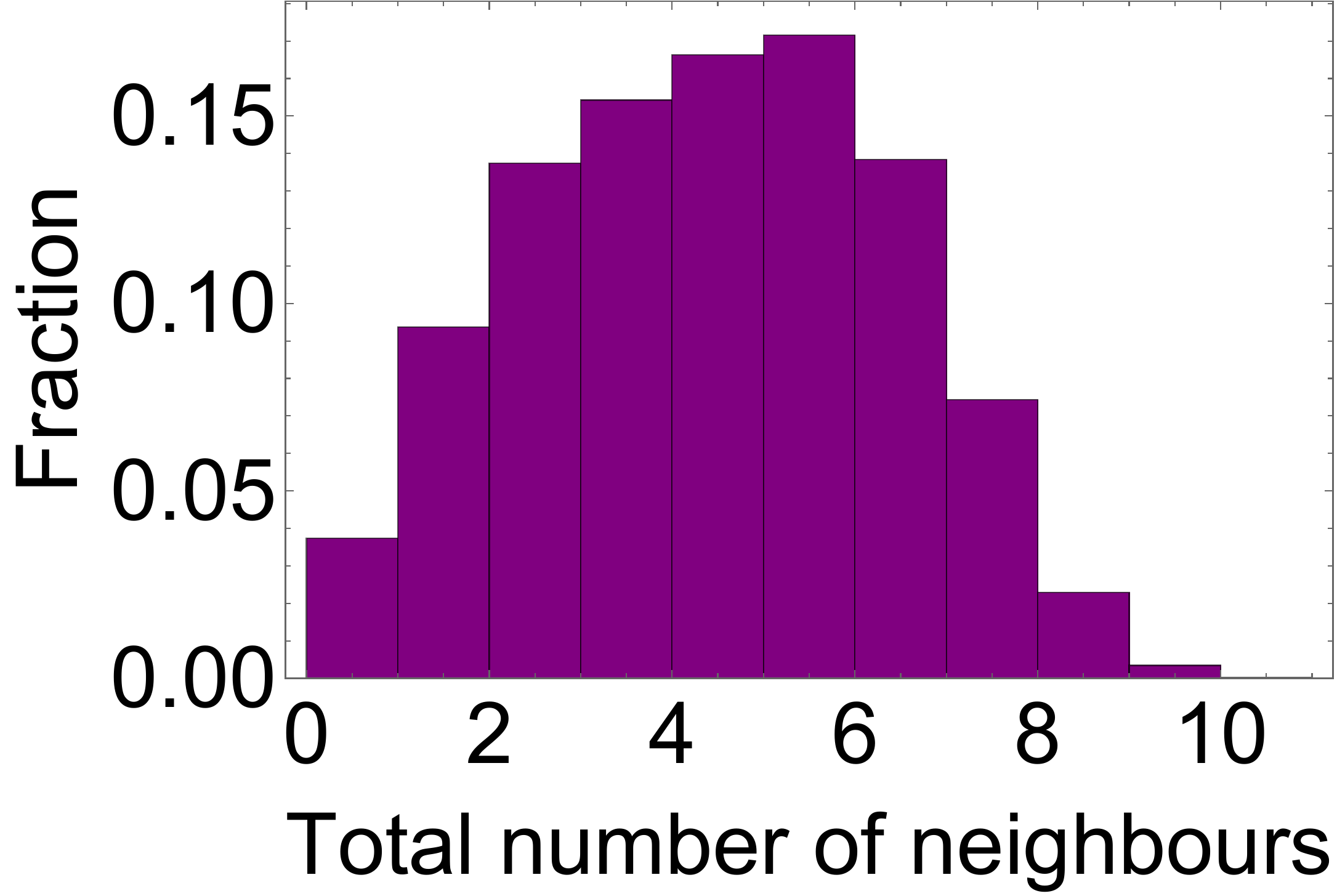}
        \caption{Ring-like SMPs}
        \label{fig:neigh-all-ring}
    \end{subfigure}
       \caption{Histogram of the total amount of non-permanent neighbours that each monomer has in a cluster. No discrimination is made: both monomers from the same SMP and from different ones are counted. The values are averaged over all production runs. Subfigures are for clusters formed by (a) chain-like SMPs; (b) Y-like SMPs; (c) X-like SMPs and (d) by ring-like SMPs.}
       \label{fig:neighbours-all}
\end{figure}

\begin{figure}[h]
    \begin{subfigure}[b]{0.49\columnwidth}
        \includegraphics[width=\textwidth]{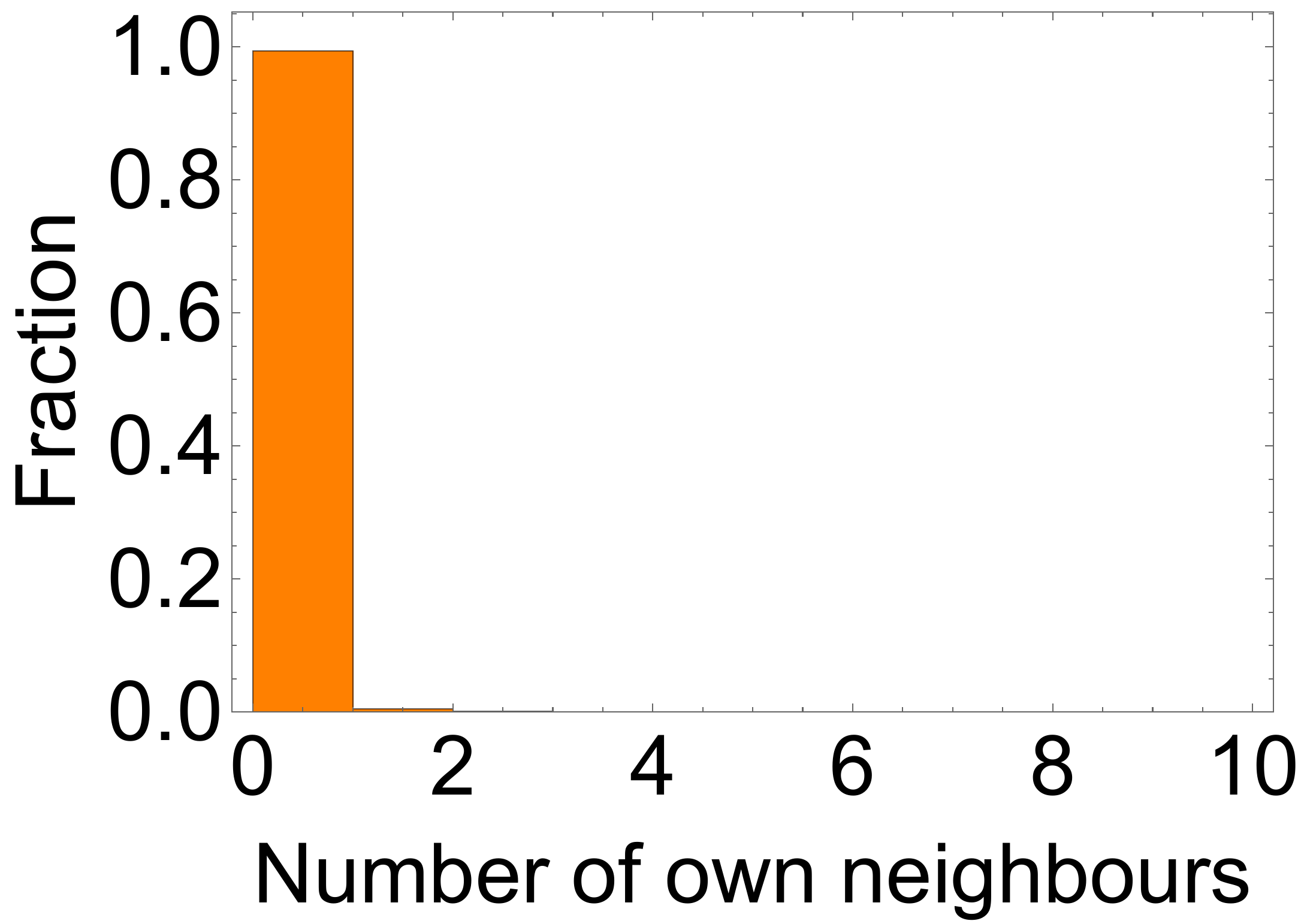}
        \caption{Chain-like SMPs}
        \label{fig:-neigh-self-chain}
    \end{subfigure}
    \begin{subfigure}[b]{0.49\columnwidth}
        \includegraphics[width=\textwidth]{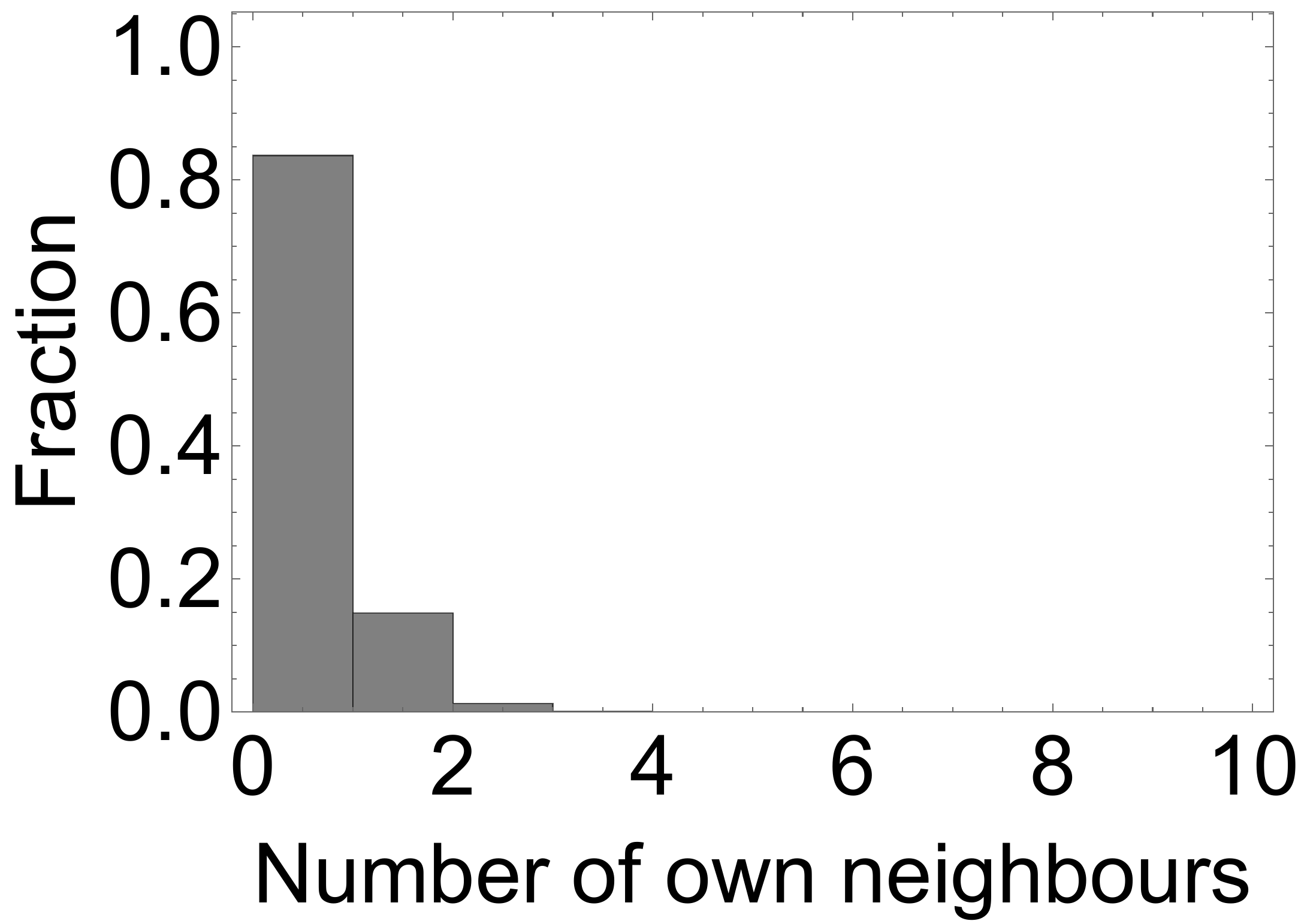}
        \caption{Y-like SMPs}
        \label{fig:-neigh-self-y}
    \end{subfigure}
     \begin{subfigure}[b]{0.49\columnwidth}
        \includegraphics[width=\textwidth]{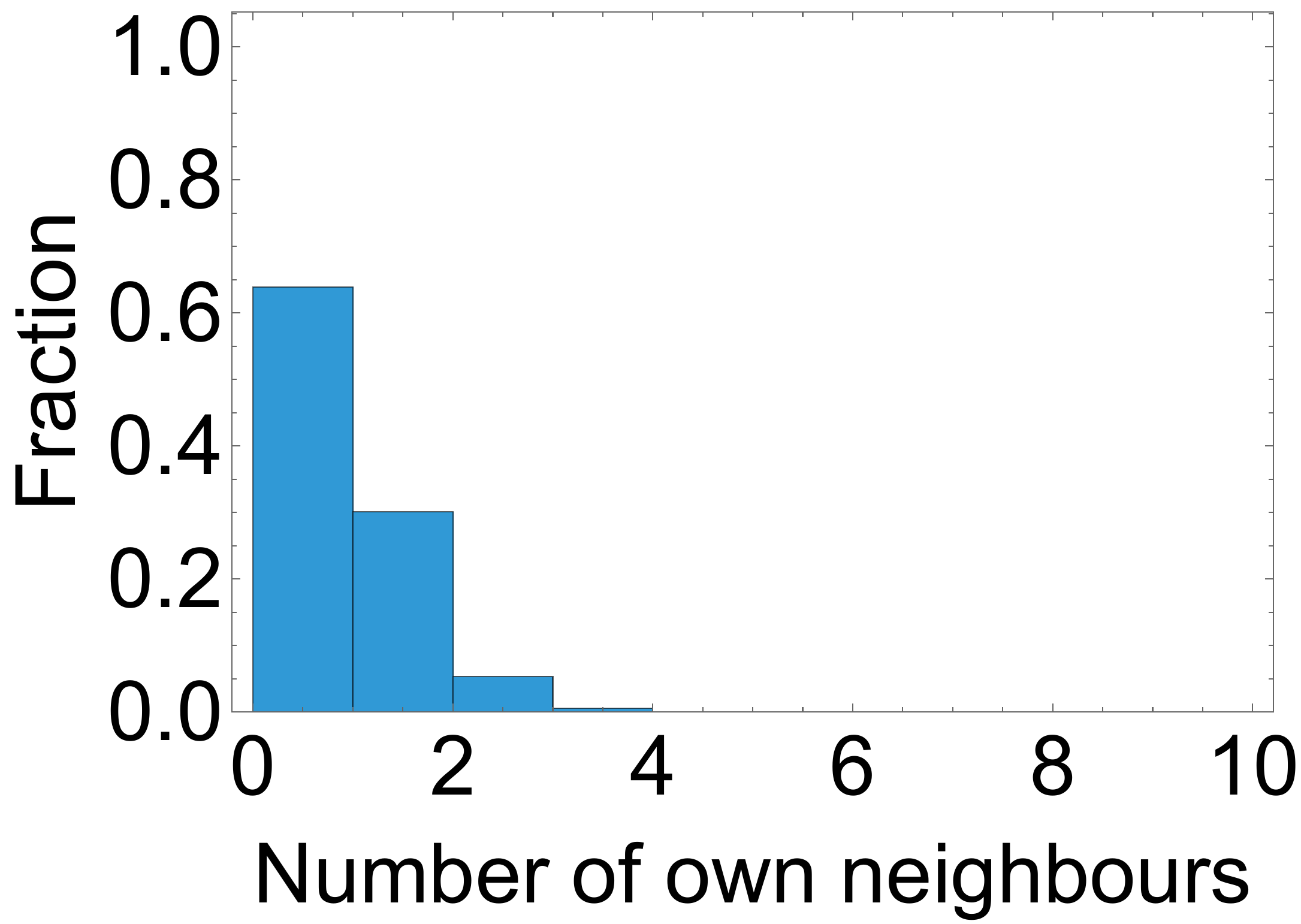}
        \caption{X-like SMPs}
        \label{fig:-neigh-self-x}
    \end{subfigure}
      \begin{subfigure}[b]{0.49\columnwidth}
        \includegraphics[width=\textwidth]{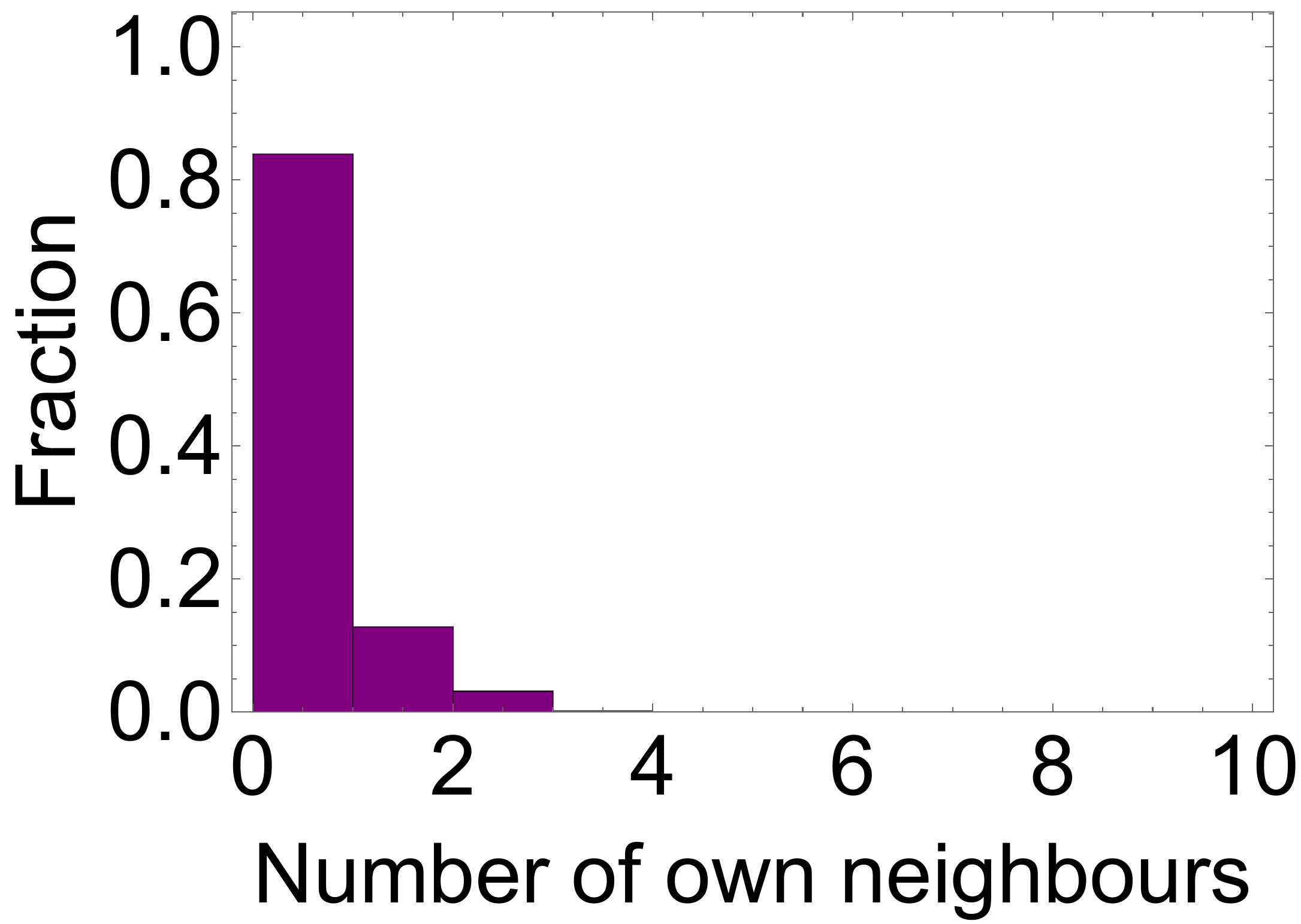}
        \caption{Ring-like SMPs}
        \label{fig:-neigh-self-ring}
    \end{subfigure}
       \caption{Histogram of the total amount of non-permanent neighbours from the same SMP that a monomer in the cluster has. The values are averaged over all production runs. Subfigures are for clusters formed by (a) chain-like SMPs; (b) Y-like SMPs; (c) X-like SMPs and (d) by ring-like SMPs.}
       \label{fig:neighbours-self}
\end{figure}

As indicated by Fig. \ref{fig:neighbours-all}, where we plot the fractions of monomers with a given number of nearest neighbours, on the level of monomers all clusters are relatively dense. Light asymmetry of the plots does not affect the main conclusion. In order to understand if monomers in various SMPs in the clusters are mainly connected to monomers from different SMPs or to monomers of the SMP they belong to, we calculated how many non-permanent bonds inside one SMP can be found. In other words, for each monomer, we calculated how many neighbours it has that belong to the same SMP and are not precrosslinked ones. The results are shown in Fig. \ref{fig:neighbours-self}. Here, one can notice again the influence of SMP topology. Thus, monomers in chain-like SMPs, see Fig. \ref{fig:-neigh-self-chain}, are not likely to have non-permanent neighbours within one SMP, whereas for Y-, X- and ring-like SMPs (Figs. \ref{fig:-neigh-self-y} --  \ref{fig:-neigh-self-ring}), the probability of self-touching is higher, albeit still not significant.  The fraction of such neighbours point to low degree of SMP folding inside the clusters, that is the result of the dipolar forces that favour head-to-tail orientations of monomer dipoles and the coupling between dipolar orientation with the structure backbone. The most likely to have self-contacts are X-like SMPs due to the fact that they have the shortest arms and the highest induced monomer proximity by design.

\begin{figure}[t!]
    \begin{subfigure}[b]{0.49\columnwidth}
        \includegraphics[width=\textwidth]{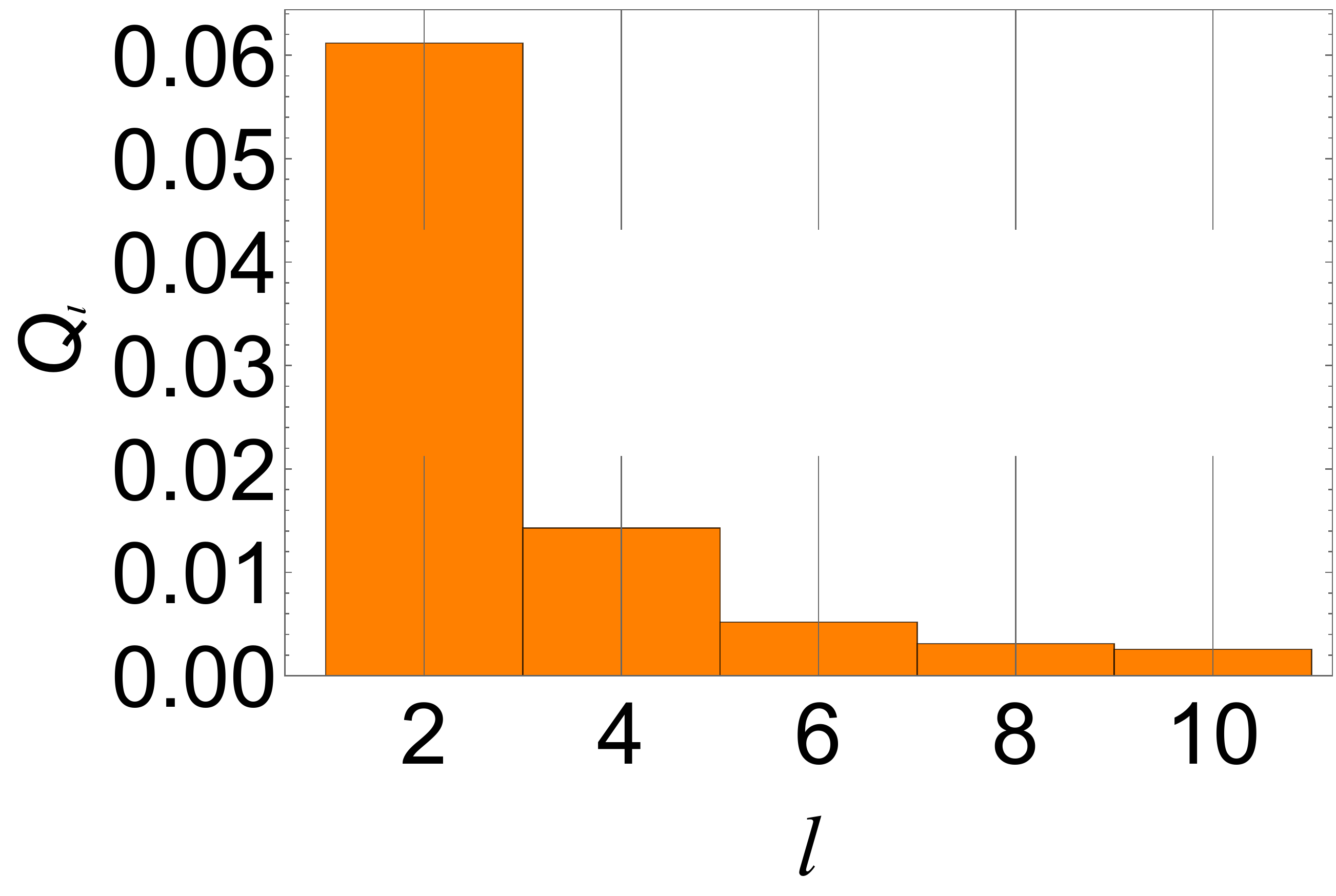}
        \caption{Chain-like SMPs}
        \label{fig:cumulant-chains}
    \end{subfigure}
    \begin{subfigure}[b]{0.49\columnwidth}
        \includegraphics[width=\textwidth]{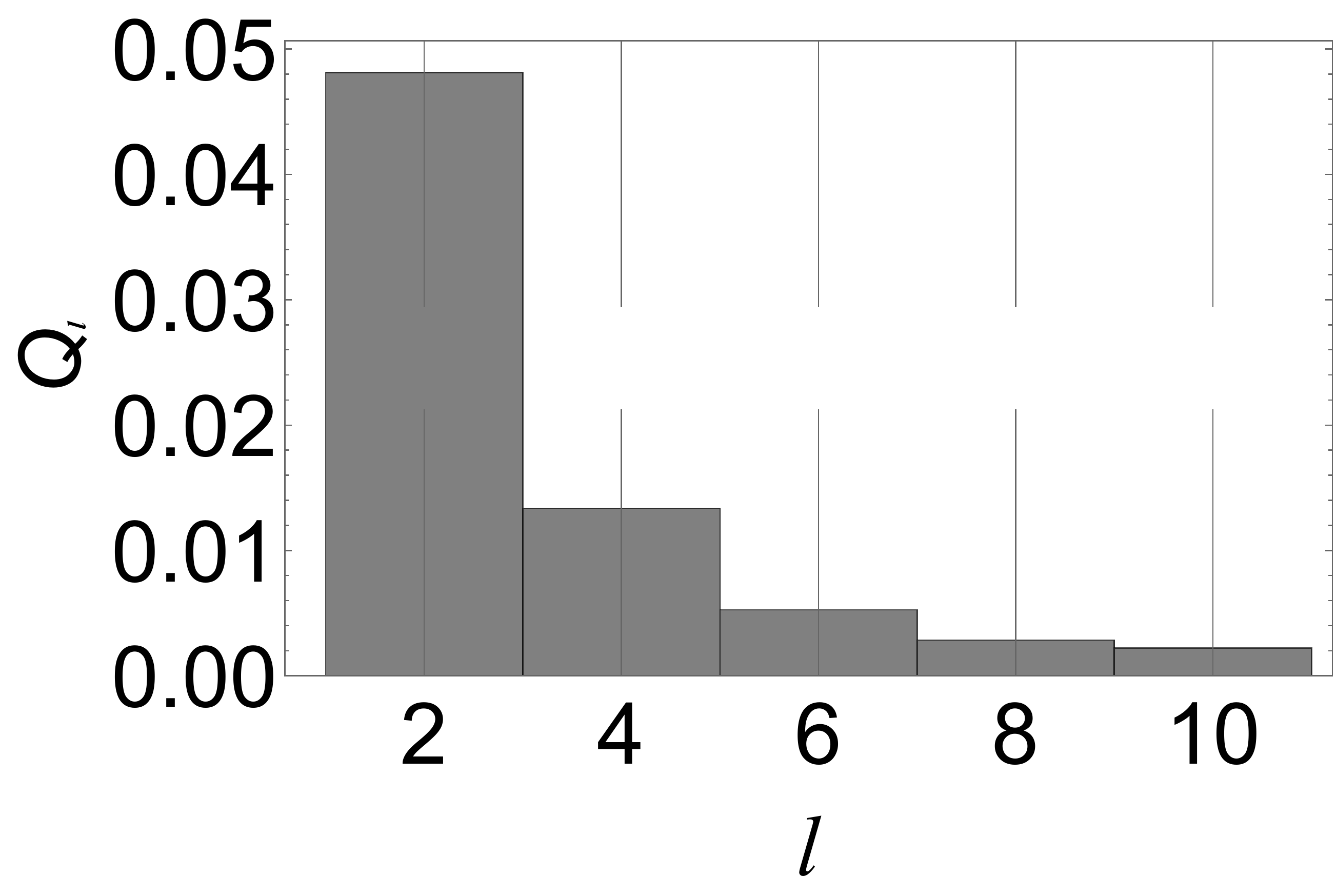}
        \caption{Y-like SMPs}
        \label{fig:cumulant-Y}
    \end{subfigure}
     \begin{subfigure}[b]{0.49\columnwidth}
        \includegraphics[width=\textwidth]{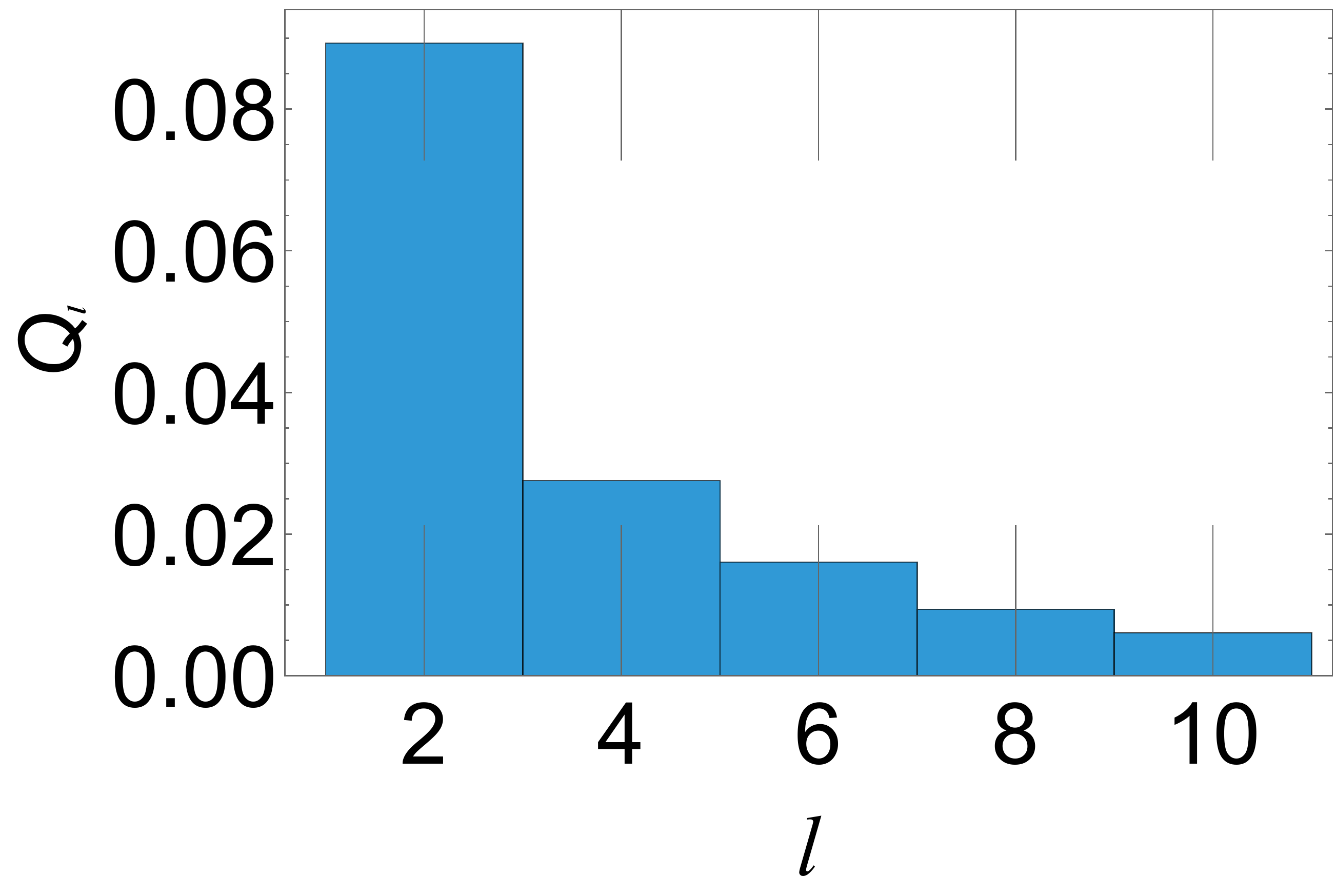}
        \caption{X-like SMPs}
        \label{fig:cumulant-X}
    \end{subfigure}
      \begin{subfigure}[b]{0.49\columnwidth}
        \includegraphics[width=\textwidth]{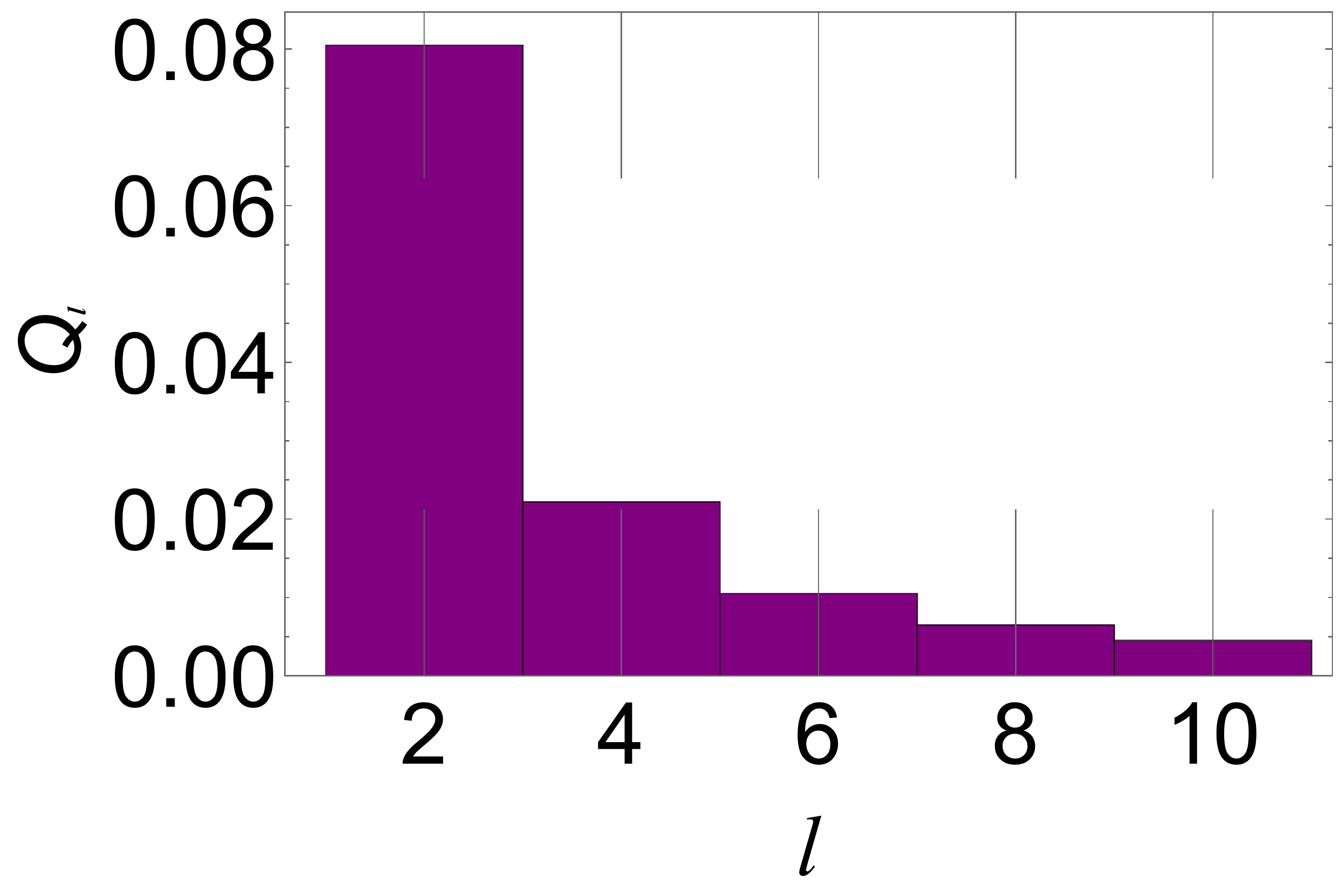}
        \caption{Ring-like SMPs}
        \label{fig:cumulant-rings}
    \end{subfigure}
       \caption{Histograms of $Q_l$, Eq. \eqref{eq:Qfinal}.  The values are averaged over all production runs. Subfigures are for clusters formed by (a) chain-like SMPs; (b) Y-like SMPs; (c) X-like SMPs and (d) by ring-like SMPs.}
       \label{fig:cumulants}
\end{figure}

In order to quantitatively characterise spatial monomer distribution and rule out the possibility of crystallisation, we calculated the order parameters for all clusters and SMPs topologies:
\begin{equation} \label{eq:order-param}
Q_{lm}(\vec r)=Y_{lm}\left(\theta(\vec r),\phi(\vec r)\right),
\end{equation}
\noindent where $Y_{lm}$ are spherical harmonics of the respective order, $lm$, and angles $\theta$ and $\phi$ are correspondingly azimuthal and polar angles of a bond between two neighbouring monomers inside the cluster, characterised by a vector $\vec{r}$,  in a lab reference frame. The function $Q_{lm}(\vec r)$ is averaged over all bonds in the cluster, then over all clusters and, finally, over all snapshots:
\begin{equation}
\overline{Q}_{lm}=\langle Q_{lm}\rangle.
\end{equation}
The resulting cumulant $Q_l$,
\begin{equation}\label{eq:Qfinal}
{Q}_{l}=\left[\frac{4\pi}{2l+1}\sum \limits_{m=-l}^{l}|\overline{Q}_{lm}|^2 \right]^{1/2},
\end{equation}
\noindent is plotted in Fig.~\ref{fig:cumulants} in the form of histograms for all four SMPs topologies.

It is known that based on $Q_l$ distribution, one can distinguish between different crystalline ordering in the system \cite{steinhardt83a}. Looking at the histograms obtained for SMP clusters, we can safely conclude that they evidence no crystalline structure, as the values of $Q_l$ are not only very low (below those found for supercooled Lennard-Jones liquid in Ref. \cite{steinhardt83a}), but also decay monotonically with $l$. 

Summarising this subsection we can say that on the monomer level the cluster structure is basically the same independently from the topology of the SMPs that form them. The only notable difference occurs on the level of SMP self-contacts inside the cluster, thus chain-like SMPs never self touch and so, the clusters will show the highest degree of mixing, whereas X-like SMPs are forming contacts inside themselves, thus presumably leading to less stable clusters, if any external force is to be applied.

\subsection{Comparison to the Stockmayer fluid with non-crosslinked monomers}

The final step in analysing the influence of permanent crosslinkers on the structure of clusters formed in Stockmayer systems, is to compare the properties of SMP clusters to that of non-crosslinked Stockmayer monomers. 
\begin{figure}[h!]
\includegraphics[width=0.65\columnwidth]{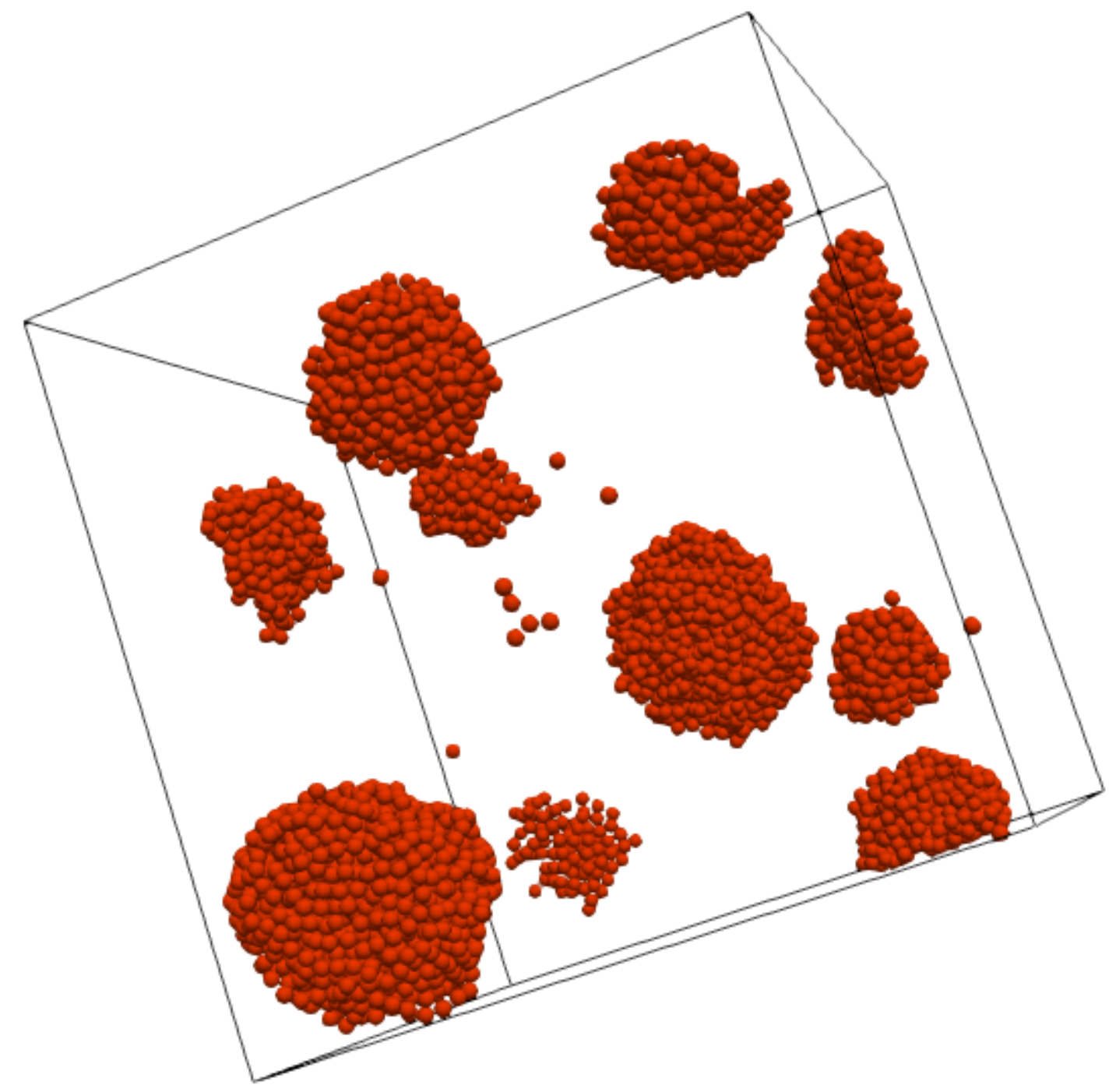}
        \caption{Typical snapshot of a Stockmayer fluid, $\mu^2 = 5$; $\rho^{\ast}=0.05$; $T=1$; $\sigma=1$.}
        \label{fig:ff-snapshot}
\end{figure}
In order to make a fair comparison, we perform simulations of the Stockmayer system of independent monomers, whose size, dipole moment and volume fraction coincide  with their counterparts in simulations of SMPs: $\sigma=1$, $\mu^2 = 5$ and $\rho^{\ast} = 0.05$. Moreover, the simulations for this system were equally long as the simulations with SMPs. It is worth mentioning that with this parameters, the system corresponds to the metastable area of $T-\rho$--phase diagrams \cite{1993-vanleeuwen,bartke07a}. Thus, the clusters we see, are indeed, only the precursors of phase separation.

In Fig. \ref{fig:ff-snapshot}, we present a characteristic snapshot of such clusters formed by non-crosslinked dipolar monomers. Even a visual comparison of this system to Fig. \ref{fig:snap_solution} shows that the clusters are bigger and they are less. In fact, the number of the clusters decreases by the factor of two once the monomer crosslinkers were removed, whereas the size distribution lays between 100 and 2300 monomers, in comparison to Y-like SMPs, in which the largest observed cluster contains 1430 monomers. As mentioned in Section \ref{sec:level1}, the largest clusters observed in simulations of Stockmayer SMPs were found for chains: the biggest observed cluster contained 270 SMPs (2700 monomers), but all others had no more than 110 SMPs.

\begin{figure}[t!]
    \begin{subfigure}[b]{0.49\columnwidth}
        \includegraphics[width=\textwidth]{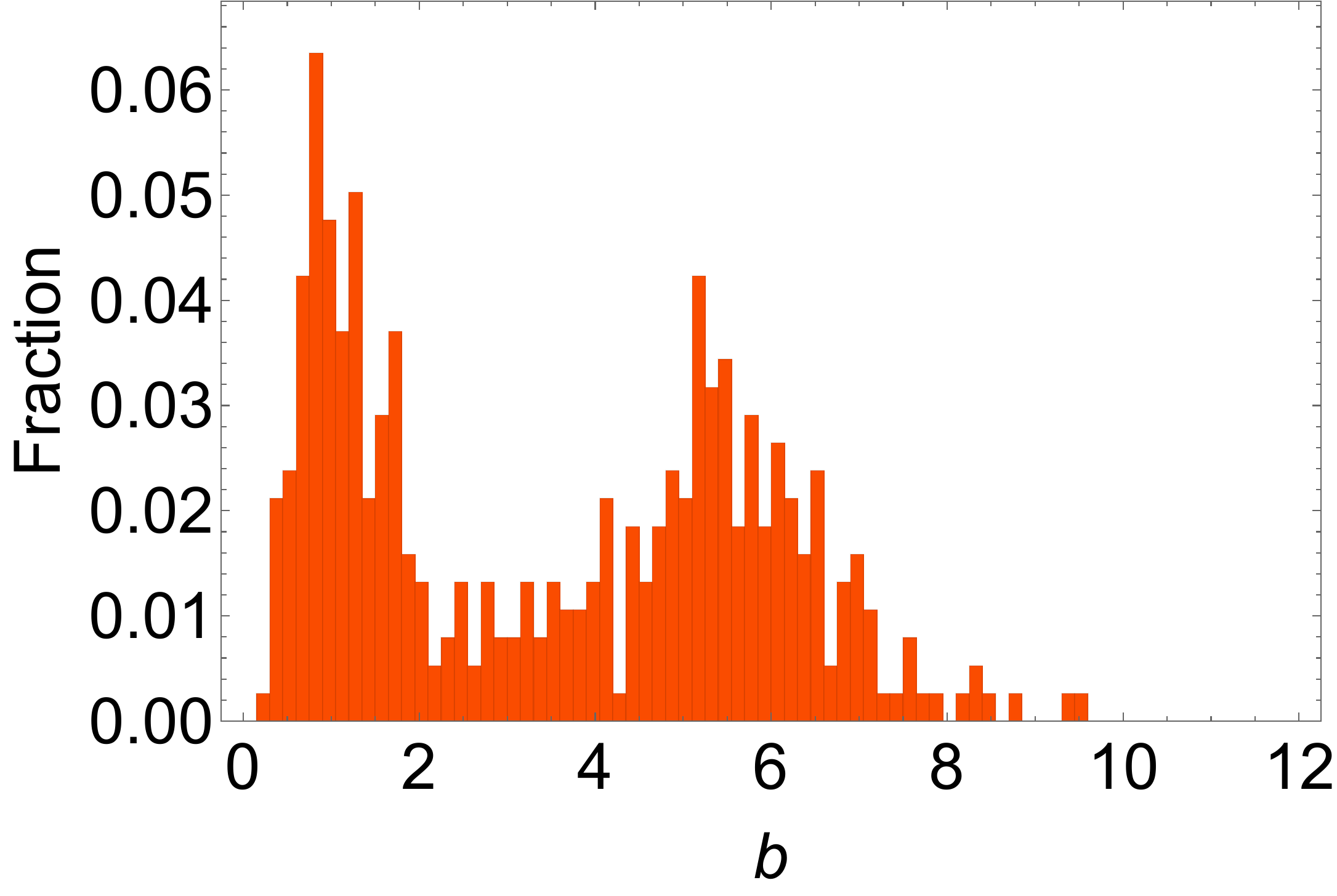}
        \caption{}
     \label{fig:aspher-ff}   
    \end{subfigure}
    \begin{subfigure}[b]{0.49\columnwidth}
        \includegraphics[width=\textwidth]{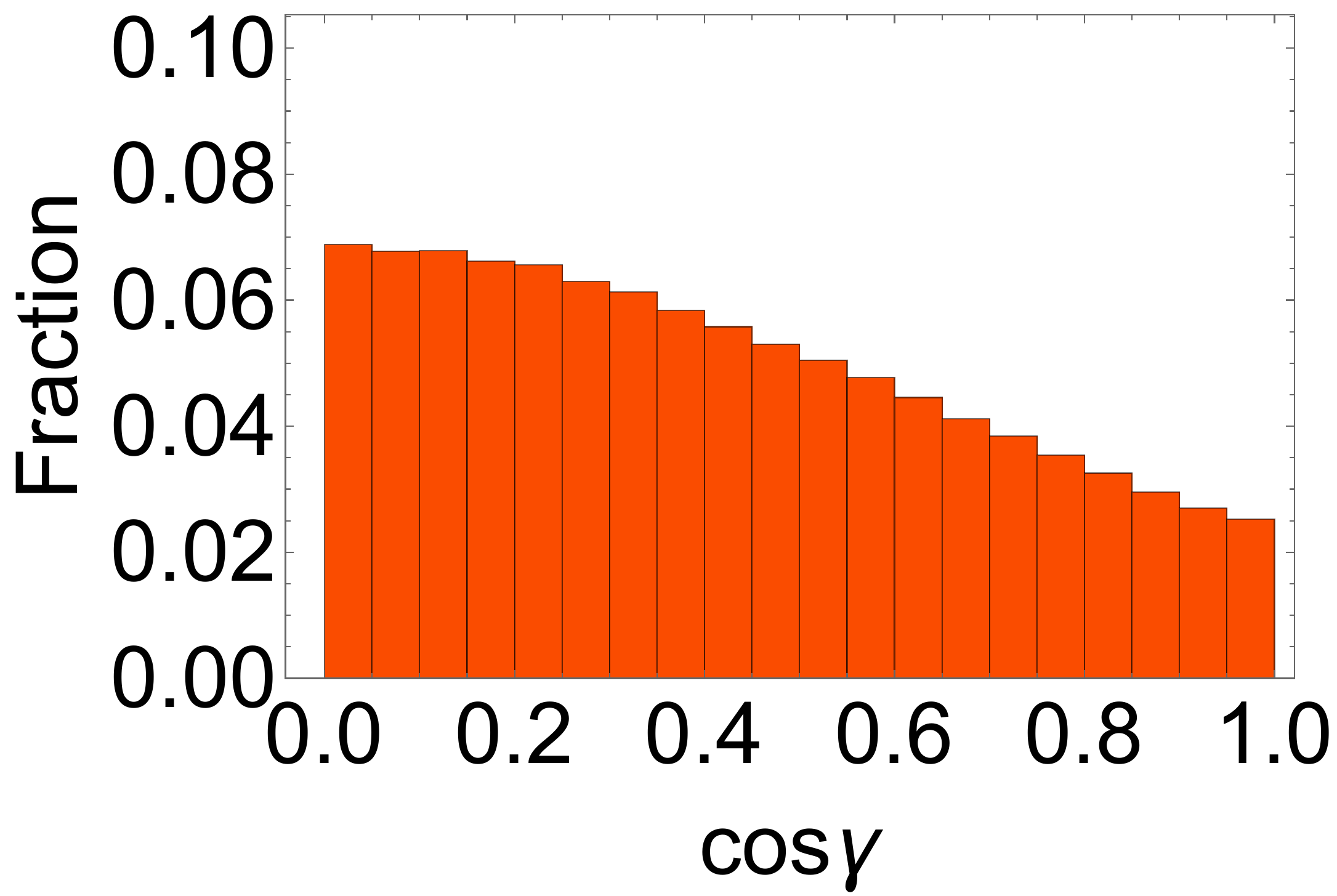}
          \caption{}
        \label{fig:cos-gam-ff}
    \end{subfigure}
     \begin{subfigure}[b]{0.49\columnwidth}
        \includegraphics[width=\textwidth]{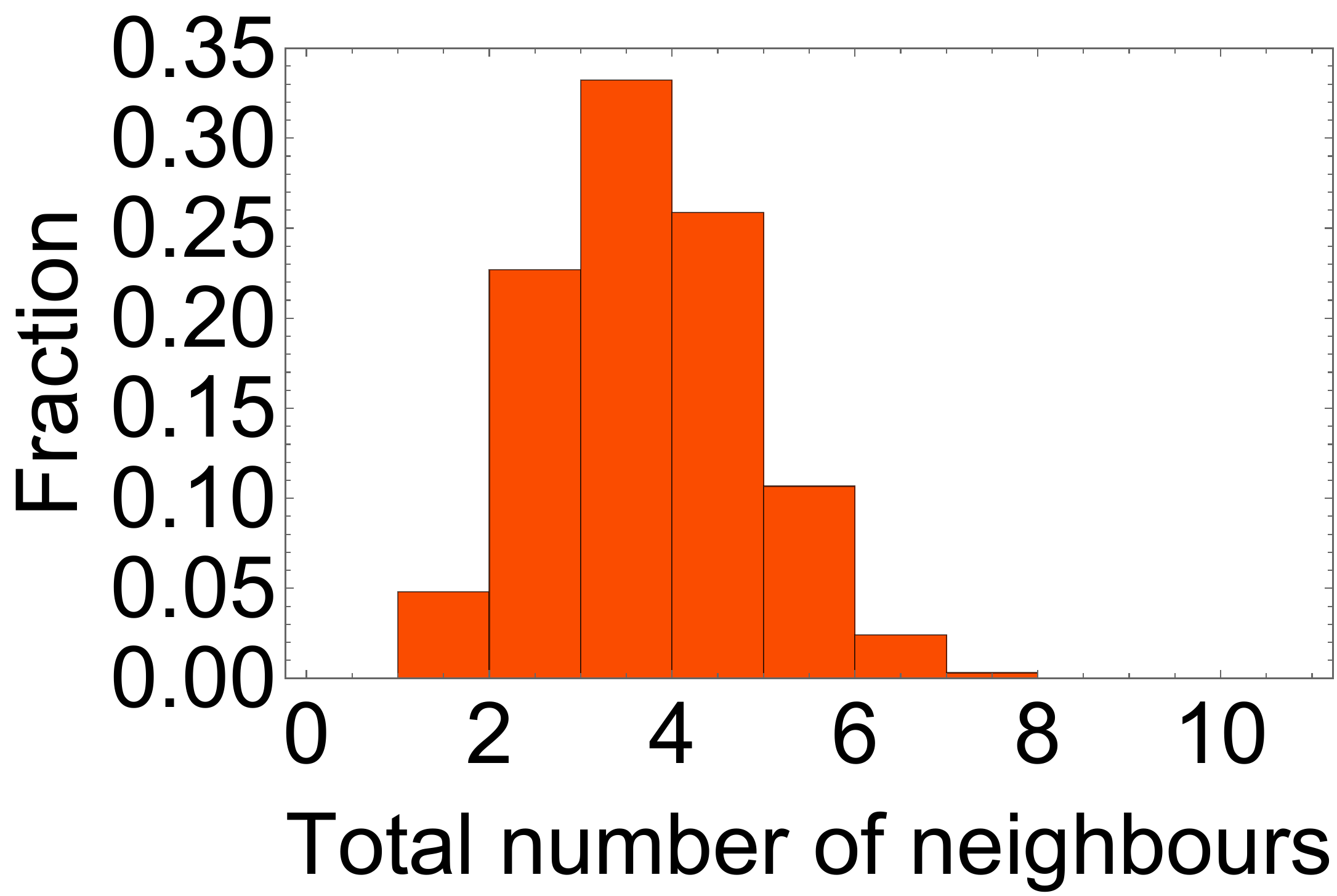}
          \caption{}
        \label{fig:neigh-all-ff}
    \end{subfigure}
       \caption{For Stockmayer fluid (a) asphericity histogram (to be compared to Fig. \ref{fig:asphericity}); (b) $\cos\gamma$ (to be compared to Fig. \ref{fig:cos-mag-mom-mon}); (c) the histogram of the total amount of nearest neighbours each monomer has (to be compared to Fig. \ref{fig:neighbours-all}).}
       \label{fig:ff-cluster-char}
\end{figure}

Looking at Fig. \ref{fig:ff-cluster-char}, where, for the clusters formed in Stockmayer fluid, we plot the asphericity histogram in (a), the histogram for $\cos \gamma$  -- in (b) and the histogram of the total amount of nearest neighbours each monomer has in the cluster -- in (c), we notice the similarities in the properties of the monomers in the clusters here and of those in clusters formed by chain-like SMPs. Thus, for example, if one compares Fig. \ref{fig:aspher-ff}  and Fig. \ref{fig:asph-chain}, the asphericity has two close but distinct maxima. On the other hand, the overwhelming tendency of monomers to align tangentially seem to be even stronger in a non-crosslinked state, compare Fig. \ref{fig:cos-gam-ff} and Fig. \ref{fig:cos-gam-chains}. Finally, the number of nearest neighbours each monomer has, Fig. \ref{fig:neigh-all-ff}, shifts towards lower values in comparison to the histogram plotted in Fig. \ref{fig:neigh-all-chain}: the absence of permanent crosslinkers results in overall loosening of the clusters.

To summarise this part, one can say that permanent crosslinkers lead to smaller and more compact clusters in Stockmayer systems, but do not prevent the monomers to orient tangentially inside the clusters.

\section{Conclusion}

In the present study we investigated the effect of the SMP topology on their self-assembly driven by anisotropic magnetic dipolar and central attraction interactions. The process of self-assembly in these systems, in contrast to non-crosslinked monomers, is a competition not only between the two aforementioned interactions, but also the result of a specific initial monomer crosslinking. 

We show that the crosslinking does not change the main tendency of a Stockmayer fluid to phase separate. Thus, the overall shape of clusters weakly depends on the topology of the constituent SMPs. After having studied asphericity and anisotropy characteristics of the aggregates we found the following tendencies. All of them have a modest degree of anisometry. The lowest asymmetry is exhibited by clusters made of Y-like and X-like SMPs. The clusters formed by ring-like SMPs, in contrast, can assume a very broad range of shapes. 

Increasing the resolution of the analysis, we studied the orientations of SMPs inside the clusters. It turned out that on this level the influence of the topology is the strongest. Chain-like SMPs mix with each other when forming clusters. This way, they might have almost any orientation inside the cluster. Ring-like SMPs, on the contrary, are preferably oriented tangentially to the cluster surface, thus, forming clusters with onion-like structure. Inside the clusters formed by Y-like and X-like SMPs one finds tangentially oriented building blocks with higher probability than radially oriented ones. However, the portion of SMPs with the latter orientation is not negligible.  More uniformity is found when looking at the orientation of the net SMP magnetic moment: in case of clusters formed by chain-, Y- and X-like SMPs more than 60 per cent of the dipoles are oriented tangentially.  The net magnetic moment of ring-like SMPs is a vanishing quantity, that is why their orientations are not well defined.  

If focusing at monomers and their neighbourhood, we found that the SMPs that have more contacts with others are chain-like ones. The induced proximity of the monomers within X-like SMPs lead to the presence of bonds inside  the same SMP, and as such slightly decreasing the overall number of non-permanent bonds a monomer forms in the cluster.

Finally, we compared the properties of clusters formed by different SMPs to those of clusters formed in a non-crosslinked Stockmayer fluid with the same dipolar strength and magnetic monomer concentration. These parameters correspond to a metastable state below the gas-liquid critical point. It turns out that non-crosslinked monomers under the same conditions as SMPs form much larger clusters. However the alignment of magnetic moments in them is similar to that in the clusters formed by chain-like SMPs. In general, the shape of the clusters and the number of nearest neighbours each monomer has in them for the non-crosslinked system are similar to their counterparts for the system of chain-like SMPs. The latter has the lowest degree of crosslinking, thus, underling the impact of SMPs topology.

The differences between the clusters formed by SMPs of different topology, observed in this study, will have an impact on magnetic response of these systems. Moreover, the rheological properties, such as the behaviour of a cluster in a shear flow, will be also affected by the internal orientations of SMPs as well as the overall cluster anisotropy. These questions form the basis for our future investigations.

\begin{acknowledgments}
EVN and ESP have been supported by the Russian Science Foundation Grant 19-72-10033. SSK is grateful to E. Sega for valuable insight concerning cluster shape and she has been supported by the FWF START-Projekt Y 627-N27.

\end{acknowledgments}

\providecommand{\noopsort}[1]{}\providecommand{\singleletter}[1]{#1}%


\begin{thebibliography}{43}%
\makeatletter
\providecommand \@ifxundefined [1]{%
 \@ifx{#1\undefined}
}%
\providecommand \@ifnum [1]{%
 \ifnum #1\expandafter \@firstoftwo
 \else \expandafter \@secondoftwo
 \fi
}%
\providecommand \@ifx [1]{%
 \ifx #1\expandafter \@firstoftwo
 \else \expandafter \@secondoftwo
 \fi
}%
\providecommand \natexlab [1]{#1}%
\providecommand \enquote  [1]{``#1''}%
\providecommand \bibnamefont  [1]{#1}%
\providecommand \bibfnamefont [1]{#1}%
\providecommand \citenamefont [1]{#1}%
\providecommand \href@noop [0]{\@secondoftwo}%
\providecommand \href [0]{\begingroup \@sanitize@url \@href}%
\providecommand \@href[1]{\@@startlink{#1}\@@href}%
\providecommand \@@href[1]{\endgroup#1\@@endlink}%
\providecommand \@sanitize@url [0]{\catcode `\\12\catcode `\$12\catcode
  `\&12\catcode `\#12\catcode `\^12\catcode `\_12\catcode `\%12\relax}%
\providecommand \@@startlink[1]{}%
\providecommand \@@endlink[0]{}%
\providecommand \url  [0]{\begingroup\@sanitize@url \@url }%
\providecommand \@url [1]{\endgroup\@href {#1}{\urlprefix }}%
\providecommand \urlprefix  [0]{URL }%
\providecommand \Eprint [0]{\href }%
\providecommand \doibase [0]{https://doi.org/}%
\providecommand \selectlanguage [0]{\@gobble}%
\providecommand \bibinfo  [0]{\@secondoftwo}%
\providecommand \bibfield  [0]{\@secondoftwo}%
\providecommand \translation [1]{[#1]}%
\providecommand \BibitemOpen [0]{}%
\providecommand \bibitemStop [0]{}%
\providecommand \bibitemNoStop [0]{.\EOS\space}%
\providecommand \EOS [0]{\spacefactor3000\relax}%
\providecommand \BibitemShut  [1]{\csname bibitem#1\endcsname}%
\let\auto@bib@innerbib\@empty
\bibitem [{\citenamefont {van Roij}(1996)}]{1996-roij}%
  \BibitemOpen
  \bibfield  {author} {\bibinfo {author} {\bibfnamefont {R.}~\bibnamefont {van
  Roij}},\ }\bibfield  {title} {\bibinfo {title} {Theory of chain association
  versus liquid condensation},\ }\href
  {https://doi.org/10.1103/PhysRevLett.76.3348} {\bibfield  {journal} {\bibinfo
   {journal} {Phys. Rev. Lett.}\ }\textbf {\bibinfo {volume} {76}},\ \bibinfo
  {pages} {3348} (\bibinfo {year} {1996})}\BibitemShut {NoStop}%
\bibitem [{\citenamefont {Sear}(1996)}]{1996-sear}%
  \BibitemOpen
  \bibfield  {author} {\bibinfo {author} {\bibfnamefont {R.~P.}\ \bibnamefont
  {Sear}},\ }\bibfield  {title} {\bibinfo {title} {Low-density fluid phase of
  dipolar hard spheres},\ }\href {https://doi.org/10.1103/PhysRevLett.76.2310}
  {\bibfield  {journal} {\bibinfo  {journal} {Phys. Rev. Lett.}\ }\textbf
  {\bibinfo {volume} {76}},\ \bibinfo {pages} {2310} (\bibinfo {year}
  {1996})}\BibitemShut {NoStop}%
\bibitem [{\citenamefont {Levin}(1999)}]{1999-levin}%
  \BibitemOpen
  \bibfield  {author} {\bibinfo {author} {\bibfnamefont {Y.}~\bibnamefont
  {Levin}},\ }\bibfield  {title} {\bibinfo {title} {What happened to the
  gas-liquid transition in the system of dipolar hard spheres?},\ }\href
  {https://doi.org/10.1103/PhysRevLett.83.1159} {\bibfield  {journal} {\bibinfo
   {journal} {Phys. Rev. Lett.}\ }\textbf {\bibinfo {volume} {83}},\ \bibinfo
  {pages} {1159} (\bibinfo {year} {1999})}\BibitemShut {NoStop}%
\bibitem [{\citenamefont {Teixeira}\ \emph {et~al.}(2000)\citenamefont
  {Teixeira}, \citenamefont {Tavares},\ and\ \citenamefont
  {da~Gama}}]{teixeira00a}%
  \BibitemOpen
  \bibfield  {author} {\bibinfo {author} {\bibfnamefont {P.~I.~C.}\
  \bibnamefont {Teixeira}}, \bibinfo {author} {\bibfnamefont {J.~M.}\
  \bibnamefont {Tavares}},\ and\ \bibinfo {author} {\bibfnamefont {M.~M.~T.}\
  \bibnamefont {da~Gama}},\ }\bibfield  {title} {\bibinfo {title} {The effect
  of dipolar forces on the structure and thermodynamics of classical fluids},\
  }\href@noop {} {\bibfield  {journal} {\bibinfo  {journal} {J. Phys.: Cond.
  Matter}\ }\textbf {\bibinfo {volume} {12}},\ \bibinfo {pages} {R411}
  (\bibinfo {year} {2000})}\BibitemShut {NoStop}%
\bibitem [{\citenamefont {Weis}\ and\ \citenamefont
  {Levesque}(1993{\natexlab{a}})}]{weis93a}%
  \BibitemOpen
  \bibfield  {author} {\bibinfo {author} {\bibfnamefont {J.~J.}\ \bibnamefont
  {Weis}}\ and\ \bibinfo {author} {\bibfnamefont {D.}~\bibnamefont
  {Levesque}},\ }\bibfield  {title} {\bibinfo {title} {Chain formation in low
  density dipolar hard shperes: a monte carlo study},\ }\href@noop {}
  {\bibfield  {journal} {\bibinfo  {journal} {Phys. Rev. Lett.}\ }\textbf
  {\bibinfo {volume} {71}},\ \bibinfo {pages} {2729} (\bibinfo {year}
  {1993}{\natexlab{a}})}\BibitemShut {NoStop}%
\bibitem [{\citenamefont {Weis}\ and\ \citenamefont
  {Levesque}(1993{\natexlab{b}})}]{weis93b}%
  \BibitemOpen
  \bibfield  {author} {\bibinfo {author} {\bibfnamefont {J.~J.}\ \bibnamefont
  {Weis}}\ and\ \bibinfo {author} {\bibfnamefont {D.}~\bibnamefont
  {Levesque}},\ }\bibfield  {title} {\bibinfo {title} {Ferroelectric phases of
  dipolar hard shperes},\ }\href@noop {} {\bibfield  {journal} {\bibinfo
  {journal} {Phys. Rev. E}\ }\textbf {\bibinfo {volume} {48}},\ \bibinfo
  {pages} {3728} (\bibinfo {year} {1993}{\natexlab{b}})}\BibitemShut {NoStop}%
\bibitem [{\citenamefont {Tlusty}\ and\ \citenamefont
  {Safran}(2000)}]{safrannew}%
  \BibitemOpen
  \bibfield  {author} {\bibinfo {author} {\bibfnamefont {T.}~\bibnamefont
  {Tlusty}}\ and\ \bibinfo {author} {\bibfnamefont {S.~A.}\ \bibnamefont
  {Safran}},\ }\bibfield  {title} {\bibinfo {title} {Defect-induced phase
  separation in dipolar fluids},\ }\href@noop {} {\bibfield  {journal}
  {\bibinfo  {journal} {Science}\ }\textbf {\bibinfo {volume} {290}},\ \bibinfo
  {pages} {1328} (\bibinfo {year} {2000})}\BibitemShut {NoStop}%
\bibitem [{\citenamefont {Rovigatti}\ \emph {et~al.}(2011)\citenamefont
  {Rovigatti}, \citenamefont {Russo},\ and\ \citenamefont
  {Sciortino}}]{rovigatti11a}%
  \BibitemOpen
  \bibfield  {author} {\bibinfo {author} {\bibfnamefont {L.}~\bibnamefont
  {Rovigatti}}, \bibinfo {author} {\bibfnamefont {J.}~\bibnamefont {Russo}},\
  and\ \bibinfo {author} {\bibfnamefont {F.}~\bibnamefont {Sciortino}},\
  }\bibfield  {title} {\bibinfo {title} {No evidence of gas-liquid coexistence
  in dipolar hard spheres},\ }\href
  {https://doi.org/10.1103/PhysRevLett.107.237801} {\bibfield  {journal}
  {\bibinfo  {journal} {Phys. Rev. Lett.}\ }\textbf {\bibinfo {volume} {107}},\
  \bibinfo {pages} {237801} (\bibinfo {year} {2011})}\BibitemShut {NoStop}%
\bibitem [{\citenamefont {Kantorovich}\ \emph {et~al.}(2013)\citenamefont
  {Kantorovich}, \citenamefont {Ivanov}, \citenamefont {Rovigatti},
  \citenamefont {Tavares},\ and\ \citenamefont
  {Sciortino}}]{2013-kantorovich-prl}%
  \BibitemOpen
  \bibfield  {author} {\bibinfo {author} {\bibfnamefont {S.}~\bibnamefont
  {Kantorovich}}, \bibinfo {author} {\bibfnamefont {A.~O.}\ \bibnamefont
  {Ivanov}}, \bibinfo {author} {\bibfnamefont {L.}~\bibnamefont {Rovigatti}},
  \bibinfo {author} {\bibfnamefont {J.~M.}\ \bibnamefont {Tavares}},\ and\
  \bibinfo {author} {\bibfnamefont {F.}~\bibnamefont {Sciortino}},\ }\bibfield
  {title} {\bibinfo {title} {Nonmonotonic magnetic susceptibility of dipolar
  hard-spheres at low temperature and density},\ }\href
  {https://doi.org/10.1103/PhysRevLett.110.148306} {\bibfield  {journal}
  {\bibinfo  {journal} {Phys. Rev. Lett.}\ }\textbf {\bibinfo {volume} {110}},\
  \bibinfo {pages} {148306} (\bibinfo {year} {2013})}\BibitemShut {NoStop}%
\bibitem [{\citenamefont {Camp}\ \emph {et~al.}(2000)\citenamefont {Camp},
  \citenamefont {Shelley},\ and\ \citenamefont {Patey}}]{2000-camp-prl}%
  \BibitemOpen
  \bibfield  {author} {\bibinfo {author} {\bibfnamefont {P.~J.}\ \bibnamefont
  {Camp}}, \bibinfo {author} {\bibfnamefont {J.~C.}\ \bibnamefont {Shelley}},\
  and\ \bibinfo {author} {\bibfnamefont {G.~N.}\ \bibnamefont {Patey}},\
  }\bibfield  {title} {\bibinfo {title} {Isotropic fluid phases of dipolar hard
  spheres},\ }\href {https://doi.org/10.1103/PhysRevLett.84.115} {\bibfield
  {journal} {\bibinfo  {journal} {Phys. Rev. Lett.}\ }\textbf {\bibinfo
  {volume} {84}},\ \bibinfo {pages} {115} (\bibinfo {year} {2000})}\BibitemShut
  {NoStop}%
\bibitem [{\citenamefont {Klokkenburg}\ \emph {et~al.}(2006)\citenamefont
  {Klokkenburg}, \citenamefont {Dullens}, \citenamefont {Kegel}, \citenamefont
  {Ern\'e},\ and\ \citenamefont {Philipse}}]{2006-klokkenburg}%
  \BibitemOpen
  \bibfield  {author} {\bibinfo {author} {\bibfnamefont {M.}~\bibnamefont
  {Klokkenburg}}, \bibinfo {author} {\bibfnamefont {R.~P.~A.}\ \bibnamefont
  {Dullens}}, \bibinfo {author} {\bibfnamefont {W.~K.}\ \bibnamefont {Kegel}},
  \bibinfo {author} {\bibfnamefont {B.~H.}\ \bibnamefont {Ern\'e}},\ and\
  \bibinfo {author} {\bibfnamefont {A.~P.}\ \bibnamefont {Philipse}},\
  }\bibfield  {title} {\bibinfo {title} {Quantitative real-space analysis of
  self-assembled structures of magnetic dipolar colloids},\ }\href
  {https://doi.org/10.1103/PhysRevLett.96.037203} {\bibfield  {journal}
  {\bibinfo  {journal} {Phys. Rev. Lett.}\ }\textbf {\bibinfo {volume} {96}},\
  \bibinfo {pages} {037203} (\bibinfo {year} {2006})}\BibitemShut {NoStop}%
\bibitem [{\citenamefont {Ronti}\ \emph {et~al.}(2017)\citenamefont {Ronti},
  \citenamefont {Rovigatti}, \citenamefont {Tavares}, \citenamefont {Ivanov},
  \citenamefont {Kantorovich},\ and\ \citenamefont {Sciortino}}]{2017-ronti}%
  \BibitemOpen
  \bibfield  {author} {\bibinfo {author} {\bibfnamefont {M.}~\bibnamefont
  {Ronti}}, \bibinfo {author} {\bibfnamefont {L.}~\bibnamefont {Rovigatti}},
  \bibinfo {author} {\bibfnamefont {J.~M.}\ \bibnamefont {Tavares}}, \bibinfo
  {author} {\bibfnamefont {A.~O.}\ \bibnamefont {Ivanov}}, \bibinfo {author}
  {\bibfnamefont {S.~S.}\ \bibnamefont {Kantorovich}},\ and\ \bibinfo {author}
  {\bibfnamefont {F.}~\bibnamefont {Sciortino}},\ }\bibfield  {title} {\bibinfo
  {title} {Accurate free energy calculations for rings and chains formed by
  dipolar hard spheres},\ }\href@noop {} {\bibfield  {journal} {\bibinfo
  {journal} {Soft Matter}\ }\textbf {\bibinfo {volume} {13}},\ \bibinfo {pages}
  {7870} (\bibinfo {year} {2017})}\BibitemShut {NoStop}%
\bibitem [{\citenamefont {Kantorovich}\ \emph {et~al.}(2015)\citenamefont
  {Kantorovich}, \citenamefont {Ivanov}, \citenamefont {Rovigatti},
  \citenamefont {Tavares},\ and\ \citenamefont
  {Sciortino}}]{2015-kantorovich-pccp1}%
  \BibitemOpen
  \bibfield  {author} {\bibinfo {author} {\bibfnamefont {S.~S.}\ \bibnamefont
  {Kantorovich}}, \bibinfo {author} {\bibfnamefont {A.~O.}\ \bibnamefont
  {Ivanov}}, \bibinfo {author} {\bibfnamefont {L.}~\bibnamefont {Rovigatti}},
  \bibinfo {author} {\bibfnamefont {J.~M.}\ \bibnamefont {Tavares}},\ and\
  \bibinfo {author} {\bibfnamefont {F.}~\bibnamefont {Sciortino}},\ }\bibfield
  {title} {\bibinfo {title} {Temperature-induced structural transitions in
  self-assembling magnetic nanocolloids},\ }\href
  {https://doi.org/10.1039/C5CP01558H} {\bibfield  {journal} {\bibinfo
  {journal} {Phys. Chem. Chem. Phys.}\ }\textbf {\bibinfo {volume} {17}},\
  \bibinfo {pages} {16601} (\bibinfo {year} {2015})}\BibitemShut {NoStop}%
\bibitem [{\citenamefont {Korth}\ \emph {et~al.}(2006)\citenamefont {Korth},
  \citenamefont {Keng}, \citenamefont {Shim}, \citenamefont {Bowles},
  \citenamefont {Tang}, \citenamefont {Kowalewski}, \citenamefont {Nebesny},\
  and\ \citenamefont {Pyun}}]{2006-korth}%
  \BibitemOpen
  \bibfield  {author} {\bibinfo {author} {\bibfnamefont {B.~D.}\ \bibnamefont
  {Korth}}, \bibinfo {author} {\bibfnamefont {P.}~\bibnamefont {Keng}},
  \bibinfo {author} {\bibfnamefont {I.}~\bibnamefont {Shim}}, \bibinfo {author}
  {\bibfnamefont {S.~E.}\ \bibnamefont {Bowles}}, \bibinfo {author}
  {\bibfnamefont {C.}~\bibnamefont {Tang}}, \bibinfo {author} {\bibfnamefont
  {T.}~\bibnamefont {Kowalewski}}, \bibinfo {author} {\bibfnamefont {K.~W.}\
  \bibnamefont {Nebesny}},\ and\ \bibinfo {author} {\bibfnamefont
  {J.}~\bibnamefont {Pyun}},\ }\bibfield  {title} {\bibinfo {title}
  {Polymer-coated ferromagnetic colloids from well-defined macromolecular
  surfactants and assembly into nanoparticle chains},\ }\href
  {https://doi.org/10.1021/ja0609147} {\bibfield  {journal} {\bibinfo
  {journal} {J Am Chem Soc}\ }\textbf {\bibinfo {volume} {128}},\ \bibinfo
  {pages} {6562} (\bibinfo {year} {2006})}\BibitemShut {NoStop}%
\bibitem [{\citenamefont {Keng}\ \emph {et~al.}(2007)\citenamefont {Keng},
  \citenamefont {Shim}, \citenamefont {Korth}, \citenamefont {Douglas},\ and\
  \citenamefont {Pyun}}]{2007-keng}%
  \BibitemOpen
  \bibfield  {author} {\bibinfo {author} {\bibfnamefont {P.~Y.}\ \bibnamefont
  {Keng}}, \bibinfo {author} {\bibfnamefont {I.}~\bibnamefont {Shim}}, \bibinfo
  {author} {\bibfnamefont {B.~D.}\ \bibnamefont {Korth}}, \bibinfo {author}
  {\bibfnamefont {J.~F.}\ \bibnamefont {Douglas}},\ and\ \bibinfo {author}
  {\bibfnamefont {J.}~\bibnamefont {Pyun}},\ }\bibfield  {title} {\bibinfo
  {title} {Synthesis and self-assembly of polymer-coated ferromagnetic
  nanoparticles},\ }\href {https://doi.org/10.1021/nn7001213} {\bibfield
  {journal} {\bibinfo  {journal} {ACS Nano}\ }\textbf {\bibinfo {volume} {1}},\
  \bibinfo {pages} {279} (\bibinfo {year} {2007})}\BibitemShut {NoStop}%
\bibitem [{\citenamefont {Bharti}\ \emph {et~al.}(2015)\citenamefont {Bharti},
  \citenamefont {Fameau}, \citenamefont {Rubinstein},\ and\ \citenamefont
  {Velev}}]{2015-bharti}%
  \BibitemOpen
  \bibfield  {author} {\bibinfo {author} {\bibfnamefont {B.}~\bibnamefont
  {Bharti}}, \bibinfo {author} {\bibfnamefont {A.-L.}\ \bibnamefont {Fameau}},
  \bibinfo {author} {\bibfnamefont {M.}~\bibnamefont {Rubinstein}},\ and\
  \bibinfo {author} {\bibfnamefont {O.~D.}\ \bibnamefont {Velev}},\ }\bibfield
  {title} {\bibinfo {title} {Nanocapillarity-mediated magnetic assembly of
  nanoparticles into ultraflexible filaments and reconfigurable networks},\
  }\href {https://doi.org/10.1038/nmat4364} {\bibfield  {journal} {\bibinfo
  {journal} {Nat. Mater.}\ }\textbf {\bibinfo {volume} {14}},\ \bibinfo {pages}
  {1104} (\bibinfo {year} {2015})}\BibitemShut {NoStop}%
\bibitem [{\citenamefont {Yuan}\ \emph {et~al.}(2017)\citenamefont {Yuan},
  \citenamefont {Zvonkina}, \citenamefont {Al-Enizi}, \citenamefont
  {Elzatahry}, \citenamefont {Pyun},\ and\ \citenamefont {Karim}}]{2017-hongy}%
  \BibitemOpen
  \bibfield  {author} {\bibinfo {author} {\bibfnamefont {H.}~\bibnamefont
  {Yuan}}, \bibinfo {author} {\bibfnamefont {I.~J.}\ \bibnamefont {Zvonkina}},
  \bibinfo {author} {\bibfnamefont {A.~M.}\ \bibnamefont {Al-Enizi}}, \bibinfo
  {author} {\bibfnamefont {A.~A.}\ \bibnamefont {Elzatahry}}, \bibinfo {author}
  {\bibfnamefont {J.}~\bibnamefont {Pyun}},\ and\ \bibinfo {author}
  {\bibfnamefont {A.}~\bibnamefont {Karim}},\ }\bibfield  {title} {\bibinfo
  {title} {Facile assembly of aligned magnetic nanoparticle chains in polymer
  nanocomposite films by magnetic flow coating},\ }\href
  {https://doi.org/10.1021/acsami.7b02186} {\bibfield  {journal} {\bibinfo
  {journal} {ACS Applied Materials \& Interfaces}\ }\textbf {\bibinfo {volume}
  {9}},\ \bibinfo {pages} {11290} (\bibinfo {year} {2017})}\BibitemShut {NoStop}%
\bibitem [{\citenamefont {Dreyfus}\ \emph {et~al.}(2005)\citenamefont
  {Dreyfus}, \citenamefont {Baudry}, \citenamefont {Roper}, \citenamefont
  {Fermigier}, \citenamefont {Stone},\ and\ \citenamefont
  {Bibette}}]{2005-dreyfus}%
  \BibitemOpen
  \bibfield  {author} {\bibinfo {author} {\bibfnamefont {R.}~\bibnamefont
  {Dreyfus}}, \bibinfo {author} {\bibfnamefont {J.}~\bibnamefont {Baudry}},
  \bibinfo {author} {\bibfnamefont {M.~L.}\ \bibnamefont {Roper}}, \bibinfo
  {author} {\bibfnamefont {M.}~\bibnamefont {Fermigier}}, \bibinfo {author}
  {\bibfnamefont {H.~A.}\ \bibnamefont {Stone}},\ and\ \bibinfo {author}
  {\bibfnamefont {J.}~\bibnamefont {Bibette}},\ }\bibfield  {title} {\bibinfo
  {title} {Microscopic artificial swimmers},\ }\href
  {https://doi.org/10.1038/nature04090} {\bibfield  {journal} {\bibinfo
  {journal} {Nature}\ }\textbf {\bibinfo {volume} {437}},\ \bibinfo {pages}
  {862} (\bibinfo {year} {2005})}\BibitemShut {NoStop}%
\bibitem [{\citenamefont {Choi}\ \emph {et~al.}(2008)\citenamefont {Choi},
  \citenamefont {Koo}, \citenamefont {Kim},\ and\ \citenamefont
  {Huck}}]{2008-choi}%
  \BibitemOpen
  \bibfield  {author} {\bibinfo {author} {\bibfnamefont {W.~S.}\ \bibnamefont
  {Choi}}, \bibinfo {author} {\bibfnamefont {H.~Y.}\ \bibnamefont {Koo}},
  \bibinfo {author} {\bibfnamefont {J.~Y.}\ \bibnamefont {Kim}},\ and\ \bibinfo
  {author} {\bibfnamefont {W.~T.~S.}\ \bibnamefont {Huck}},\ }\bibfield
  {title} {\bibinfo {title} {Collective behavior of magnetic nanoparticles in
  polyelectrolyte brushes},\ }\href {https://doi.org/10.1002/adma.200801423}
  {\bibfield  {journal} {\bibinfo  {journal} {Adv. Mater.}\ }\textbf {\bibinfo
  {volume} {20}},\ \bibinfo {pages} {4504} (\bibinfo {year}
  {2008})}\BibitemShut {NoStop}%
\bibitem [{\citenamefont {\={E}rglis}\ \emph {et~al.}(2008)\citenamefont
  {\={E}rglis}, \citenamefont {Zhulenkovs}, \citenamefont {Sharipo},\ and\
  \citenamefont {C\={e}bers}}]{2008-erglis-jpcm}%
  \BibitemOpen
  \bibfield  {author} {\bibinfo {author} {\bibfnamefont {K.}~\bibnamefont
  {\={E}rglis}}, \bibinfo {author} {\bibfnamefont {D.}~\bibnamefont
  {Zhulenkovs}}, \bibinfo {author} {\bibfnamefont {A.}~\bibnamefont
  {Sharipo}},\ and\ \bibinfo {author} {\bibfnamefont {A.}~\bibnamefont
  {C\={e}bers}},\ }\bibfield  {title} {\bibinfo {title} {Elastic properties of
  dna linked flexible magnetic filaments},\ }\href
  {https://doi.org/10.1088/0953-8984/20/20/204107} {\bibfield  {journal}
  {\bibinfo  {journal} {J Phys-Condens Mat}\ }\textbf {\bibinfo {volume}
  {20}},\ \bibinfo {pages} {204107} (\bibinfo {year} {2008})}\BibitemShut
  {NoStop}%
\bibitem [{\citenamefont {Zhou}\ \emph {et~al.}(2008)\citenamefont {Zhou},
  \citenamefont {Biesheuvel}, \citenamefont {Choi}, \citenamefont {Shu},
  \citenamefont {Poetes}, \citenamefont {Steiner}, ,\ and\ \citenamefont
  {Huck}}]{2008-zhou}%
  \BibitemOpen
  \bibfield  {author} {\bibinfo {author} {\bibfnamefont {F.}~\bibnamefont
  {Zhou}}, \bibinfo {author} {\bibfnamefont {P.~M.}\ \bibnamefont
  {Biesheuvel}}, \bibinfo {author} {\bibfnamefont {E.-Y.}\ \bibnamefont
  {Choi}}, \bibinfo {author} {\bibfnamefont {W.}~\bibnamefont {Shu}}, \bibinfo
  {author} {\bibfnamefont {R.}~\bibnamefont {Poetes}}, \bibinfo {author}
  {\bibfnamefont {U.}~\bibnamefont {Steiner}}, ,\ and\ \bibinfo {author}
  {\bibfnamefont {W.~T.~S.}\ \bibnamefont {Huck}},\ }\bibfield  {title}
  {\bibinfo {title} {Polyelectrolyte brush amplified electroactuation of
  microcantilevers},\ }\href {https://doi.org/10.1021/nl073157z} {\bibfield
  {journal} {\bibinfo  {journal} {Nano Letters}\ }\textbf {\bibinfo {volume}
  {8}},\ \bibinfo {pages} {725} (\bibinfo {year} {2008})}\BibitemShut {NoStop}%
\bibitem [{\citenamefont {C\={e}bers}(2005)}]{2005-cebers}%
  \BibitemOpen
  \bibfield  {author} {\bibinfo {author} {\bibfnamefont {A.}~\bibnamefont
  {C\={e}bers}},\ }\bibfield  {title} {\bibinfo {title} {Flexible magnetic
  filaments},\ }\href {https://doi.org/10.1016/j.cocis.2005.07.002} {\bibfield
  {journal} {\bibinfo  {journal} {Curr Opin Colloid Interface Sci}\ }\textbf
  {\bibinfo {volume} {10}},\ \bibinfo {pages} {167} (\bibinfo {year}
  {2005})}\BibitemShut {NoStop}%
\bibitem [{\citenamefont {Belovs}\ and\ \citenamefont
  {C\=ebers}(2009)}]{2009-belovs-pre}%
  \BibitemOpen
  \bibfield  {author} {\bibinfo {author} {\bibfnamefont {M.}~\bibnamefont
  {Belovs}}\ and\ \bibinfo {author} {\bibfnamefont {A.}~\bibnamefont
  {C\=ebers}},\ }\bibfield  {title} {\bibinfo {title} {Ferromagnetic
  microswimmer},\ }\href {https://doi.org/10.1103/PhysRevE.79.051503}
  {\bibfield  {journal} {\bibinfo  {journal} {Phys Rev E}\ }\textbf {\bibinfo
  {volume} {79}},\ \bibinfo {pages} {051503} (\bibinfo {year}
  {2009})}\BibitemShut {NoStop}%
\bibitem [{\citenamefont {Javaitis}\ and\ \citenamefont
  {Zilgalve}(2011)}]{2011-javaitis}%
  \BibitemOpen
  \bibfield  {author} {\bibinfo {author} {\bibfnamefont {I.}~\bibnamefont
  {Javaitis}}\ and\ \bibinfo {author} {\bibfnamefont {V.}~\bibnamefont
  {Zilgalve}},\ }\bibfield  {title} {\bibinfo {title} {Physics of flexible
  magnetic filaments},\ }\href
  {https://doi.org/10.4028/www.scientific.net/AMR.222.221} {\bibfield
  {journal} {\bibinfo  {journal} {Adv Mat Res}\ }\textbf {\bibinfo {volume}
  {222}},\ \bibinfo {pages} {221} (\bibinfo {year} {2011})}\BibitemShut
  {NoStop}%
\bibitem [{\citenamefont {S\'anchez}\ \emph {et~al.}(2011)\citenamefont
  {S\'anchez}, \citenamefont {Cerd\`a}, \citenamefont {Ballenegger},
  \citenamefont {Sintes}, \citenamefont {Piro},\ and\ \citenamefont
  {Holm}}]{2011-sanchez-sm}%
  \BibitemOpen
  \bibfield  {author} {\bibinfo {author} {\bibfnamefont {P.~A.}\ \bibnamefont
  {S\'anchez}}, \bibinfo {author} {\bibfnamefont {J.~J.}\ \bibnamefont
  {Cerd\`a}}, \bibinfo {author} {\bibfnamefont {V.}~\bibnamefont
  {Ballenegger}}, \bibinfo {author} {\bibfnamefont {T.}~\bibnamefont {Sintes}},
  \bibinfo {author} {\bibfnamefont {O.}~\bibnamefont {Piro}},\ and\ \bibinfo
  {author} {\bibfnamefont {C.}~\bibnamefont {Holm}},\ }\bibfield  {title}
  {\bibinfo {title} {Semiflexible magnetic filaments near attractive flat
  surfaces: a {L}angevin dynamics study},\ }\href
  {https://doi.org/10.1039/C0SM00772B} {\bibfield  {journal} {\bibinfo
  {journal} {Soft Matter}\ }\textbf {\bibinfo {volume} {7}},\ \bibinfo {pages}
  {1809} (\bibinfo {year} {2011})}\BibitemShut {NoStop}%
\bibitem [{\citenamefont {Cerd\`{a}}\ \emph {et~al.}(2016)\citenamefont
  {Cerd\`{a}}, \citenamefont {S\'{a}nchez}, \citenamefont {L\"{u}sebrink},
  \citenamefont {Kantorovich},\ and\ \citenamefont {Sintes}}]{2016-cerda-pccp}%
  \BibitemOpen
  \bibfield  {author} {\bibinfo {author} {\bibfnamefont {J.~J.}\ \bibnamefont
  {Cerd\`{a}}}, \bibinfo {author} {\bibfnamefont {P.~A.}\ \bibnamefont
  {S\'{a}nchez}}, \bibinfo {author} {\bibfnamefont {D.}~\bibnamefont
  {L\"{u}sebrink}}, \bibinfo {author} {\bibfnamefont {S.~S.}\ \bibnamefont
  {Kantorovich}},\ and\ \bibinfo {author} {\bibfnamefont {T.}~\bibnamefont
  {Sintes}},\ }\bibfield  {title} {\bibinfo {title} {Flexible magnetic
  filaments under the influence of external magnetic fields in the limit of
  infinite dilution},\ }\href {https://doi.org/10.1039/C6CP00923A} {\bibfield
  {journal} {\bibinfo  {journal} {Phys. Chem. Chem. Phys.}\ }\textbf {\bibinfo
  {volume} {18}},\ \bibinfo {pages} {12616} (\bibinfo {year}
  {2016})}\BibitemShut {NoStop}%
\bibitem [{\citenamefont {Wei}\ \emph {et~al.}(2016)\citenamefont {Wei},
  \citenamefont {Song},\ and\ \citenamefont {Dobnikar}}]{2016-wei}%
  \BibitemOpen
  \bibfield  {author} {\bibinfo {author} {\bibfnamefont {J.}~\bibnamefont
  {Wei}}, \bibinfo {author} {\bibfnamefont {F.}~\bibnamefont {Song}},\ and\
  \bibinfo {author} {\bibfnamefont {J.}~\bibnamefont {Dobnikar}},\ }\bibfield
  {title} {\bibinfo {title} {Assembly of superparamagnetic filaments in
  external field},\ }\href {https://doi.org/10.1021/acs.langmuir.6b02268}
  {\bibfield  {journal} {\bibinfo  {journal} {Langmuir}\ }\textbf {\bibinfo
  {volume} {32}},\ \bibinfo {pages} {9321} (\bibinfo {year}
  {2016})}\BibitemShut {NoStop}%
\bibitem [{\citenamefont {Kuznetsov}(2019)}]{2019-kuznetsov}%
  \BibitemOpen
  \bibfield  {author} {\bibinfo {author} {\bibfnamefont {A.~A.}\ \bibnamefont
  {Kuznetsov}},\ }\bibfield  {title} {\bibinfo {title} {Equilibrium properties
  of magnetic filament suspensions},\ }\href
  {https://doi.org/https://doi.org/10.1016/j.jmmm.2017.10.091} {\bibfield
  {journal} {\bibinfo  {journal} {Journal of Magnetism and Magnetic Materials}\
  }\textbf {\bibinfo {volume} {470}},\ \bibinfo {pages} {28 } (\bibinfo {year}
  {2019})},\ \bibinfo {note} {international Baltic Conference on Magnetism:
  focus on functionalized magnetic structures for energy and
  biotechnology}\BibitemShut {NoStop}%
\bibitem [{\citenamefont {de~Vicente}\ \emph {et~al.}(2011)\citenamefont
  {de~Vicente}, \citenamefont {Klingenberg},\ and\ \citenamefont
  {Hidalgo-Alvarez}}]{2011-devicente}%
  \BibitemOpen
  \bibfield  {author} {\bibinfo {author} {\bibfnamefont {J.}~\bibnamefont
  {de~Vicente}}, \bibinfo {author} {\bibfnamefont {D.~J.}\ \bibnamefont
  {Klingenberg}},\ and\ \bibinfo {author} {\bibfnamefont {R.}~\bibnamefont
  {Hidalgo-Alvarez}},\ }\bibfield  {title} {\bibinfo {title}
  {Magnetorheological fluids: a review},\ }\href
  {https://doi.org/10.1039/C0SM01221A} {\bibfield  {journal} {\bibinfo
  {journal} {Soft Matter}\ }\textbf {\bibinfo {volume} {7}},\ \bibinfo {pages}
  {3701} (\bibinfo {year} {2011})}\BibitemShut {NoStop}%
\bibitem [{\citenamefont {Park}\ \emph {et~al.}(2010)\citenamefont {Park},
  \citenamefont {Fang},\ and\ \citenamefont {Choi}}]{2010-park}%
  \BibitemOpen
  \bibfield  {author} {\bibinfo {author} {\bibfnamefont {B.~J.}\ \bibnamefont
  {Park}}, \bibinfo {author} {\bibfnamefont {F.~F.}\ \bibnamefont {Fang}},\
  and\ \bibinfo {author} {\bibfnamefont {H.~J.}\ \bibnamefont {Choi}},\
  }\bibfield  {title} {\bibinfo {title} {Magnetorheology: materials and
  application},\ }\href {https://doi.org/10.1039/C0SM00014K} {\bibfield
  {journal} {\bibinfo  {journal} {Soft Matter}\ }\textbf {\bibinfo {volume}
  {6}},\ \bibinfo {pages} {5246} (\bibinfo {year} {2010})}\BibitemShut
  {NoStop}%
\bibitem [{\citenamefont {Rozhkov}\ \emph {et~al.}(2018)\citenamefont
  {Rozhkov}, \citenamefont {Pyanzina}, \citenamefont {Novak}, \citenamefont
  {Cerd{\`a}}, \citenamefont {Sintes}, \citenamefont {Ronti}, \citenamefont
  {S{\'a}nchez},\ and\ \citenamefont {Kantorovich}}]{rozhkov17a}%
  \BibitemOpen
  \bibfield  {author} {\bibinfo {author} {\bibfnamefont {D.~A.}\ \bibnamefont
  {Rozhkov}}, \bibinfo {author} {\bibfnamefont {E.~S.}\ \bibnamefont
  {Pyanzina}}, \bibinfo {author} {\bibfnamefont {E.~V.}\ \bibnamefont {Novak}},
  \bibinfo {author} {\bibfnamefont {J.~J.}\ \bibnamefont {Cerd{\`a}}}, \bibinfo
  {author} {\bibfnamefont {T.}~\bibnamefont {Sintes}}, \bibinfo {author}
  {\bibfnamefont {M.}~\bibnamefont {Ronti}}, \bibinfo {author} {\bibfnamefont
  {P.~A.}\ \bibnamefont {S{\'a}nchez}},\ and\ \bibinfo {author} {\bibfnamefont
  {S.~S.}\ \bibnamefont {Kantorovich}},\ }\bibfield  {title} {\bibinfo {title}
  {Self-assembly of polymer-like structures of magnetic colloids: Langevin
  dynamics study of basic topologies},\ }\href
  {https://doi.org/10.1080/08927022.2017.1378815} {\bibfield  {journal}
  {\bibinfo  {journal} {Molecular Simulation}\ }\textbf {\bibinfo {volume}
  {44}},\ \bibinfo {pages} {507} (\bibinfo {year} {2018})}\BibitemShut
  {NoStop}%
\bibitem [{\citenamefont {Novak}\ \emph {et~al.}(2018)\citenamefont {Novak},
  \citenamefont {Pyanzina}, \citenamefont {Rozhkov}, \citenamefont {Ronti},
  \citenamefont {Cerdà}, \citenamefont {Sintes}, \citenamefont {Sánchez},\
  and\ \citenamefont {Kantorovich}}]{2018-novak}%
  \BibitemOpen
  \bibfield  {author} {\bibinfo {author} {\bibfnamefont {E.}~\bibnamefont
  {Novak}}, \bibinfo {author} {\bibfnamefont {E.}~\bibnamefont {Pyanzina}},
  \bibinfo {author} {\bibfnamefont {D.}~\bibnamefont {Rozhkov}}, \bibinfo
  {author} {\bibfnamefont {M.}~\bibnamefont {Ronti}}, \bibinfo {author}
  {\bibfnamefont {J.}~\bibnamefont {Cerdà}}, \bibinfo {author} {\bibfnamefont
  {T.}~\bibnamefont {Sintes}}, \bibinfo {author} {\bibfnamefont
  {P.}~\bibnamefont {Sánchez}},\ and\ \bibinfo {author} {\bibfnamefont
  {S.}~\bibnamefont {Kantorovich}},\ }\bibfield  {title} {\bibinfo {title}
  {Suspensions of supracolloidal magnetic polymers: Self-assembly properties
  from computer simulations},\ }\href
  {https://doi.org/https://doi.org/10.1016/j.molliq.2018.08.145} {\bibfield
  {journal} {\bibinfo  {journal} {Journal of Molecular Liquids}\ }\textbf
  {\bibinfo {volume} {271}},\ \bibinfo {pages} {631 } (\bibinfo {year}
  {2018})}\BibitemShut {NoStop}%
\bibitem [{\citenamefont {Bianchi}\ \emph {et~al.}(2011)\citenamefont
  {Bianchi}, \citenamefont {Blaak},\ and\ \citenamefont
  {Likos}}]{bianchi2011patchy}%
  \BibitemOpen
  \bibfield  {author} {\bibinfo {author} {\bibfnamefont {E.}~\bibnamefont
  {Bianchi}}, \bibinfo {author} {\bibfnamefont {R.}~\bibnamefont {Blaak}},\
  and\ \bibinfo {author} {\bibfnamefont {C.~N.}\ \bibnamefont {Likos}},\
  }\bibfield  {title} {\bibinfo {title} {Patchy colloids: state of the art and
  perspectives},\ }\href {https://doi.org/10.1039/C0CP02296A} {\bibfield
  {journal} {\bibinfo  {journal} {Phys. Chem. Chem. Phys.}\ }\textbf {\bibinfo
  {volume} {13}},\ \bibinfo {pages} {6397} (\bibinfo {year}
  {2011})}\BibitemShut {NoStop}%
\bibitem [{\citenamefont {Smit}\ \emph {et~al.}(1989)\citenamefont {Smit},
  \citenamefont {Williams}, \citenamefont {Hendriks},\ and\ \citenamefont
  {Leeuw}}]{1989-smit}%
  \BibitemOpen
  \bibfield  {author} {\bibinfo {author} {\bibfnamefont {B.}~\bibnamefont
  {Smit}}, \bibinfo {author} {\bibfnamefont {C.}~\bibnamefont {Williams}},
  \bibinfo {author} {\bibfnamefont {E.}~\bibnamefont {Hendriks}},\ and\
  \bibinfo {author} {\bibfnamefont {S.~D.}\ \bibnamefont {Leeuw}},\ }\bibfield
  {title} {\bibinfo {title} {Vapour-liquid equilibria for stockmayer fluids},\
  }\href {https://doi.org/10.1080/00268978900102531} {\bibfield  {journal}
  {\bibinfo  {journal} {Molecular Physics}\ }\textbf {\bibinfo {volume} {68}},\
  \bibinfo {pages} {765} (\bibinfo {year} {1989})}\BibitemShut {NoStop}%
\bibitem [{\citenamefont {van Leeuwen}\ and\ \citenamefont
  {Smit}(1993)}]{1993-vanleeuwen}%
  \BibitemOpen
  \bibfield  {author} {\bibinfo {author} {\bibfnamefont {M.~E.}\ \bibnamefont
  {van Leeuwen}}\ and\ \bibinfo {author} {\bibfnamefont {B.}~\bibnamefont
  {Smit}},\ }\bibfield  {title} {\bibinfo {title} {What makes a polar liquid a
  liquid?},\ }\href {https://doi.org/10.1103/PhysRevLett.71.3991} {\bibfield
  {journal} {\bibinfo  {journal} {Phys. Rev. Lett.}\ }\textbf {\bibinfo
  {volume} {71}},\ \bibinfo {pages} {3991} (\bibinfo {year}
  {1993})}\BibitemShut {NoStop}%
\bibitem [{\citenamefont {Panagiotopoulos}(1992)}]{1992-panagiotopoulos}%
  \BibitemOpen
  \bibfield  {author} {\bibinfo {author} {\bibfnamefont {A.~Z.}\ \bibnamefont
  {Panagiotopoulos}},\ }\bibfield  {title} {\bibinfo {title} {Direct
  determination of fluid phase equilibria by simulation in the gibbs ensemble:
  A review},\ }\href {https://doi.org/10.1080/08927029208048258} {\bibfield
  {journal} {\bibinfo  {journal} {Molecular Simulation}\ }\textbf {\bibinfo
  {volume} {9}},\ \bibinfo {pages} {1} (\bibinfo {year} {1992})}\BibitemShut {NoStop}%
\bibitem [{\citenamefont {Stevens}\ and\ \citenamefont
  {Grest}(1995)}]{1995-stevens}%
  \BibitemOpen
  \bibfield  {author} {\bibinfo {author} {\bibfnamefont {M.~J.}\ \bibnamefont
  {Stevens}}\ and\ \bibinfo {author} {\bibfnamefont {G.~S.}\ \bibnamefont
  {Grest}},\ }\bibfield  {title} {\bibinfo {title} {Structure of soft-sphere
  dipolar fluids},\ }\href {https://doi.org/10.1103/PhysRevE.51.5962}
  {\bibfield  {journal} {\bibinfo  {journal} {Phys. Rev. E}\ }\textbf {\bibinfo
  {volume} {51}},\ \bibinfo {pages} {5962} (\bibinfo {year}
  {1995})}\BibitemShut {NoStop}%
\bibitem [{\citenamefont {Adams}\ and\ \citenamefont
  {Adams}(1981)}]{1981-adams}%
  \BibitemOpen
  \bibfield  {author} {\bibinfo {author} {\bibfnamefont {D.}~\bibnamefont
  {Adams}}\ and\ \bibinfo {author} {\bibfnamefont {E.}~\bibnamefont {Adams}},\
  }\bibfield  {title} {\bibinfo {title} {Static dielectric properties of the
  stockmayer fluid from computer simulation},\ }\href
  {https://doi.org/10.1080/00268978100100701} {\bibfield  {journal} {\bibinfo
  {journal} {Molecular Physics}\ }\textbf {\bibinfo {volume} {42}},\ \bibinfo
  {pages} {907} (\bibinfo {year} {1981})}\BibitemShut {NoStop}%
\bibitem [{\citenamefont {Novak}\ \emph {et~al.}(2017)\citenamefont {Novak},
  \citenamefont {Rozhkov}, \citenamefont {S\'{a}nchez},\ and\ \citenamefont
  {Kantorovich}}]{2017-novak-jmmmb}%
  \BibitemOpen
  \bibfield  {author} {\bibinfo {author} {\bibfnamefont {E.}~\bibnamefont
  {Novak}}, \bibinfo {author} {\bibfnamefont {D.}~\bibnamefont {Rozhkov}},
  \bibinfo {author} {\bibfnamefont {P.}~\bibnamefont {S\'{a}nchez}},\ and\
  \bibinfo {author} {\bibfnamefont {S.}~\bibnamefont {Kantorovich}},\
  }\bibfield  {title} {\bibinfo {title} {Self-assembly of designed
  supramolecular magnetic filaments of different shapes},\ }\href
  {https://doi.org/10.1016/j.jmmm.2016.10.046} {\bibfield  {journal} {\bibinfo
  {journal} {J. Magn. Magn. Mater.}\ }\textbf {\bibinfo {volume} {431}},\
  \bibinfo {pages} {152} (\bibinfo {year} {2017})}\BibitemShut {NoStop}%
\bibitem [{\citenamefont {Limbach}\ \emph {et~al.}(2006)\citenamefont
  {Limbach}, \citenamefont {Arnold}, \citenamefont {Mann},\ and\ \citenamefont
  {Holm}}]{2006-limbach}%
  \BibitemOpen
  \bibfield  {author} {\bibinfo {author} {\bibfnamefont {H.~J.}\ \bibnamefont
  {Limbach}}, \bibinfo {author} {\bibfnamefont {A.}~\bibnamefont {Arnold}},
  \bibinfo {author} {\bibfnamefont {B.~A.}\ \bibnamefont {Mann}},\ and\
  \bibinfo {author} {\bibfnamefont {C.}~\bibnamefont {Holm}},\ }\bibfield
  {title} {\bibinfo {title} {{ESPResSo} -- an extensible simulation package for
  research on soft matter systems},\ }\href
  {https://doi.org/10.1016/j.cpc.2005.10.005} {\bibfield  {journal} {\bibinfo
  {journal} {Comput Phys Commun}\ }\textbf {\bibinfo {volume} {174}},\ \bibinfo
  {pages} {704} (\bibinfo {year} {2006})}\BibitemShut {NoStop}%
\bibitem [{\citenamefont {Cerd\`{a}}\ \emph {et~al.}(2008)\citenamefont
  {Cerd\`{a}}, \citenamefont {Ballenegger}, \citenamefont {Lenz},\ and\
  \citenamefont {Holm}}]{2008-cerda-jcp}%
  \BibitemOpen
  \bibfield  {author} {\bibinfo {author} {\bibfnamefont {J.~J.}\ \bibnamefont
  {Cerd\`{a}}}, \bibinfo {author} {\bibfnamefont {V.}~\bibnamefont
  {Ballenegger}}, \bibinfo {author} {\bibfnamefont {O.}~\bibnamefont {Lenz}},\
  and\ \bibinfo {author} {\bibfnamefont {C.}~\bibnamefont {Holm}},\ }\bibfield
  {title} {\bibinfo {title} {P3m algorithm for dipolar interactions},\ }\href
  {https://doi.org/10.1063/1.3000389} {\bibfield  {journal} {\bibinfo
  {journal} {J Chem Phys}\ }\textbf {\bibinfo {volume} {129}},\ \bibinfo
  {pages} {234104} (\bibinfo {year} {2008})}\BibitemShut {NoStop}%
\bibitem [{\citenamefont {Steinhardt}\ \emph {et~al.}(1983)\citenamefont
  {Steinhardt}, \citenamefont {Nelson},\ and\ \citenamefont
  {Ronchetti}}]{steinhardt83a}%
  \BibitemOpen
  \bibfield  {author} {\bibinfo {author} {\bibfnamefont {P.~J.}\ \bibnamefont
  {Steinhardt}}, \bibinfo {author} {\bibfnamefont {D.~R.}\ \bibnamefont
  {Nelson}},\ and\ \bibinfo {author} {\bibfnamefont {M.}~\bibnamefont
  {Ronchetti}},\ }\bibfield  {title} {\bibinfo {title} {Bond-orientational
  order in liquids and glasses},\ }\href
  {https://doi.org/10.1103/PhysRevB.28.784} {\bibfield  {journal} {\bibinfo
  {journal} {Phys. Rev. B}\ }\textbf {\bibinfo {volume} {28}},\ \bibinfo
  {pages} {784} (\bibinfo {year} {1983})}\BibitemShut {NoStop}%
\bibitem [{\citenamefont {Bartke}\ and\ \citenamefont
  {Hentschke}(2007)}]{bartke07a}%
  \BibitemOpen
  \bibfield  {author} {\bibinfo {author} {\bibfnamefont {J.}~\bibnamefont
  {Bartke}}\ and\ \bibinfo {author} {\bibfnamefont {R.}~\bibnamefont
  {Hentschke}},\ }\bibfield  {title} {\bibinfo {title} {Phase behavior of the
  stockmayer fluid via molecular dynamics simulation},\ }\href
  {https://doi.org/10.1103/PhysRevE.75.061503} {\bibfield  {journal} {\bibinfo
  {journal} {Phys. Rev. E}\ }\textbf {\bibinfo {volume} {75}},\ \bibinfo
  {pages} {061503} (\bibinfo {year} {2007})}\BibitemShut {NoStop}%
\end{thebibliography}
\end{document}